\documentclass{aa}  
\usepackage{graphicx}
\usepackage{txfonts}
\usepackage{hyperref}

\usepackage{graphicx}	
\usepackage{amsmath}	
\usepackage{amssymb}	

\usepackage{comment}

\newcommand{\drm}{$\Delta$RM}
\newcommand{\drmq}{$\Delta$RM$^2$}
\newcommand{\df}{D$_{1.4 \ \rm GHz}^{144 \ \rm MHz}$}

\begin{document}

   \title{The LOFAR view of intergalactic magnetic fields \\
   with giant radio galaxies}
   \titlerunning{Polarized GRGs with LOFAR}
   \authorrunning{C. Stuardi}

   \author{C. Stuardi
              \inst{1,2,3}\fnmsep\thanks{E-mail: \texttt{chiara.stuardi2@unibo.it}}, S.P. O'Sullivan\inst{3,4}, A. Bonafede\inst{1,2,3}, M. Br\"uggen\inst{3}, P. Dabhade\inst{5,6}, C. Horellou\inst{7}, R. Morganti\inst{8,9}, E. Carretti\inst{2}, G. Heald\inst{10} \and M. Iacobelli\inst{8} \and V. Vacca\inst{11}
          }

   \institute{Dipartimento di Fisica e Astronomia, Universit\`a di Bologna, via Gobetti 93/2, 40122 Bologna, Italy
         \and INAF - Istituto di Radioastronomia di Bologna, Via Gobetti 101, 40129 Bologna, Italy
         \and Hamburger Sternwarte, Universit\"at Hamburg, Gojenbergsweg 112, 21029 Hamburg, Germany
         \and School of Physical Sciences and Centre for Astrophysics \& Relativity, Dublin City University, Glasnevin, D09 W6Y4, Ireland
         \and Leiden Observatory, Leiden University, P.O. Box 9513, NL-2300 RA, Leiden, The Netherlands 
         \and Inter University Centre for Astronomy and Astrophysics (IUCAA), Pune 411007, India
         \and Chalmers University of Technology, Dept of Space, Earth and Environment, Onsala Space Observatory, 439 92 Onsala, Sweden
         \and ASTRON, the Netherlands Institute for Radio Astronomy, Oude Hoogeveensedijk 4, 7991 PD Dwingeloo, The Netherlands 
         \and Kapteyn Astronomical Institute, University of Groningen, P.O. Box 800,9700 AV Groningen, The Netherlands
         \and CSIRO Astronomy and Space Science, PO Box 1130, Bentley, WA, 6012, Australia
         \and INAF - Osservatorio Astronomico di Cagliari, Via della Scienza 5, I-09047 Selargius (CA), Italy
         }

   \date{Received XX; accepted YY}

 
  \abstract
   {Giant radio galaxies (GRGs) are physically large radio sources that extend well beyond their host galaxy environment. Their polarization properties are affected by the poorly constrained magnetic field that permeates the intergalactic medium on Mpc scales. A low frequency (< 200 MHz) polarization study of this class of radio sources is now possible with LOFAR.}
   {Here we investigate the polarization properties and Faraday rotation measure (RM) of a catalog of GRGs detected in the LOFAR Two-metre Sky Survey. This is the first low frequency polarization study of a large sample of radio galaxies selected on their physical size. We explore the magneto-ionic properties of their under-dense environment and probe intergalactic magnetic fields using the Faraday rotation properties of their radio lobes. LOFAR is a key instrument for this kind of analysis because it can probe small amounts of Faraday dispersion ($<$ 1 rad m$^{-2}$) which are associated with weak magnetic fields and low thermal gas densities.}
   {We use RM synthesis in the 120-168 MHz band to search for polarized emission and to derive the RM and fractional polarization of each detected source component. We study the depolarization between 1.4 GHz and 144 MHz using images from the NRAO VLA Sky Survey. We investigate the correlation of the detection rate, the RM difference between the lobes and the depolarization with different parameters: the angular and linear size of the sources and the projected distance from the closest foreground galaxy cluster. We included in our sample also 3C\,236, one of the largest radio galaxies known.}
   {From a sample of 240 GRGs, we detected 37 sources in polarization, all with a total flux density above 56 mJy. We detected significant RM differences between the lobes which would be inaccessible at GHz frequencies, with a median value of $\sim$1 rad m$^{-2}$. The fractional polarization of the detected GRGs at 1.4 GHz and 144 MHz is consistent with a small amount of Faraday depolarization (a Faraday dispersion $<0.3$~rad~m$^{-2}$). Our analysis shows that the lobes are expanding into a low-density (<10$^{-5}$ cm$^{-3}$) local environment permeated by weak magnetic fields (<0.1 $\mu$G) with fluctuations on scales of 3 to 25 kpc. The presence of foreground galaxy clusters appears to influence the polarization detection rate up to 2R$_{500}$. In general, this work demonstrates the ability of LOFAR to quantify the rarefied environments in which these GRGs exist and highlights them as an excellent statistical sample to use as high precision probes of magnetic fields in the intergalactic medium and the Milky Way.}
   {}
   
%
%



\maketitle
 
\section{Introduction}

Radio galaxies that extend to Mpc scales are often defined as giant radio galaxies \citep[GRGs,][]{Willis74}. While earlier authors adopted a lower limit of 1 Mpc to define GRGs assuming $H_0$ = 50 km s$^{-1}$, nowadays the general consensus is to use a limiting size of 0.7 Mpc in order to maintain the classification within the revised cosmology \citep[e.g.,][]{Dabhade17,Kuzmicz18}. GRGs are mostly Fanaroff-Riley type 2 radio galaxies \citep[FR\,II,][]{Fanaroff74}, with the lobes extending well beyond the host galaxy and local environment, and expanding into the surrounding intergalactic medium (IGM). They are particularly interesting objects for the study of different astrophysical problems, ranging from the evolution of radio sources \citep{Ishwara99} to the ambient gas density \citep{Mack98,Malarecki15,Subrahmanyan08}. In particular, Faraday rotation and polarization properties of the lobes/hotspots emission can be used to study the nature of the intergalactic magnetic field \citep[IGMF,][]{O'Sullivan19}. In the future, giant radio galaxies will also be targeted with the Square Kilometre Array (SKA) to probe the warm-hot intergalactic medium \citep[WHIM,][]{Peng15}. 

GRGs are a small subclass of radio galaxies: they constitute about $6\ \%$ of the complete sample of 3CR radio sources \citep{Laing83}. Until recently only a few hundred GRGs have been reported \citep[e.g.,][and references therein]{Kuzmicz18}. The LOFAR Two-metre Sky Survey \citep[LoTSS,][]{Shimwell17,Shimwell19} is one of the best surveys to identify GRGs thanks to its high sensitivity to low surface brightness sources, the high angular resolution, and the high quality associations with optical counterparts including redshifts. Recently, \citet{Dabhade20} reported a large catalog of 239 GRGs, of which 225 were new findings from the LoTSS first data release (DR1). Optical/infrared identifications and redshift estimates are available for all the sample  \citep{Williams19,Duncan19}. 

Polarization observations in the 120-168 MHz band provide exceptional Faraday rotation measure (RM) accuracy due to the large wavelength-square coverage \citep{Brentjens18,VanEck18b}. Despite the technical challenges, preliminary efforts to build a polarization catalog with LOFAR were successfully performed \citep{Mulcahy14,VanEck18a,Neld18}. LOFAR polarization capabilities have been recently shown to be well suited for the study of magnetic fields for different science cases: from the interstellar medium \citep{VanEck19} to the cosmic web \citep{O'Sullivan19,O'Sullivan20}. However, at these low frequencies most of the sources remain undetected in polarization, largely because of Faraday depolarization effects \citep[][]{Burn66,Farnsworth11}. Depolarization is less severe in low-density ionised environments characterized by weak magnetic fields with large fluctuation scales (compared to the resolution of the observations), since it depends on the magnetic field and thermal electron density along the line of sight, and on their spatial gradient within the synthesized beam.

Previous work probed the strong polarization of the lobes of GRGs at low frequencies \citep[e.g.,][]{Willis78,Bridle79,Tsien82,Mack97}. One of the first objects observed in polarization by LOFAR was the double-double giant radio galaxy B1834+620 \citep{Orru15} and, recently, a polarization study of the giant radio galaxy NGC\,6251 was performed with LOFAR (Cantwell et al. 2020, submitted). \citet{Machalski06} also showed that GRGs are less depolarized at 1.4 GHz than normal-sized radio galaxies, indicating the presence of less dense gas surrounding their lobes. Hence, the lobes of GRGs are probably one of the best targets for polarization studies at low frequencies \citep{O'Sullivan18a}. While previous GRGs polarization studies were based on single sources, or at most tens of objects, observed with different facilities, LOFAR allows us to perform the first study on a large sample of hundreds of GRGs that were selected and analyzed consistently.

A low density ($\sim10^{-5}-10^{-6}$ cm$^{-3}$) WHIM permeate the large scale structure of the Universe, from the extreme outskirts of galaxy clusters to filaments \citep{Dave01}. Previous studies demonstrated that lobes of GRGs evolve and interact with the WHIM  \citep{Mack98,Chen11}. In these regions, the IGMF is expected to range from 1 to 100 nG, with the true value being important to discriminate between different magneto-genesis scenarios \citep{Bruggen05,Vazza17,Vernstrom19}. While the detection of both thermal and non-thermal emission of the WHIM is still an observational challenge \citep{Vazza19}, GRGs are potentially indirect probes of these poorly constrained regions of the Universe \citep{Subrahmanyan08}. RM and depolarization information derived from polarized emission of GRGs can yield tomographic information about this extremely rarefied environment \citep{O'Sullivan19}.

While in this work GRGs are mainly exploited for the study of the IGM, the polarization properties of radio galaxies, in general, are crucial for the study of magnetic field structures in lobes and jets. A preliminary census of polarized sources in the LoTSS field was performed by \citet{VanEck18a}. They produced a catalog of 92 point-like sources with a resolution of $4.3\arcmin$ and a sensitivity of 1 mJy/beam within a region of 570 deg$^{2}$. \citet{O'Sullivan18a} analyzed 76 out of the 92 sources residing in the DR1 area with an improved resolution of $20\arcsec$ and \citet{O'Sullivan19} performed a detailed study of the largest radio galaxy in the sample. A complete statistical study of the bulk polarization properties of radio galaxies in the LoTSS DR1 will be presented in Mahatma et al. (2020, in prep.). The aim of our study based on the selection of radio galaxies with large physical size is twofold: on one hand it allows us to complement the work by \citet{Dabhade20} with polarization information on the GRG sample, and, on the other hand, this selection is particularly interesting for the study of IGMF. Small size radio galaxies would be more affected by the host galaxy halo and local environment than GRGs and the detection rate would be strongly reduced by the Faraday depolarization.

Recently, \citet{O'Sullivan20} presented a study of the magnetization properties of the cosmic web comparing the RM difference between lobes of radio galaxies (i.e., physical pairs) and pairs of physically unrelated sources. This work made use of the exceptional RM accuracy of LOFAR and applied the same strategy that \citet{Vernstrom19} implemented to analyze the data at 1.4 GHz of the NRAO VLA Sky Survey \citep[NVSS,][]{Condon98}. The difference in the results obtained by these works is attributed mainly to the Faraday depolarization which made the higher RM variance, detected by the NVSS, undetectable by LOFAR. Here, we can deeply investigate the origin of such depolarization on a well defined sample of sources.

In this paper, we present a polarization and RM analysis of the GRGs detected in the LoTSS DR1 \citep{Dabhade20}, plus one of the largest radio galaxies (3C\,236) observed with LOFAR as part of the ongoing LoTSS \citep{Shulevski19}. The specific nature of the sample analyzed here is that all sources have a physical size larger than 0.7 Mpc. In Sec.~\ref{sec:analysis}, we describe the data reduction, polarization and Faraday rotation analysis, the source identification, and the depolarization study. In Sec.~\ref{sec:results} we present the main properties of the detected sources and we investigate the origins of Faraday rotation and depolarization. In Sec.~\ref{sec:discussion} we discuss the results and their implications for the study of the IGMF. We conclude with a summary in Sec.~\ref{sec:conclusions}. The images of all the detected sources are shown in Appendix~\ref{appendix}. Throughout this paper, we assume a $\Lambda$CDM cosmological model, with $H_0$ = 67.8 km s$^{-1}$ Mpc$^{-1}$, $\Omega_{\rm M}$ = 0.308, $\Omega_{\Lambda}$ = 0.692 \citep{Planck16}. 

\section{Data analysis}
\label{sec:analysis}

Our work is based on the data from the LoTSS, fully described by \citet{Shimwell17,Shimwell19}. This ongoing survey is covering the entire northern sky with the LOFAR High-Band Antenna (HBA) at frequencies from 120 to 168 MHz. The LoTSS DR1 consists of images at 6$\arcsec$ resolution and a sensitivity of $\sim$70 $\mu$Jy/beam. It covers 424 deg$^{2}$ in the region of the HETDEX Spring field (i.e., 2$\%$ of the northern sky). The observing time for each pointing is $\sim$8 hours and the FWHM of the primary beam is $\sim4^{\circ}$. Although our work is mainly based on the GRG catalog by \citet{Dabhade20} which is located in the DR1 region, we make use of the updated data products from the upcoming LoTSS second data release (DR2).

\subsection{Calibration and Data Reduction}
\label{sec:reduction}

We refer the reader to \citet{Shimwell17} for the full details on the calibration and data reduction. Here we summarize only the main steps.

For our analysis we used images at 20$\arcsec$ and 45$\arcsec$ resolution. The choice of a restoring beam larger than $6\arcsec$ (used for the LoTSS DR1) was meant to maximize the sensitivity to the extended emission of the lobes. The 20$\arcsec$ resolution images from the upcoming LoTSS DR2 pipeline (Tasse et al. 2020, in prep.) were used to identify polarized sources and record the position, polarized flux density, fractional polarization, and RM of the pixels with the highest signal-to-noise ratio (see Sec.~\ref{sec:crossmatch}). The $45\arcsec$ resolution images of the detected sources were instead necessary to compare with images at 1.4 GHz and perform the depolarization analysis (see Sec.~\ref{sec:depol}). We used two different strategies for calibration and imaging at the two resolutions to cross-check the reliability from the \texttt{ddf-pipeline}\footnote{https://github.com/mhardcastle/ddf-pipeline} \citep{Tasse14,Tasse18,Shimwell19} output and also to enable deconvolution in Stokes $Q$ and $U$ at 45$\arcsec$. We obtained reliable calibration and imaging performance with both procedures, described in the following.

Direction-dependent calibration was performed using the \texttt{ddf-pipeline}. Direction-dependent calibrated data were used for the total intensity images at 20$\arcsec$ resolution in order to better resolve the morphological properties of the sources. These data were also used to image Stokes $Q$ and $U$ frequency channel cubes at 20$\arcsec$ resolution. 

We made low resolution 45$\arcsec$ images of the GRGs that were detected in polarization at 20$\arcsec$ (see Sec.~\ref{sec:crossmatch}). Only direction-independent calibration was performed using \texttt{PREFACTOR\,1.0}\footnote{https://github.com/lofar-astron/prefactor} \citep{vanWeeren16a,Williams16}. This procedure is robust, because of the absence of any large direction-dependent artifacts in the Q and U images, and allows us to deconvolve the emission at $45\arcsec$ without re-running the entire calibration on the full LoTSS field where a GRG has been detected. The rms noise level was on average one order of magnitude larger at $45\arcsec$ than at $20\arcsec$ due to uv-cut and down-weighting of data on the longer baselines. The direction-independent calibrated data were phase-shifted to the source location and averaged to 40 s (from 8 s) to speed up the imaging and deconvolution process \citep[as in, e.g.,][]{Neld18,O'Sullivan19}.

The ionospheric RM correction was applied with \texttt{RMextract}\footnote{https://github.com/lofar-astron/RMextract} \citep{Mevius18}. Residual ionospheric RM correction errors are estimated to be $\sim0.05$ rad m$^{-2}$ between observations and $\sim0.1-0.3$ rad m$^{-2}$ across the 8h observations \citep{Sotomayor-Beltran13,VanEck18a}.

\subsection{Polarization and Faraday rotation imaging}
\label{sec:polimaging}

The $Q$ and $U$ images at $20\arcsec$ resolution were not deconvolved because this procedure is not yet implemented in the \texttt{ddf-pipeline}. Although some of the RM structure for the brightest polarized sources is dominated by spurious structure, this should not affect our analysis since we used the RM value at the peak of the polarized emission. We used \texttt{WSCLEAN\,2.4}\footnote{\url{https://sourceforge.net/p/wsclean/wiki/Home/}} \citep{Offringa14} to deconvolve the $Q$ and $U$ images at 45$\arcsec$ resolution, in order to directly compare with polarization images from the NVSS at 1.4 GHz \citep{Condon98}. In $90\ \%$ of the cases, we obtained consistent RMs at $45\arcsec$ and $20\arcsec$. We found a larger scatter in the values obtained at low resolution, as expected due to the larger beam and higher noise.

We created 480 $Q$ and $U$ frequency channel images with 0.1 MHz resolution between 120 and 168 MHz with a fixed restoring beam (20$\arcsec$ or 45$\arcsec$). The primary beam correction was applied to each channel. The total intensity ($I$) image was created using the entire band at the central frequency of 144 MHz and then corrected for the primary beam. All pixels below 1 mJy/beam in total intensity (for which no fractional polarization $<50\ \%$ can be detected due to the LoTSS sensitivity) were masked out to speed up the subsequent analysis.

\begin{figure}
\includegraphics[width=0.5\textwidth]{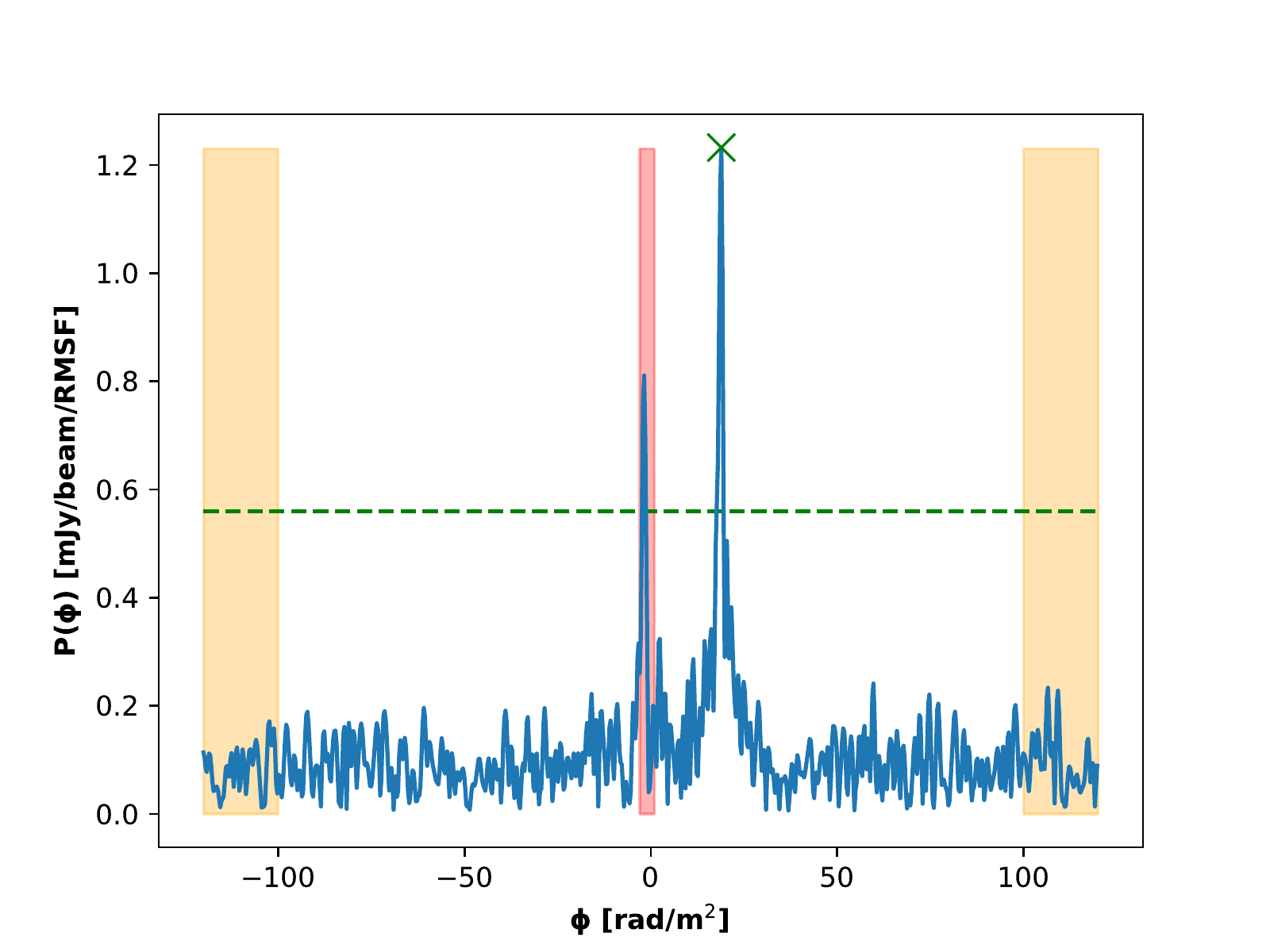}
  \caption{Example Faraday spectrum. In particular, this is the Faraday spectrum obtained at the polarized peak position of the lobe ‘‘b’’ of GRG\,2 (see Tab.~\ref{tab:pol_double}). The red shadowed area shows the region of the spectrum excluded due to the instrumental leakage contamination. The orange areas show the range used to compute the rms noise from the $Q$ and $U$ Faraday spectra. The green dashed line highlights the 8$\sigma$ detection threshold. The green ‘‘X’’ marks the position of the peak from which we derived the RM and $P$ values of the pixel.}
  \label{fig:spectrum}
\end{figure}

We performed RM synthesis \citep{Brentjens05} on the $Q$ and $U$ per-channel cubes using \texttt{PYRMSYNTH}\footnote{\url{https://github.com/mrbell/pyrmsynth}} to obtain the cubes in the Faraday depth ($\phi$) space. In these cubes every pixel contains the Faraday spectrum along the line of sight, i.e., the polarized intensity at each Faraday depth \citep[see, e.g.,][for the used terminology]{Stuardi19}.  An example Faraday spectrum extracted from the peak of polarized intensity of a source is shown in Fig.~\ref{fig:spectrum}. RM clean was also performed on the $45\arcsec$ cubes \citep{Heald09}.  

Considering the LoTSS bandwidth and the adopted channelization, using \citet{Brentjens05} we can estimate our resolution in Faraday space, $\delta\phi$ = 1.16 rad m$^{-2}$, the maximum observable Faraday depth, $|\phi_{\rm max}|$ = 168 rad m$^{-2}$, and the largest observable scale in Faraday space, $\Delta\phi_{\rm max}$ = 0.97 rad m$^{-2}$. As a consequence, with the LoTSS we can detect only emission that is unresolved in Faraday depth. Faraday cubes were created between -120 and 120 rad m$^{-2}$ and sampled at 0.3 rad m$^{-2}$. The Faraday range was chosen considering that RM values for sources at high Galactic latitude (above $b > 55^{\circ}$) and outside galaxy cluster environments are a few tens of rad m$^{-2}$ \citep[see, e.g.,][]{Bohringer16}.

The LOFAR calibration software (i.e. \texttt{PREFACTOR\,1.0}) does not allow instrumental polarization leakage correction so that peaks appear in the Faraday spectrum at the level of $\sim1.5\ \%$ of the total intensity in the range $-3<\phi<1$ rad m$^{-2}$ (see Fig.~\ref{fig:spectrum}). This asymmetric range is due to the ionospheric RM correction that shifts the leakage peak along the Faraday spectrum \citep{VanEck18a}. We thus excluded this range in order to avoid a contamination from the instrumental leakage as done by other authors \citep[e.g.,][]{Neld18,O'Sullivan19}. This method systematically excludes from this analysis all real polarized sources within this Faraday depth range. We fitted pixel-by-pixel a parabola around the main peak of the Faraday spectrum outside the excluded range. We obtained the RM and polarized intensity ($P$) images from the position of the parabola vertex in each pixel. For each pixel we computed the noise, $\sigma_{QU}$, as the standard deviation in the outer $20\ \%$ of the $Q$ and $U$ Faraday spectra and we imposed an initial 6$\sigma_{QU}$ detection threshold, which ensures an equivalent 5$\sigma$ Gaussian significance \citep{Hales12}. We also computed the fractional polarization ($p$) images by dividing the polarization image $P$ obtained from the RM-synthesis by the full-band total intensity image $I$ (with a 3$\sigma$ detection threshold, where $\sigma$ is the local rms noise). We computed the fractional polarization error map by propagating the uncertainties on $P$ and $I$ images. 

The RM error map was computed as $\delta\phi$ divided by twice the signal-to-noise of the detection \citep{Brentjens05}. This formula is computed for zero spectral index and equal rms noise in Stokes $Q$ and $U$ and it can be used as a reference value. Furthermore, the computed error does not include the systematic error from the ionospheric RM correction \citep[$\sim$0.1 rad m$^{-2}$,][]{VanEck18a}.

\subsection{Source identification}
\label{sec:crossmatch}

\begin{table*}
\caption{Polarized GRGs.}
  \centering
\begin{tabular}{rllrcccc}
\hline
  \multicolumn{1}{c}{GRG} &
  \multicolumn{1}{c}{R.A.} &
  \multicolumn{1}{c}{Dec} &
  \multicolumn{1}{c}{$z$} &
  \multicolumn{1}{c}{Ang. size} &
  \multicolumn{1}{c}{Lin. size} &
  \multicolumn{1}{c}{FR} &
  \multicolumn{1}{c}{Remark} \\
  \multicolumn{1}{c}{} &
  \multicolumn{1}{c}{(deg)} &
  \multicolumn{1}{c}{(deg)} &
  \multicolumn{1}{c}{} &
  \multicolumn{1}{c}{(arcsec)} &
  \multicolumn{1}{c}{(Mpc)} &
  \multicolumn{1}{c}{} &
  \multicolumn{1}{c}{} \\
\hline
1 & 164.273 & 53.440 & 0.460\tablefootmark{a} & 153 & 0.92 & II & d\\
  2 & 164.289 & 48.678 & 0.276\tablefootmark{a} & 439 & 1.9 & II & d\\
  7 & 164.575 & 51.672 & 0.415\tablefootmark{a} & 330 & 1.86 & II & s\\
  19 & 167.402 & 53.230 & 0.288\tablefootmark{b} & 230 & 1.03 & II & d\\
  22 & 168.381 & 46.371 & 0.589\tablefootmark{b} & 112 & 0.76 & II & d\\
  44 & 174.882 & 47.357 & 0.518\tablefootmark{a} & 312 & 2.0 & II & s\\
  47 & 178.000 & 49.849 & 0.891\tablefootmark{a} & 96 & 0.77 & II & s\\
  51 & 180.345 & 49.427 & 0.205\tablefootmark{b} & 345 & 1.2 & I & d\\
  57 & 182.692 & 53.490 & 0.448\tablefootmark{a} & 119 & 0.71 & I & s\\
  64 & 184.576 & 53.456 & 0.568\tablefootmark{c} & 183 & 1.23 & II & d\\
  65 & 184.708 & 50.438 & 0.199\tablefootmark{a} & 210 & 0.71 & II & d\\
  77 & 186.493 & 53.161 & 0.811\tablefootmark{c} & 147 & 1.14 & II & d\\
  80 & 187.498 & 53.546 & 0.523\tablefootmark{c} & 137 & 0.88 & II & s\\
  83 & 188.210 & 49.107 & 0.690\tablefootmark{a} & 256 & 1.87 & II & s\\
  85 & 188.756 & 53.299 & 0.345\tablefootmark{d} & 683 & 3.44 & II & d\\
  87 & 189.202 & 46.068 & 0.615\tablefootmark{b} & 125 & 0.87 & II & d\\
  91 & 190.052 & 53.577 & 0.293\tablefootmark{a} & 164 & 0.74 & II & d\\
  103 & 195.396 & 54.136 & 0.313\tablefootmark{b} & 168 & 0.79 & II & d\\
  112 & 197.620 & 52.228 & 0.650\tablefootmark{b} & 197 & 1.41 & II & s\\
  117 & 199.144 & 49.544 & 0.563\tablefootmark{b} & 126 & 0.84 & II & core\\
  120 & 200.124 & 49.280 & 0.684\tablefootmark{a} & 113 & 0.82 & II & d\\
  122 & 200.902 & 47.497 & 0.440\tablefootmark{b} & 180 & 1.05 & II & s\\
  136 & 203.345 & 53.547 & 0.354\tablefootmark{b} & 173 & 0.88 & - & s\\
  137 & 203.549 & 55.024 & 1.245\tablefootmark{a} & 91 & 0.78 & II & s\\
  144 & 204.845 & 50.963 & 0.316\tablefootmark{b} & 174 & 0.83 & II & d\\
  145 & 205.263 & 49.267 & 0.747\tablefootmark{c} & 113 & 0.85 & II & d\\
  148 & 206.065 & 48.764 & 0.725\tablefootmark{b} & 202 & 1.51 & II & s\\
  149 & 206.174 & 50.383 & 0.763\tablefootmark{a} & 123 & 0.93 & II & s\\
  165 & 210.731 & 51.458 & 0.518\tablefootmark{c} & 135 & 0.87 & II & d\\
  166 & 210.813 & 51.746 & 0.485\tablefootmark{c} & 228 & 1.41 & II & d\\
  168 & 211.421 & 54.182 & 0.761\tablefootmark{c} & 116 & 0.88 & II & d\\
  177 & 213.535 & 48.699 & 1.361\tablefootmark{b} & 107 & 0.92 & II & d\\
  207 & 220.033 & 55.452 & 0.584\tablefootmark{c} & 238 & 1.62 & II & s\\
  222 & 222.739 & 53.002 & 0.918\tablefootmark{a} & 184 & 1.48 & II & d\\
  233 & 226.190 & 50.502 & 0.652\tablefootmark{c} & 201 & 1.44 & II & d\\
  234 & 226.553 & 51.619 & 0.611\tablefootmark{a} & 262 & 1.82 & II & s\\
  0$^{*}$ & 151.507 & 34.903 & 0.1005\tablefootmark{e} & 2491 & 4.76 & II & d\\
\hline
\end{tabular}
\tablefoot{
Column 1: progressive GRG identification number from Tab. 2 in \citet{Dabhade20}; Column 2 and 3: J2000 celestial coordinates of the host galaxy. The reference is \citet{Dabhade20} for all the GRGs apart from GRG\,0 for which we refer to \citet{Becker95}; Column 4: redshift ($z$); Column 5 and 6: angular and projected linear size; Column 7: Fanaroff-Riley type \citep{Fanaroff74}. GRG\,136 has a peculiar morphology and thus it is not classified; Column 8: the letter indicates if the GRG is detected as a double (``d'') or a single (``s'') source in polarization. Polarized emission was detected from the core/inner jets region only in the case of GRG\,117.\\
\tablefoottext{a}{Spectroscopic redshifts from the Sloan Digital Sky Survey (SDSS, \citep{York00}}
\tablefoottext{b}{Redshifts from the LoTSS DR1 value-added catalog \citep{Williams19,Duncan19}}
\tablefoottext{c}{Photometric redshifts from the SDSS.}
\tablefoottext{d}{Spectroscopic redshift from \citet{O'Sullivan19}}
\tablefoottext{e}{Spectroscopic redshift from \citet{Hill96}. }
\tablefoottext{*}{GRG\,0 is 3C\,236 that was added to the \citet{Dabhade20} catalog for this analysis.}
}
  \label{tab:GRG_all}
\end{table*} 

Using the $20\arcsec$ images we compiled a catalog of polarized sources in the LoTSS. Each source is represented by the pixel with the highest signal-to-noise ratio within a $\sim$5-beam-size region above the 6$\sigma_{QU}$ threshold. For each source we listed the sky coordinates, the polarization signal-to-noise level, the fractional polarization, the RM value, and the separation from the pointing center in degree. When the same source was detected in several pointings of the survey, we selected the image with the highest signal-to-noise and closest to the pointing center.

We cross-matched our catalog with the catalog of 239 GRGs in the LoTSS DR1 compiled by \citet{Dabhade20} choosing different radii to match the angular size of the sources. The cross-match resulted in 51 GRGs showing radio emission coincident with at least one entry in the polarization catalog. Through a careful visual inspection, we excluded 15 sources for which polarization was detected in less than four pixels with signal-to-noise lower than 8 and only in one pointing of the survey (or in two pointings but with different RM values). The final detection threshold in polarization is thus 8$\sigma_{QU}$: this conservative choice is motivated, both, by the literature \citep[see, e.g.,][]{George12,Hales12} and by our experience with RM synthesis data. The 36 GRGs clearly detected in polarization are listed in Tab.~\ref{tab:GRG_all}. The GRG numbers refer to the source numbers in the \cite{Dabhade20} catalog. In Tab.~\ref{tab:GRG_all} we also added 3C\,236: it is one of the largest radio galaxies known \citep{Willis74} and, although it was not present in the LoTSS DR1, it was recently observed by LOFAR \citep{Shulevski19}. Hereafter we will refer to this source as GRG\,0.

\subsection{Faraday depolarization}
\label{sec:depol}

We used the images of the NVSS in order to estimate the amount of Faraday depolarization between 1.4 GHz and 144 MHz. To match the NVSS resolution, we used the 144 MHz images at 45$\arcsec$. We find that $8.5\ \%$ of the sources detected at 144 MHz are not detected by the NVSS due to the lower sensitivity of this survey compared to the LoTSS. For some sources, the polarized emission is not exactly co-spatially located at the two frequencies but always separated by less than a single beam-width of $45\arcsec$ (see Appendix~\ref{appendix}). 

For each component (i.e., lobes and hotspots of single and double detections, and the core/inner jets of GRG\,117), we estimated the depolarization factor, \df, as the ratio between the degree of polarization at 144 MHz (at the peak polarized intensity location at 45$\arcsec$) and the degree of polarization in the NVSS image at the same location. When there was an offset between LOFAR and NVSS detection, we chose the brightest LOFAR pixel in the overlapping region to compute the depolarization factor. With this definition, \df=1 means no depolarization while lower values of \df indicate stronger depolarization.

\section{Results}
\label{sec:results}

\begin{figure}
\centering
\includegraphics[width=0.45\textwidth]{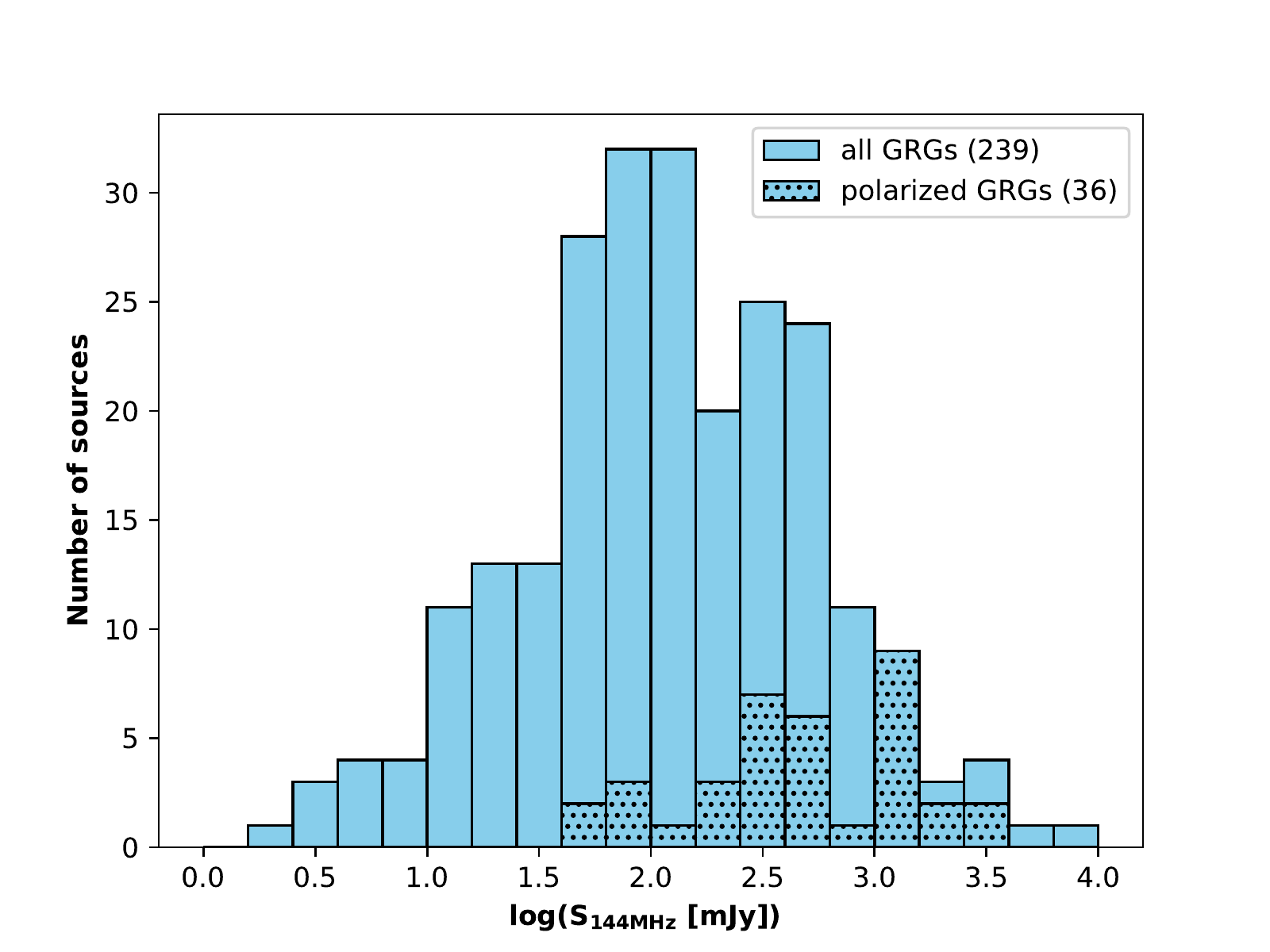}
\includegraphics[width=0.45\textwidth]{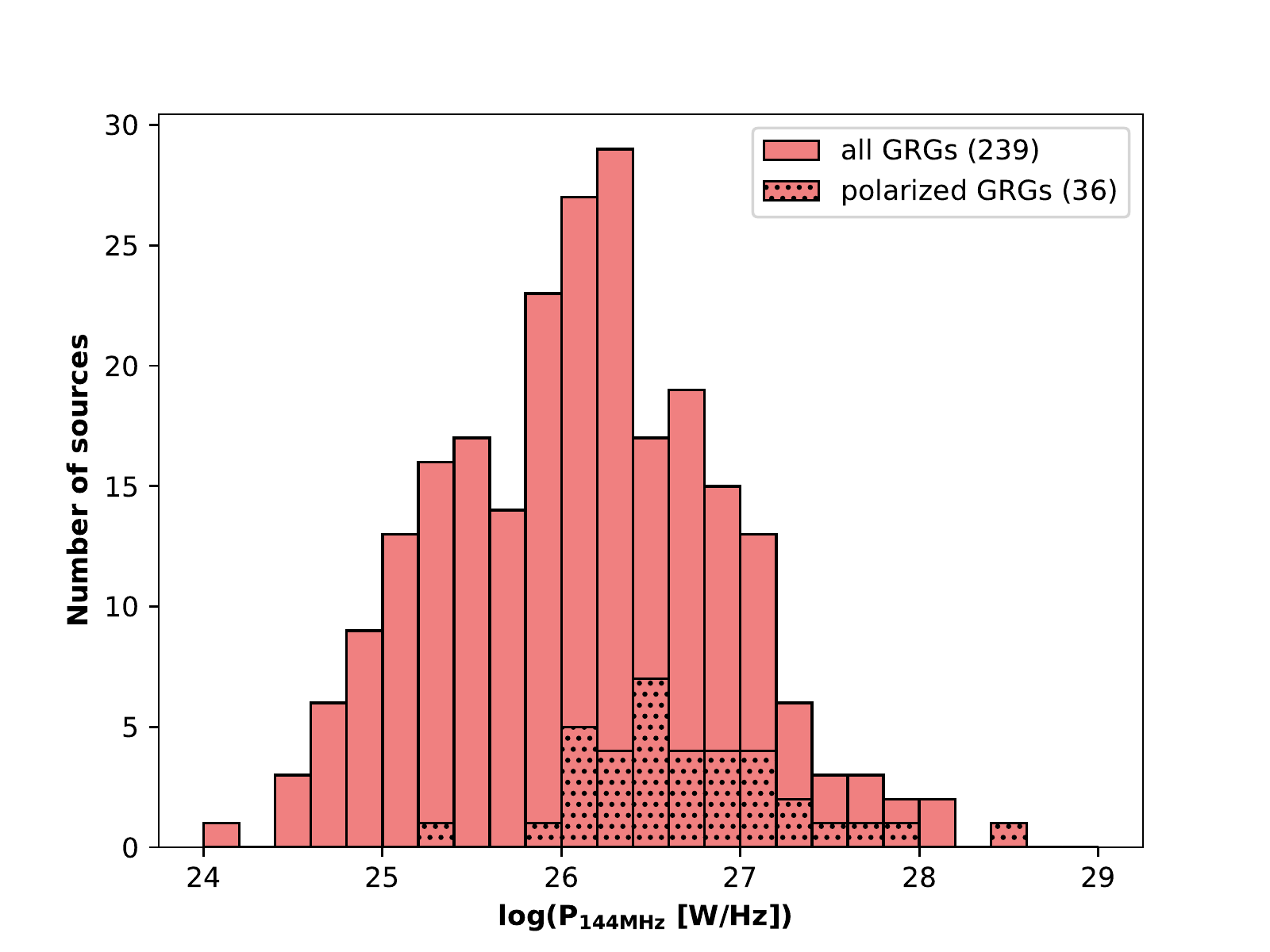}
\includegraphics[width=0.45\textwidth]{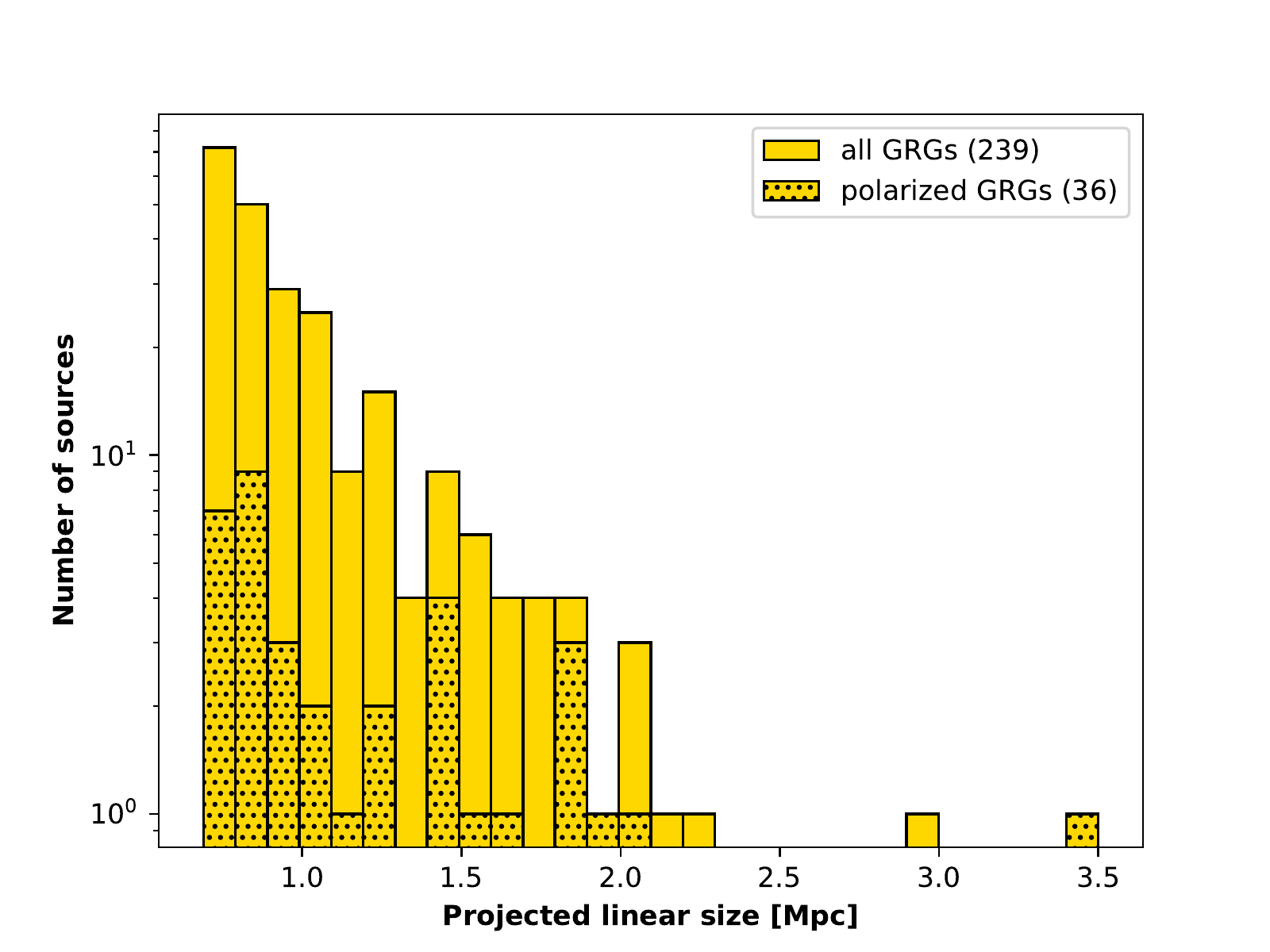}
  \caption{Flux density (top), radio power (center), and projected linear scale (bottom) distributions of the LoTSS DR1 GRG catalog \citep{Dabhade20} compared with the 36 GRGs detected in polarization at 144 MHz within this sample.}
  \label{fig:all_hist}
\end{figure}

\begin{figure}
\centering
\includegraphics[width=0.45\textwidth]{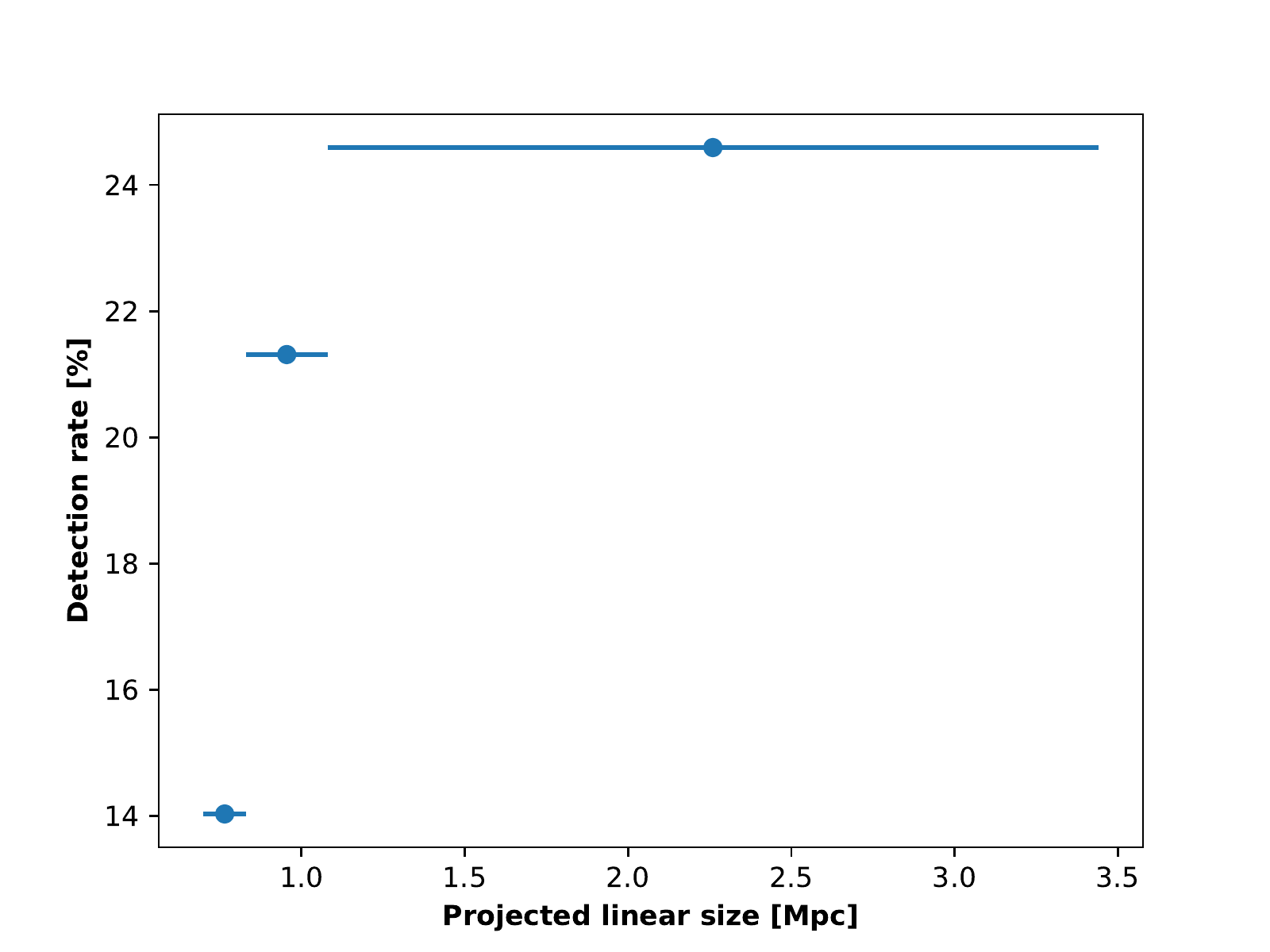} \caption{Detection rate as function of the projected linear size of the GRGs from the distribution shown in the bottom panel of Fig.~\ref{fig:all_hist}. The widths of the bins are computed to contain the same total number of sources ($\sim$60). Markers are positioned at center of each bin and the  error bars show the bin width.}
 \label{fig:detrate_linsize}
\end{figure}

The 37 GRGs are displayed in Fig.~\ref{fig:GRG1}. Contours show the total intensity. The left-hand panel is the total intensity image at 20$\arcsec$ resolution, the central panel is the LOFAR fractional polarization at 45$\arcsec$ resolution, the right-hand panel is the NVSS fractional polarization at 45$\arcsec$. The color scale and limits are the same per source for both fractional polarization images. In all three images the cyan squares mark the component detected at 20$\arcsec$, cyan points mark the peak of polarized intensity at 20$\arcsec$ (where we derived the RM and fractional polarization values) while magenta points mark the position where we extracted the depolarization factors. The separation between these two points is always within the 45$\arcsec$ resolution element.

3C\,236 (GRG\,0) was not present in the original GRG catalog by \citet{Dabhade20}. Since it was selected only because its polarization at low frequencies was studied in previous work \citep[e.g.,][]{Mack97}, it is not included in the following paragraphs where we compute the polarization detection rates.

Out of the 36 polarized sources in the GRG catalog, 33 are FR\,II type sources, 2 are FR\,I (i.e., GRG\,51 and GRG\,57), and GRG\,136 has a peculiar morphology (see Tab.~\ref{tab:GRG_all}). Only 6 of them have a quasar host, while all the others are radio galaxies \citep{Dabhade20}. In 75$\%$ of cases the detection is coincident with the hotspots of FR\,II radio galaxies. This is consistent with the fact that compact emission regions probe smaller Faraday depth volumes and are thus less depolarized. In 19$\%$ of cases, the polarized emission is detected from the more diffuse lobe regions. In these cases, the hotspots may have a lower intrinsic fractional polarization than the lobes. In one case (GRG\,117) we detected polarization coincident with the core within our spatial resolution. Since the core of a radio galaxy is not expected to be significantly polarized, this may be a restarted radio galaxy \citep[e.g.,][]{Mahatma19} with polarized emission arising from the unresolved inner jets. The other detections are from the outer edge of FR\,I type galaxies and from the extended lobe of the peculiar GRG\,136.

The histogram distributions of total radio flux density, total radio power, and projected linear size of the whole sample of 239 GRGs are shown in Fig.~\ref{fig:all_hist}, together with the distribution of polarized ones. The GRGs detected in polarization have S$_{\rm 144  MHz}\geq$ 56 mJy in total intensity, suggesting a selection effect due to the sensitivity of the survey. Out of the 239 GRGs in the parent sample, 179 sources have S$_{\rm 144  MHz}>$ 50 mJy: above this threshold the detection rate is thus the $20.1\ \%$. With a lower flux density limit of 10 mJy (i.e., 223 GRGs), the detection rate is  $16.1\ \%$. 

The preliminary LoTSS polarized point-source catalog compiled by \citet{VanEck18a} obtained a $\ll 1 \%$ polarization detection rate for all the sources in the DR1 with total flux densities above 10 mJy \citep[see also][]{O'Sullivan18a}. Our results cannot be directly compared with this work because of the different resolution and the peculiar nature of GRGs. While the majority of the sources in our sample has a large physical and also angular extent, the detection rate computed by \citet{VanEck18a} takes into account more compact sources. Furthermore, \citet{VanEck18a} used preliminary LoTSS images with 4.3$\arcmin$ angular resolution. In-beam depolarization, due to the mixing of different lines-of-sight into the same resolution element, can substantially affect the detection rate. Despite their large physical size, only 29 GRGs out of 239 are larger than 4.3$\arcmin$. All the others are unresolved in the \citet{VanEck18a} catalog, and thus suffer from the same in-beam depolarization as more compact AGN. To better compare our work with \citet{VanEck18a} we cross-matched the position of the 195 GRGs with angular size lower than 4.3$\arcmin$ with the point source catalog compiled by \citet{VanEck18a}. The cross-match resulted in 11 sources, which are also detected in polarization in this work with $20\arcsec$ resolution. The polarization detection rate of the unresolved GRGs in the \citet{VanEck18a} catalog is thus $5.6\ \%$ (11/195). A parent population with large physical size has a higher polarization detection rate than the overall AGN population, even if not resolved. The high detection rate within the GRGs sample suggests the presence of a small amount of depolarization (see also Sec.~\ref{sec:depol}). Out of the 29 GRGs larger than $4.3\arcmin$, and thus also resolved in the \citet{VanEck18a} catalog, four are cataloged as point-sources while only GRG\,85 has both lobes detected in polarization. We refer the reader to Mahatma et al. (2020, in prep.) for a more complete statistical study of the polarization properties and detection rate of radio galaxies within the LoTSS DR1.

The central panel of Fig.~\ref{fig:all_hist} shows a clear selection effect for GRGs with high total radio power. The median radio power of GRGs detected in polarization is 4.07$\times$10$^{26}$ W/Hz while it is 1.03$\times$10$^{26}$ W/Hz for undetected sources (1.8$\times$10$^{26}$ W/Hz considering only sources with flux density above 50 mJy). 

The fraction of GRGs detected in polarization increases with the linear size of the source (see Fig.~\ref{fig:detrate_linsize}), being $31\%$ for the GRGs with physical sizes larger than 1.5 Mpc. This points to a possible decrease in the amount of Faraday depolarization far from the local environment of the host galaxy. In fact, Faraday depolarization decreases far away from the host galaxy and possible groups or clusters of galaxies \citep{Strom88,Machalski06}. However, this effect is conflated with the fact that the majority of sources with linear sizes larger than 1.5 Mpc have high radio power. Using the Kolmogorov-Smirnov (KS) test to compare the linear sizes we found a marginal difference between the samples of detected and undetected GRGs with S$_{\rm 144  MHz}>$ 50 mJy ($p$-value of 0.08). Although beam depolarization may also have a role, the KS test between the angular sizes of detected and undetected sources with S$_{\rm 144  MHz}>$ 50 mJy suggests that they are drawn from a similar distribution ($p$-value of 0.29).

\cite{Dabhade20} found 21/239 GRGs to be associated with brightest cluster galaxies (BCGs) by cross-matching their catalog with the \citet{Wen12} and \citet{Hao10} clusters catalogs. None of them are detected in polarization apart from GRG\,85, whose polarization properties were already studied \citep{O'Sullivan19}. It has a linear size of 3.4 Mpc and probably resides in a small group of galaxies. The localization of the sources in galaxy group or cluster environments seems to be an exclusion criterion for polarization detection at 144 MHz, and this is likely due to the effect of Faraday depolarization.

\begin{figure*}
\centering
\includegraphics[width=0.9\textwidth]{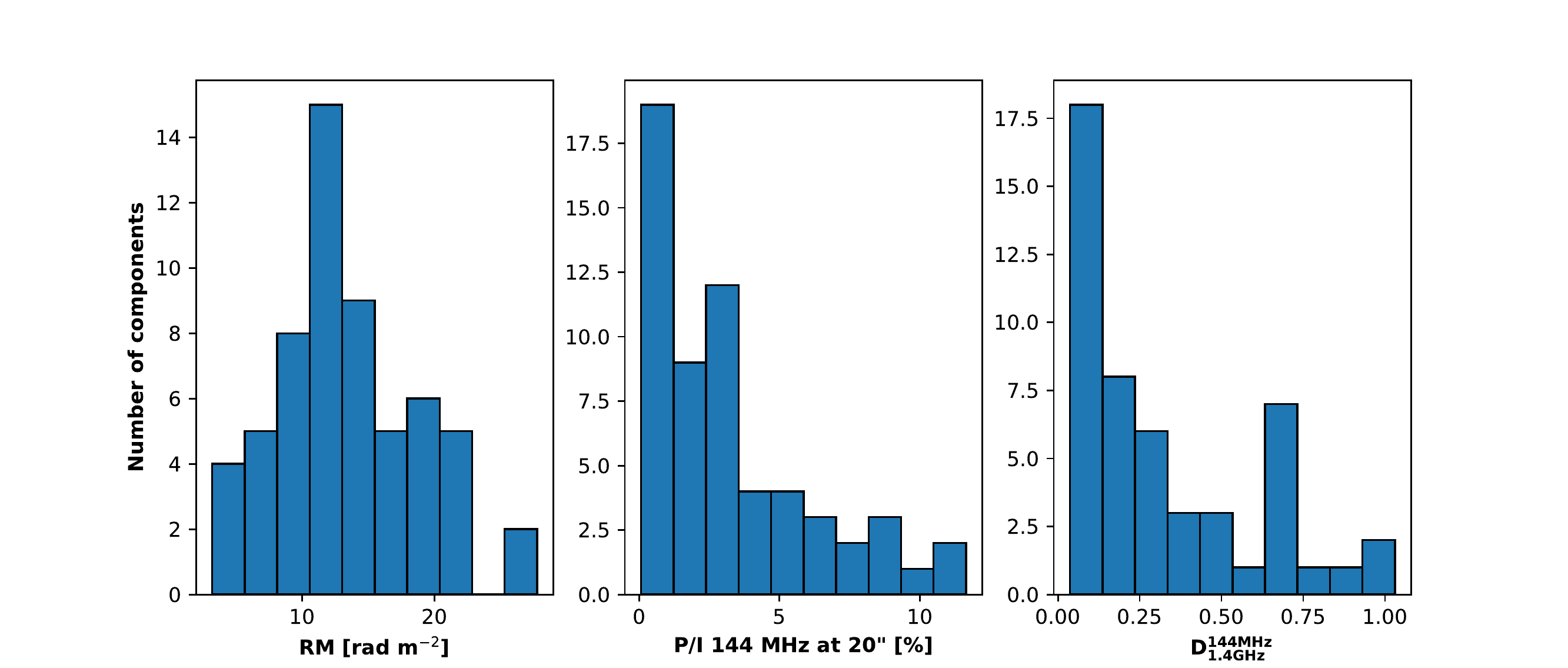}
  \caption{Distribution of Faraday rotation measure (left), fractional polarization (center), and depolarization factor between 1.4 GHz and 144 MHz (right) of the 59 components (lobes, hotspots, and core) detected in polarization.}
  \label{fig:pol_hist}
\end{figure*}

Polarization, Faraday rotation, and depolarization information for all sources are reported in Tab.~\ref{tab:pol_double} when both the lobes are detected, and in Tab.~\ref{tab:pol_single} when only one source component is detected. The histograms of RM and fractional polarization of the detected components (considering both lobes and hotspots of single and double detections) are shown in Fig.~\ref{fig:pol_hist}.

\begin{table*}
\centering
\caption{Results of the polarized intensity study of detected double-lobed sources.}
\begin{tabular}{lllccrrr}
\hline
\multicolumn{1}{c}{GRG} &
  \multicolumn{1}{c}{R.A.} &
  \multicolumn{1}{c}{Dec.} &
  \multicolumn{1}{c}{P} &
  \multicolumn{1}{c}{$\sigma_{QU}$} &
  \multicolumn{1}{c}{$p$} &
  \multicolumn{1}{c}{RM} &
  \multicolumn{1}{c}{\df} \\
\multicolumn{1}{c}{} &
  \multicolumn{1}{c}{(deg)} &
  \multicolumn{1}{c}{(deg)} &
  \multicolumn{1}{c}{(mJy)} &
  \multicolumn{1}{c}{(mJy/beam)} &
  \multicolumn{1}{c}{($\%$)} &
  \multicolumn{1}{c}{(rad/m$^{2}$)} &
  \multicolumn{1}{c}{} \\
\hline
 0a & 151.228 & 35.026 & 4.5 & 0.2 & 11.7 $\pm$ 0.5 & 3.23 $\pm$ 0.02 & 0.7 $\pm$ 0.2\\
  0b & 151.918 & 34.687 & 26.2 & 0.3 & 5.40 $\pm$ 0.06 & 9.071 $\pm$ 0.006 & 0.126 $\pm$ 0.007\\
  1a & 164.276 & 53.430 & 44.0 & 0.2 & 5.28 $\pm$ 0.02 & 12.855 $\pm$ 0.002 & 0.83 $\pm$ 0.02\\
  1b & 164.264 & 53.448 & 4.69 & 0.08 & 2.57 $\pm$ 0.05 & 12.20 $\pm$ 0.01 & 0.167 $\pm$ 0.008\\
  2a & 164.257 & 48.613 & 14.83 & 0.09 & 8.56 $\pm$ 0.05 & 16.940 $\pm$ 0.003 & 0.70 $\pm$ 0.02\\
  2b & 164.339 & 48.725 & 1.23 & 0.07 & 0.67 $\pm$ 0.04 & 19.01 $\pm$ 0.04 & 0.072 $\pm$ 0.007\\
  19a & 167.363 & 53.255 & 1.5 & 0.2 & 3.2 $\pm$ 0.3 & 11.18 $\pm$ 0.06 & 0.5 $\pm$ 0.1\\
  19b & 167.422 & 53.211 & 1.3 & 0.2 & 0.75 $\pm$ 0.09 & 11.39 $\pm$ 0.07 & 0.088 $\pm$ 0.007\\
  22a & 168.399 & 46.381 & 0.87 & 0.09 & 1.4 $\pm$ 0.2 & 4.04 $\pm$ 0.06 & \\
  22b & 168.381 & 46.364 & 0.48 & 0.07 & 0.9 $\pm$ 0.1 & 4.57 $\pm$ 0.09 & \\
  51a & 180.311 & 49.384 & 0.96 & 0.09 & 7.4 $\pm$ 0.7 & 22.03 $\pm$ 0.05 & 0.13 $\pm$ 0.04\\
  51b & 180.380 & 49.458 & 3.1 & 0.1 & 10.3 $\pm$ 0.3 & 22.70 $\pm$ 0.02 & 0.40 $\pm$ 0.09\\
  64a & 184.574 & 53.441 & 1.9 & 0.1 & 0.32 $\pm$ 0.02 & 15.30 $\pm$ 0.03 & 0.062 $\pm$ 0.007\\
  64b & 184.569 & 53.477 & 1.8 & 0.1 & 1.28 $\pm$ 0.07 & 14.57 $\pm$ 0.03 & 0.21 $\pm$ 0.07\\
  65a & 184.659 & 50.431 & 33.6 & 0.2 & 3.21 $\pm$ 0.02 & 27.784 $\pm$ 0.003 & 0.72 $\pm$ 0.02\\
  65b & 184.742 & 50.445 & 17.0 & 0.1 & 3.00 $\pm$ 0.02 & 26.682 $\pm$ 0.005 & 0.43 $\pm$ 0.01\\
  77a & 186.468 & 53.153 & 0.8 & 0.1 & 0.73 $\pm$ 0.09 & 13.10 $\pm$ 0.08 & 0.07 $\pm$ 0.02\\
  77b & 186.514 & 53.168 & 1.25 & 0.09 & 3.5 $\pm$ 0.3 & 11.90 $\pm$ 0.04 & \\
  85a & 188.648 & 53.376 & 5.95 & 0.1 & 4.41 $\pm$ 0.09 & 7.51 $\pm$ 0.01 & 0.64 $\pm$ 0.07\\
  85b & 188.853 & 53.247 & 1.0 & 0.1 & 4.5 $\pm$ 0.4 & 10.08 $\pm$ 0.06 & 0.12 $\pm$ 0.01\\
  87a & 189.208 & 46.064 & 1.6 & 0.1 & 3.1 $\pm$ 0.2 & 21.44 $\pm$ 0.04 & 0.18 $\pm$ 0.03\\
  87b & 189.190 & 46.083 & 0.8 & 0.1 & 1.4 $\pm$ 0.2 & 16.92 $\pm$ 0.08 & 0.08 $\pm$ 0.02\\
  91a & 190.090 & 53.581 & 11.2 & 0.1 & 2.86 $\pm$ 0.03 & 17.952 $\pm$ 0.006 & 0.185 $\pm$ 0.006\\
  91b & 190.027 & 53.573 & 10.35 & 0.09 & 3.02 $\pm$ 0.03 & 19.353 $\pm$ 0.005 & 0.88 $\pm$ 0.09\\
  103a & 195.379 & 54.130 & 4.53 & 0.07 & 1.28 $\pm$ 0.02 & 13.676 $\pm$ 0.009 & 0.097 $\pm$ 0.002\\
  103b & 195.441 & 54.145 & 13.85 & 0.09 & 1.71 $\pm$ 0.01 & 14.017 $\pm$ 0.004 & 0.61 $\pm$ 0.03\\
  120a & 200.110 & 49.284 & 0.61 & 0.07 & 4.1 $\pm$ 0.4 & 10.85 $\pm$ 0.06 & \\
  120b & 200.127 & 49.277 & 0.48 & 0.06 & 6.9 $\pm$ 0.9 & 10.90 $\pm$ 0.08 & \\
  144a & 204.835 & 50.982 & 0.93 & 0.09 & 8.4 $\pm$ 0.8 & 9.05 $\pm$ 0.06 & \\
  144b & 204.847 & 50.937 & 0.57 & 0.08 & 4.3 $\pm$ 0.6 & 8.22 $\pm$ 0.08 & \\
  145a & 205.259 & 49.278 & 3.18 & 0.07 & 2.33 $\pm$ 0.05 & 10.52 $\pm$ 0.01 & 0.32 $\pm$ 0.02\\
  145b & 205.266 & 49.258 & 5.27 & 0.07 & 6.68 $\pm$ 0.09 & 10.002 $\pm$ 0.008 & 0.71 $\pm$ 0.06\\
  165a & 210.762 & 51.456 & 2.91 & 0.07 & 7.0 $\pm$ 0.2 & 19.41 $\pm$ 0.01 & 1.0 $\pm$ 0.3\\
  165b & 210.714 & 51.458 & 0.97 & 0.07 & 1.01 $\pm$ 0.07 & 17.62 $\pm$ 0.04 & 0.4 $\pm$ 0.1\\
  166a & 210.770 & 51.749 & 1.47 & 0.07 & 0.87 $\pm$ 0.04 & 11.38 $\pm$ 0.03 & 0.096 $\pm$ 0.007\\
  166b & 210.851 & 51.744 & 1.48 & 0.07 & 0.27 $\pm$ 0.01 & 12.87 $\pm$ 0.03 & 0.25 $\pm$ 0.03\\
  168a & 211.414 & 54.197 & 7.6 & 0.09 & 8.9 $\pm$ 0.1 & 14.998 $\pm$ 0.007 & 1.0 $\pm$ 0.3\\
  168b & 211.428 & 54.173 & 0.84 & 0.07 & 0.27 $\pm$ 0.02 & 13.34 $\pm$ 0.05 & 0.13 $\pm$ 0.03\\
  177a & 213.511 & 48.707 & 2.14 & 0.07 & 0.79 $\pm$ 0.02 & 19.94 $\pm$ 0.02 & 0.7 $\pm$ 0.2\\
  177b & 213.545 & 48.694 & 0.51 & 0.07 & 0.14 $\pm$ 0.02 & 19.18 $\pm$ 0.08 & 0.31 $\pm$ 0.07\\
  222a & 222.690 & 53.000 & 4.86 & 0.09 & 0.80 $\pm$ 0.02 & 16.91 $\pm$ 0.01 & 0.45 $\pm$ 0.07\\
  222b & 222.761 & 53.005 & 1.35 & 0.08 & 0.29 $\pm$ 0.02 & 15.19 $\pm$ 0.04 & 0.12 $\pm$ 0.02\\
  233a & 226.152 & 50.501 & 3.0 & 0.2 & 3.5 $\pm$ 0.2 & 6.16 $\pm$ 0.03 & 0.044 $\pm$ 0.003\\
  233b & 226.225 & 50.505 & 2.4 & 0.2 & 0.88 $\pm$ 0.06 & 5.71 $\pm$ 0.04 & 0.25 $\pm$ 0.04\\
  \hline\end{tabular}
  \tablefoot{Column 1: as in Tab.~\ref{tab:GRG_all} with a letter to distinguish the two lobes; Column 2 and 3: J2000 celestial coordinates of the highest signal-to-noise pixel; Column 4: polarized flux density  of the detected source component; Column 5: polarization noise derived from the Faraday $Q$ and $U$ spectra; Column 6: fractional polarization at the position of the most significant pixel. The uncertainty is derived from the propagation of the rms noise in the polarized and total intensity images; Column 7: Faraday rotation derived from the main peak of the Faraday spectrum of the most significant pixel. The uncertainty is computed as the resolution of the Faraday spectrum divided by two times the signal-to-noise of the detection. This does not include the systematic error from the ionospheric RM correction \citep[of the order of $\sim$0.1 rad m$^{-2}$,][]{VanEck18a}; Column 8: depolarization factor. The uncertainties are derived with standard propagation from the rms noise of the images. The values reported in Column 2 to 7 are derived from the $20\arcsec$ images, while the depolarization factor in Column 8 is obtained using $45\arcsec$ resolution images.}
\label{tab:pol_double}
\end{table*}

\begin{table*}
\centering
\caption{Results of the polarized intensity study for sources with a single polarized detection.}
\begin{tabular}{lllccrrr}
\hline
\multicolumn{1}{c}{GRG} &
  \multicolumn{1}{c}{R.A.} &
  \multicolumn{1}{c}{Dec.} &
  \multicolumn{1}{c}{P} &
  \multicolumn{1}{c}{$\sigma_{QU}$} &
  \multicolumn{1}{c}{$p$} &
  \multicolumn{1}{c}{RM} &
  \multicolumn{1}{c}{\df} \\
\multicolumn{1}{c}{} &
  \multicolumn{1}{c}{(deg)} &
  \multicolumn{1}{c}{(deg)} &
  \multicolumn{1}{c}{(mJy)} &
  \multicolumn{1}{c}{(mJy/beam)} &
  \multicolumn{1}{c}{($\%$)} &
  \multicolumn{1}{c}{(rad/m$^{2}$)} &
  \multicolumn{1}{c}{} \\
\hline
  7 & 164.634 & 51.687 & 0.81 & 0.07 & 2.5 $\pm$ 0.2 & 21.67 $\pm$ 0.05 & 0.19 $\pm$ 0.06\\
  44 & 174.908 & 47.332 & 0.54 & 0.06 & 5.3 $\pm$ 0.6 & 22.20 $\pm$ 0.07 & 0.19 $\pm$ 0.05\\
  47 & 177.991 & 49.837 & 0.59 & 0.07 & 0.16 $\pm$ 0.02 & 16.53 $\pm$ 0.07 & 0.052 $\pm$ 0.007\\
  57 & 182.675 & 53.485 & 4.69 & 0.07 & 5.81 $\pm$ 0.09 & 12.214 $\pm$ 0.009 & 0.70 $\pm$ 0.09\\
  80 & 187.512 & 53.531 & 0.57 & 0.06 & 1.0 $\pm$ 0.1 & 10.71 $\pm$ 0.07 & 0.13 $\pm$ 0.04\\
  83 & 188.252 & 49.119 & 1.14 & 0.08 & 1.19 $\pm$ 0.08 & 13.56 $\pm$ 0.04 & 0.037 $\pm$ 0.004\\
  112 & 197.578 & 52.222 & 0.86 & 0.09 & 1.8 $\pm$ 0.2 & 3.19 $\pm$ 0.06 & \\
  117 & 199.144 & 49.544 & 1.2 & 0.07 & 3.0 $\pm$ 0.2 & 13.00 $\pm$ 0.03 & 0.19 $\pm$ 0.02\\
  122 & 200.906 & 47.511 & 0.61 & 0.079 & 3.4 $\pm$ 0.4 & 7.47 $\pm$ 0.07 & 0.21 $\pm$ 0.05\\
  136 & 203.374 & 53.521 & 1.1 & 0.1 & 11.0 $\pm$ 1.0 & 10.91 $\pm$ 0.07 & 0.050 $\pm$ 0.009\\
  137 & 203.561 & 55.013 & 0.76 & 0.08 & 0.073 $\pm$ 0.008 & 8.05 $\pm$ 0.06 & \\
  148 & 206.071 & 48.787 & 0.8 & 0.1 & 0.8 $\pm$ 0.1 & 12.50 $\pm$ 0.07 & 0.045 $\pm$ 0.004\\
  149 & 206.178 & 50.395 & 1.1 & 0.08 & 7.1 $\pm$ 0.5 & 10.45 $\pm$ 0.04 & 0.4 $\pm$ 0.2\\
  207 & 220.024 & 55.487 & 0.56 & 0.06 & 2.0 $\pm$ 0.2 & 11.64 $\pm$ 0.07 & 0.26 $\pm$ 0.07\\
  234 & 226.541 & 51.591 & 0.93 & 0.09 & 2.1 $\pm$ 0.2 & 9.74 $\pm$ 0.06 & 0.27 $\pm$ 0.06\\
\hline\end{tabular}
\tablefoot{Column headings are the same as in Tab.~\ref{tab:pol_double}.}
\label{tab:pol_single}
\end{table*}

\subsection{RM difference between lobes}
\label{sec:rmdiff}

The observed RM is derived from the main peak of the Faraday spectrum at each pixel because all the detected components show a simple Faraday spectrum (i.e., with a single and isolated peak, contrary to the complex Faraday spectrum where multiple peaks are observed, e.g., in \citealt{Stuardi19}). In this case, the RM is equal to the Faraday depth, a physical quantity given by:

\begin{equation}
    \phi = 0.812 \int_{\rm{source}}^{\rm{observer}} {n_e B_{\parallel} {\rm d}l} \quad \rm{[rad \ m^{-2}]}~,
    \label{eq:RM}
\end{equation}

where $n_e$ is the thermal electron density in cm$^{-3}$, $B_{\parallel}$ is the magnetic field component parallel to the line of sight in $\mu$G, and ${\rm d}l$ is the infinitesimal path length in parsecs. 

The values of RM obtained are between 3 and 28 rad m$^{-2}$ with a median value of 12.8 rad m$^{-2}$ (see left panel of Fig.~\ref{fig:pol_hist}). The fact that they are all positive points out that in the sampled 424 deg$^{2}$ sky region the magnetic field of our Galaxy is pointing toward us and it is the dominant source of the mean Faraday rotation. This implies a smooth Galactic magnetic field on scales of $\sim$10 deg (i.e., the median distance between the sources).

Among the 36 detected sources, 21 GRGs have both lobes detected in polarization (at least one above the 8$\sigma$ significance level). For these sources, plus GRG\,0, we computed the RM difference between the two lobes (\drm). This quantity indicates a difference in the intervening magneto-ionic medium on large scales (of the order of 1 Mpc at the redshifts of the sources). \drm\ can be caused by variations in the Galactic RM (GRM), in addition to a different line-of-sight path length between the two lobes in the local environment and/or differences in the IGM on large scales.

\begin{figure*}
\centering
\includegraphics[width=\textwidth]{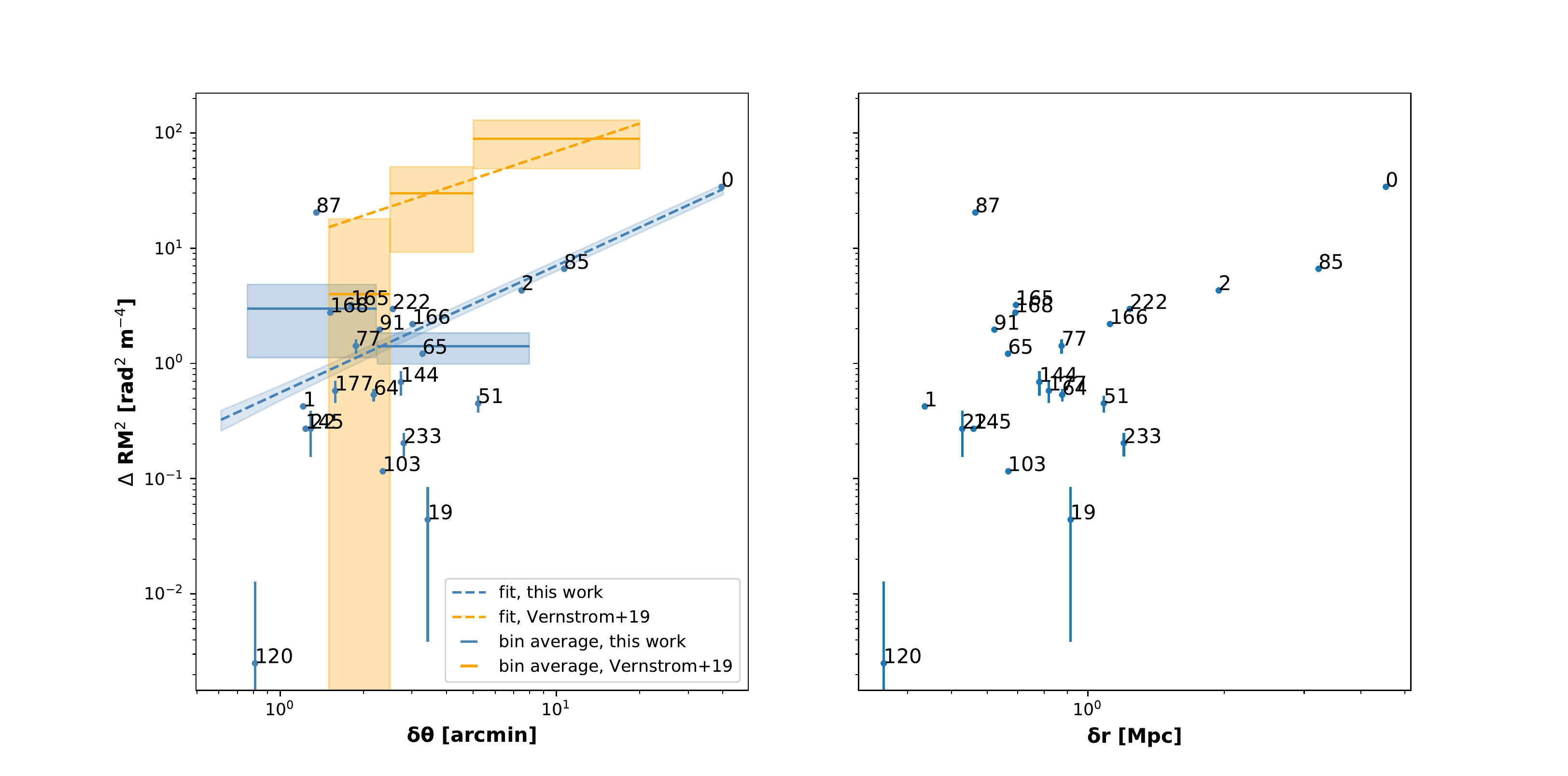}
  \caption{Squared RM difference versus angular (left) and physical (right) separation between the detected lobes. A number corresponds to each GRG and the numbers are listed in Tab.~\ref{tab:pol_double}. The blue dashed line is the power-law fit to the data with 1$\sigma$ uncertainty (see Sec.~\ref{sec:rmdiff}). Orange bars shows binned averages between 1.5$\arcmin$ and 20$\arcmin$  obtained by \citet{Vernstrom19} for physical pairs observed at 1.4 GHz and the dashed orange line shows the derived structure function. Blue bars show the binned averages of the sources in this work with angular separation lower than $10\arcmin$: each bin contains 10 sources, the uncertainty is computed as the standard deviation on the mean. Shadowed areas show the uncertainties.}
  \label{fig:delta_RM2_vs_sep}
\end{figure*}

The reconstruction of the GRM by \citet{Oppermann15} has a resolution of 1$^{\circ}$ (i.e., the typical spacing of extra-galactic sources in the \citealt{Taylor09} catalog) so that most of our double-lobed GRGs lie in the same resolution element of the reconstruction. All the measured RMs are within the 3$\sigma$ error of the estimated GRM, with the exception of GRG\,144, for which the difference is within the 4$\sigma$ error. The average of the GRM values at the position of the detected components (i.e., on scales of of $\sim$10 deg) is 13$\pm$1 rad m$^{-2}$: consistent with the one found from our measurements. Due its low angular resolution this map cannot be used to probe RM variations on scales smaller than 1$^{\circ}$ for selected sources. However, RM structure function studies (i.e., $<\Delta{\rm RM}^2>$ versus angular separation) have probed the RM variance on scales below 1$^{\circ}$, but with large uncertainties \citep{Stil11, Vernstrom19}. The GRM variance was found to have a strong dependence on angular separation, in particular at low Galactic latitude. The 22 double GRGs have angular separations ($\delta\theta$) ranging between $\sim$1.8$\arcmin$ and $\sim$40$\arcmin$ and they all have Galactic latitude above 50$^{\circ}$, with GRG\,0 being the largest in size and closest to the Galactic plane. The study of \drmq\ as a function of angular separation in our sample can be used to understand if the RM difference is dominated by the turbulence in the Galactic interstellar medium. 

\drmq\ is plotted against the angular separation of the lobes in the left panel of Fig.~\ref{fig:delta_RM2_vs_sep}. Despite the large scatter at low angular separation, a general increasing trend of \drmq\ with $\delta\theta$ is observed. We computed the average \drmq\ for the sources with $\delta\theta<10\arcmin$ (thus excluding GRG\,85 and GRG\,0) divided in two bins with 10 sources each (uncertainties are computed as the standard deviation on the mean). The binned averages are over-plotted in Fig.~\ref{fig:delta_RM2_vs_sep}. We fitted a power law of the form:

\begin{equation}
    \Delta{\rm RM}(\delta\theta)^2 = A \delta\theta^B \ ,
\label{eq:strfunc}
\end{equation}

and we obtained: $A=0.56\pm0.06$ rad$^2$ m$^{-4}$ and $B=1.1\pm0.1$ with $\chi^{2}=515$ (the blue line in  Fig.~\ref{fig:delta_RM2_vs_sep}). The fit suggests an increasing influence of the Milky Way foreground with angular size. However, it is dominated by a few GRG with the largest angular sizes and more sources at large $\delta\theta$ would be required to confirm this behavior. Conversely, the binned average for sources at low angular separation shows a large scatter and points to a flattening of the power-law slope for $\delta\theta<2\arcmin$. This could be related to an increasing influence of the extra-galactic contribution over the Galactic one at small angular scales.  

We can compare our result with the structure function studies of \citet{Stil11} and \citet{Vernstrom19}. While \citet{Stil11} considered together all kinds of source pairs (physical and non-physical), \citet{Vernstrom19} separated physical and non-physical pairs. The latter is thus best suited for a direct comparison with our work where all pairs are physical. \citet{Vernstrom19} made use of the \citet{Taylor09} catalog of polarized sources observed at 1.4 GHz. For a sample of 317 physical pairs with angular separations between $1.5\arcmin$ and $20\arcmin$ they obtained $A=11\pm15$ rad$^2$ m$^{-4}$ and $B=0.8\pm0.2$. The fit is shown as a comparison in the left hand panel of Fig.~\ref{fig:delta_RM2_vs_sep}. The slopes are consistent within the 2$\sigma$ uncertainty. The slightly steeper power-law compared to the one obtained by \citet{Vernstrom19} can be attributed to the presence of GRG\,0 in our sample. In both cases the trend is dominated by pairs of sources at $\delta\theta>10\arcmin$, indicating an increasing contribution from the GRM. 

Due to their large size, GRGs are expected to lie at large angles to the line of sight and to extend well beyond the group/cluster environment so that the differential Faraday rotation effect originating in the local environment should be minimal \citep{Laing88,Garrington88}. Furthermore, none of our sources show a prominent one-sided large-scale jet that would indicate motion toward the line of sight, not even the six sources with a quasar host (i.e., GRG\,1, GRG\,47, GRG\,91, GRG\,120, GRG\,137, GRG\,222\,). Thus, \drm\ is not expected to strongly correlate with the source physical size. However, to investigate the local contribution, we plotted the RM difference squared against the physical separation between the two lobes (Fig.~\ref{fig:delta_RM2_vs_sep}, right panel). Notable is the similarity between the right-hand and left-hand panel of Fig.~\ref{fig:delta_RM2_vs_sep}. If the main contribution was due to the local environment, we would typically expect a larger RM difference between the lobes at smaller physical separations. Conversely, the similarity between the panels of Fig.~\ref{fig:delta_RM2_vs_sep} suggests that this trend is dominated by the angular separation trend, which is driven by Galactic structures. This points out that the local environment is sub-dominant in determining \drm. 

Asymmetries in the foreground large-scale structures could also contribute to the RM difference between the two lobes. We expect much more large-scale asymmetries close to galaxy clusters \citep{Bohringer16}. We note that, according to the environment analysis of \citet{Dabhade20}, none of the GRGs detected in polarization are associated with the BCG of a dense cluster of galaxies. However, foreground galaxy clusters are Faraday screens for all the sources that are in the background. Therefore, we cross-matched the position of the 22 GRGs with the cluster catalog of \citet{Wen15} in order to find the foreground galaxy cluster at the smallest projected distance from each GRG. This catalog is based on photometric redshifts from the SDSS III and lists clusters in the redshift range 0.05 $< z <$ 0.8. In the redshift range 0.05 $< z <$ 0.42 it is 95$\%$ complete for clusters with mass M$_{200} > 10^{14}$ M$_{\odot}$. Taking into account the uncertainty on the photometric redshift estimates, $\Delta z=0.04(1+z)$, we considered a cluster as being in the foreground of a particular GRG for all clusters with $z-\Delta z$ lower than the redshift of the GRG plus its uncertainty.

We computed the angular separation between each GRG lobe and the closest foreground galaxy cluster ($\delta\theta^{\rm min}_{\rm cluster}$ and $\delta\theta^{\rm max}_{\rm cluster}$, for the closest and farthest lobe, respectively). \drmq\ is plotted against $\delta\theta^{\rm min}_{\rm cluster}$ divided by angle subtended by $R_{500}$ of the cluster ($\theta_{R_{500}}$, in arcminutes) in the top panel of Fig.~\ref{fig:vs_cluster}. Most of the GRGs lie at projected distances larger than $R_{500}$ and the trend does not show a clear dependence of \drm\ on the distance from the closest foreground cluster. Asymmetries in the foreground large-scale structures are thus probably sub-dominant compared to the ones caused by the GRM. However, this will be further discussed in Sec.~\ref{sec:discussion}.
 
\subsection{Faraday depolarization}
\label{sec:depolorig}

RM fluctuations within group and cluster environments can be caused by turbulent magnetic field fluctuations over a range of scales. While large scale fluctuation are mostly responsible for the RM difference between the lobes, fluctuation on the smallest scale may be at the origin of Faraday depolarization. The mixing of different polarization vector orientations within the observing beam and along the line of sight reduces the fractional polarization. The RM dispersion for a simple single-scale model of randomly orientated magnetic field is

\begin{equation}
    \sigma_{\rm RM}^2 = 0.812^2 \Lambda_{c} \int_{\rm{source}}^{\rm{observer}} {(n_e B_{\parallel})^2 {\rm d}l} \quad \rm{[rad^2 \ m^{-4}]}~, 
\label{eq:sigma_rm}
\end{equation}

where $\Lambda_{c}$ is the correlation length of the magnetic field in parsecs \citep[e.g.,][]{Felten96,Murgia04}. The RM dispersion is responsible for the Faraday depolarization which in the case of an external screen \citep{Burn66} is expressed as:

\begin{equation}
    p(\lambda)=p(\lambda =0)e^{-2 \sigma_{\rm RM}^2 \lambda^4}~.
\label{eq:depol}
\end{equation}

In the GRGs sample, the fractional polarization at 20$\arcsec$ resolution ranges between 0.07 and $11.7\ \%$ with a median value of $2.6\ \%$ (see central panel of Fig.~\ref{fig:pol_hist}). LOFAR has a unique capability to reliably detect very low fractional polarization values (i.e., $<0.5\ \%$) when RM is outside the range $-3< \phi <1$ rad m$^{-2}$ because of the high resolution in Faraday space that allows a clear separation from the leakage contribution. 

Four components detected at 20$\arcsec$ are under the detection threshold at 45$\arcsec$. This is due to the lower sensitivity at 45$\arcsec$ resolution. Only in one case (GRG\,112) is the non-detection likely caused by beam depolarization on scales between 20$\arcsec$ and 45$\arcsec$ (i.e., 140 and 315 kpc at the source redshift). Instead, five sources are not detected in the NVSS due to the lower sensitivity of this survey. Hence, there are 28 sources with depolarization measurements. The distribution of depolarization factors computed at 45$\arcsec$ is shown in the right panel of Fig.~\ref{fig:pol_hist}. All the sources have \df>0.03 and the median value is 0.2. 

Our measurements enable us to probe magnetic field fluctuations on scales below the 45$\arcsec$ restoring beam, which for the redshift range of our sample corresponds to physical scales of 80-480 kpc. Faraday depolarization can occur internally to the source or can be due to the small-scale fluctuation of the magnetic field in the medium external to the source. 

With LoTSS data, we are not able to observe internal depolarization, that would appear as a thick Faraday component through RM-synthesis. This is because the largest observable Faraday scale is smaller than the resolution in Faraday space (see Sec.~\ref{sec:polimaging}). Broad-band polarization studies at higher frequencies and/or detailed modeling of internal Faraday screens would be needed to distinguish between these two scenarios.

In the case of external depolarization, Eq.~\ref{eq:depol} implies that the effect of a $\sigma_{\rm RM}\leq1$ rad m$^{-2}$ is only observable at very large wavelengths. For this reason, by comparing measurements at 1.4 GHz and at 144 MHz it is possible to study the depolarization caused by low $\sigma_{\rm RM}$. On the other hand, $\sigma_{\rm RM}\geq1$ rad m$^{-2}$ can completely depolarize the emission and make it undetectable by LOFAR. Within galaxy clusters, where $B\sim0.1-10 \ \mu$G, $n_{e}\sim10^{-3}$ cm$^{-3}$ and the magnetic field is tangled on a range of scales, the RM dispersion is clearly above this level \citep[e.g.,][]{Murgia04,Bonafede10}. 

\begin{figure}
 \includegraphics[width=0.5\textwidth]{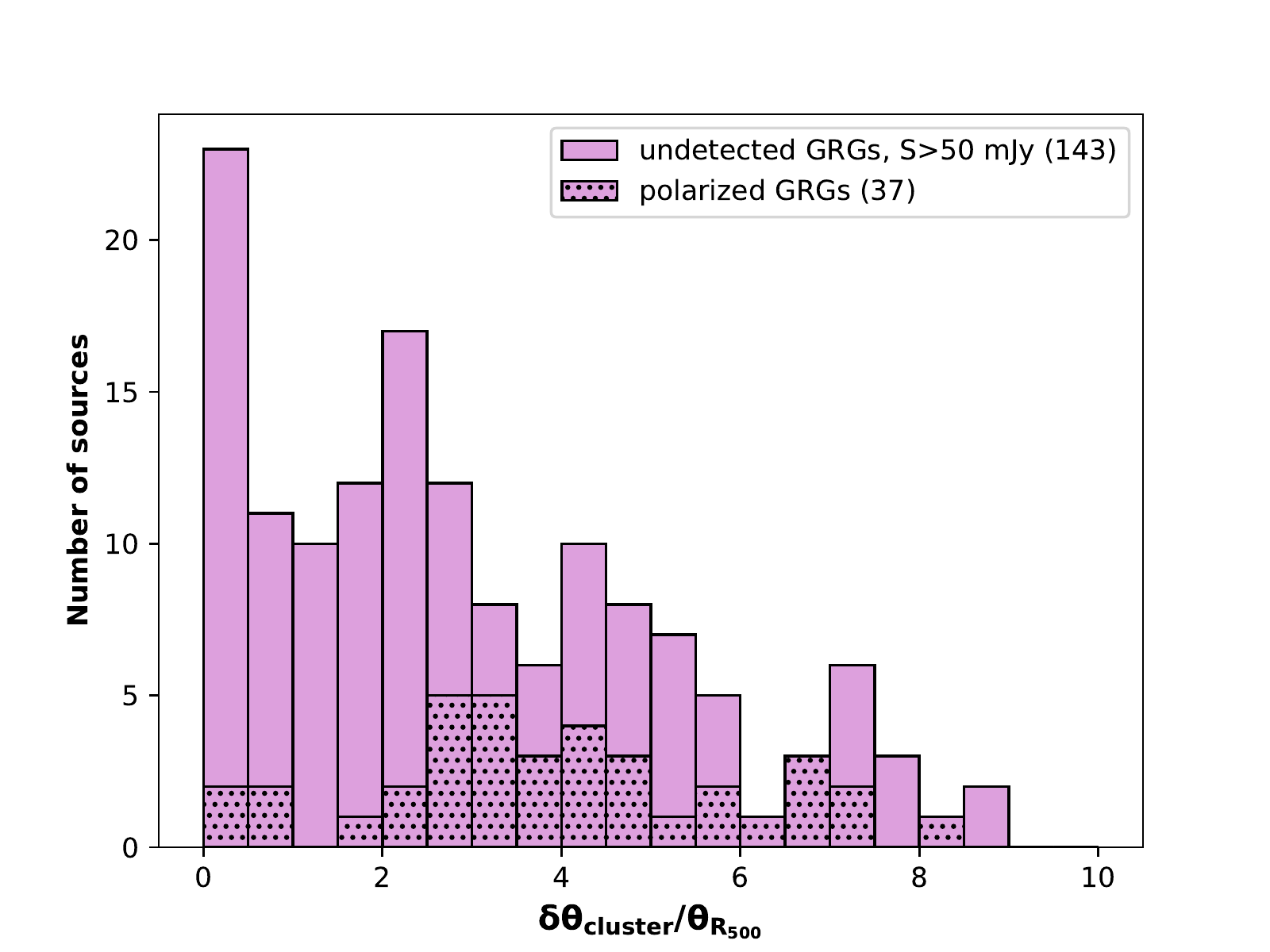}
 \includegraphics[width=0.5\textwidth]{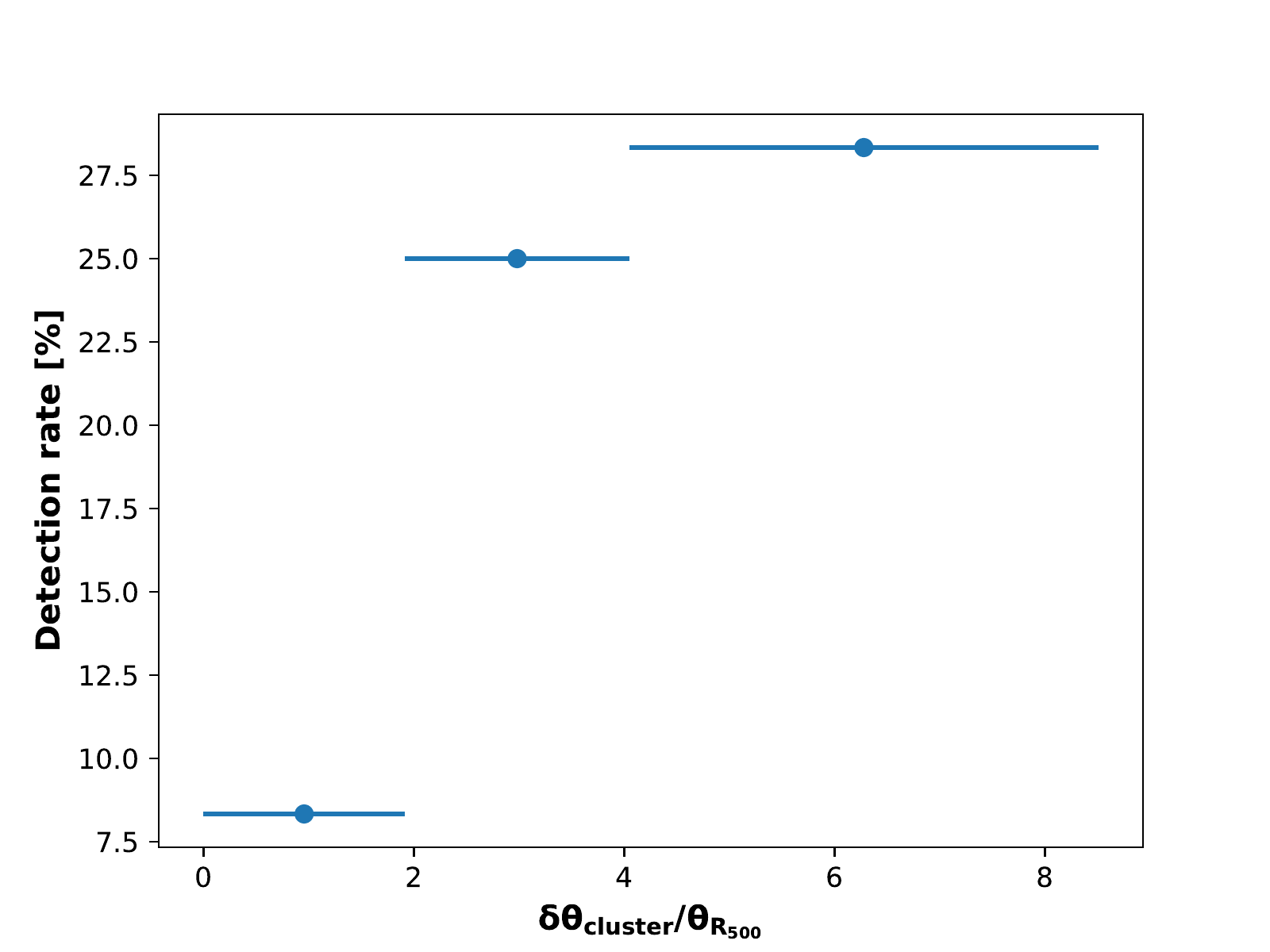}
 \caption{Distribution of minimum distance from the closest foreground cluster for detected and undetected sources in polarization (top panel), and detection rate as function of the minimum distance from foreground clusters (bottom panel). The widths of the bins are computed to contain the same total number of sources (i.e., 60). Markers are positioned at center of each bin and the  error bars show the width of the bins.}
 \label{fig:cluster_hist}
\end{figure}

The distribution of distances from the closest foreground cluster is compared for detected and undetected GRGs in polarization in the top panel of Fig.~\ref{fig:cluster_hist} while the detection rate is computed as function of the distance from the foreground cluster in the bottom panel (for GRGs with S$_{\rm 144  MHz}>$ 50 mJy). We find that $8\ \%$ of the GRGs observed within 2$R_{500}$ of the closest foreground cluster are detected in polarization, while the detection rate increases to $27\ \%$ outside 2$R_{500}$. The Kolmogorov-Smirnov test indicates a  significant difference between the samples of detected and undetected GRGs with S$_{\rm 144  MHz}>$ 50 mJy ($p$-value of 2$\times10^{-3}$). Together with the non detection of the GRGs at the center of clusters (Sec.~\ref{sec:results}), this shows that in general, to be detected by the LoTSS, sources need to avoid locations both within and in the background of galaxy clusters where the RM dispersion is too high.
 
 Only four GRGs are detected within $R_{500}$: GRG\,2, GRG\,91, GRG\,120, and GRG\,136. Among them, GRG\,2 ($z=0.27627\pm0.00005$) and GRG\,136 ($z=0.354\pm0.034$) have similar redshifts with respect to the clusters (at redshifts $0.27\pm0.05$ and $0.37\pm0.05$, respectively). They have been considered in the background due to the uncertainties on the photometric redshift estimates, but it is also possible that these GRGs are cluster members or instead lie in the foreground of the clusters. GRG\,91 and GRG\,120 are associated with compact foreground clusters with $R_{500}$ equal to 570 kpc and 650 kpc respectively.

\begin{figure}
\includegraphics[width=0.5\textwidth]{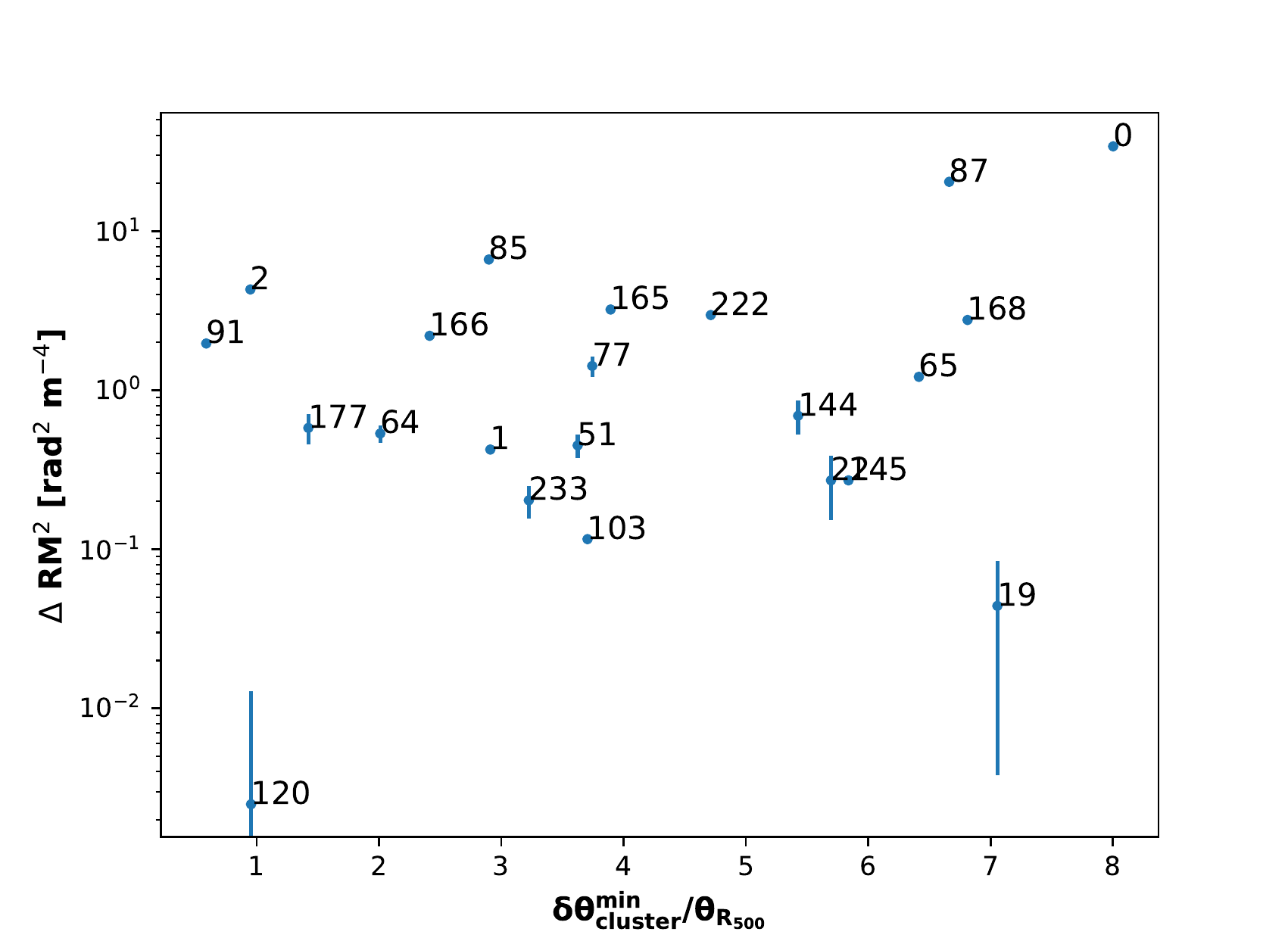}
\includegraphics[width=0.5\textwidth]{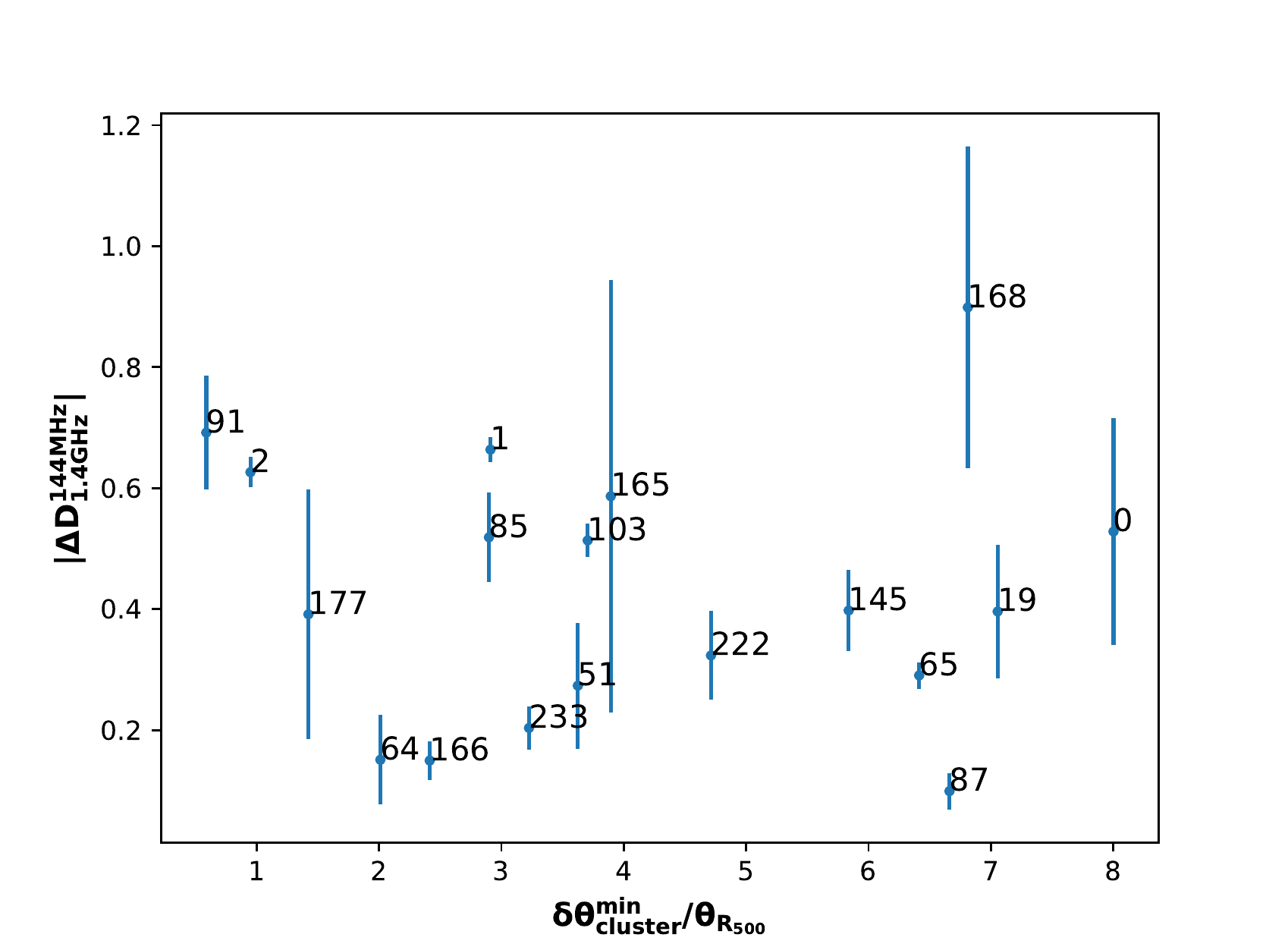}
\caption{Squared RM difference (top panel) and depolarization factor difference between the two lobes (bottom panel) versus the minimum distance from the closest foreground galaxy cluster scaled by $R_{500}$ of the cluster.}
\label{fig:vs_cluster}
\end{figure}

\begin{figure}
 \includegraphics[width=0.5\textwidth]{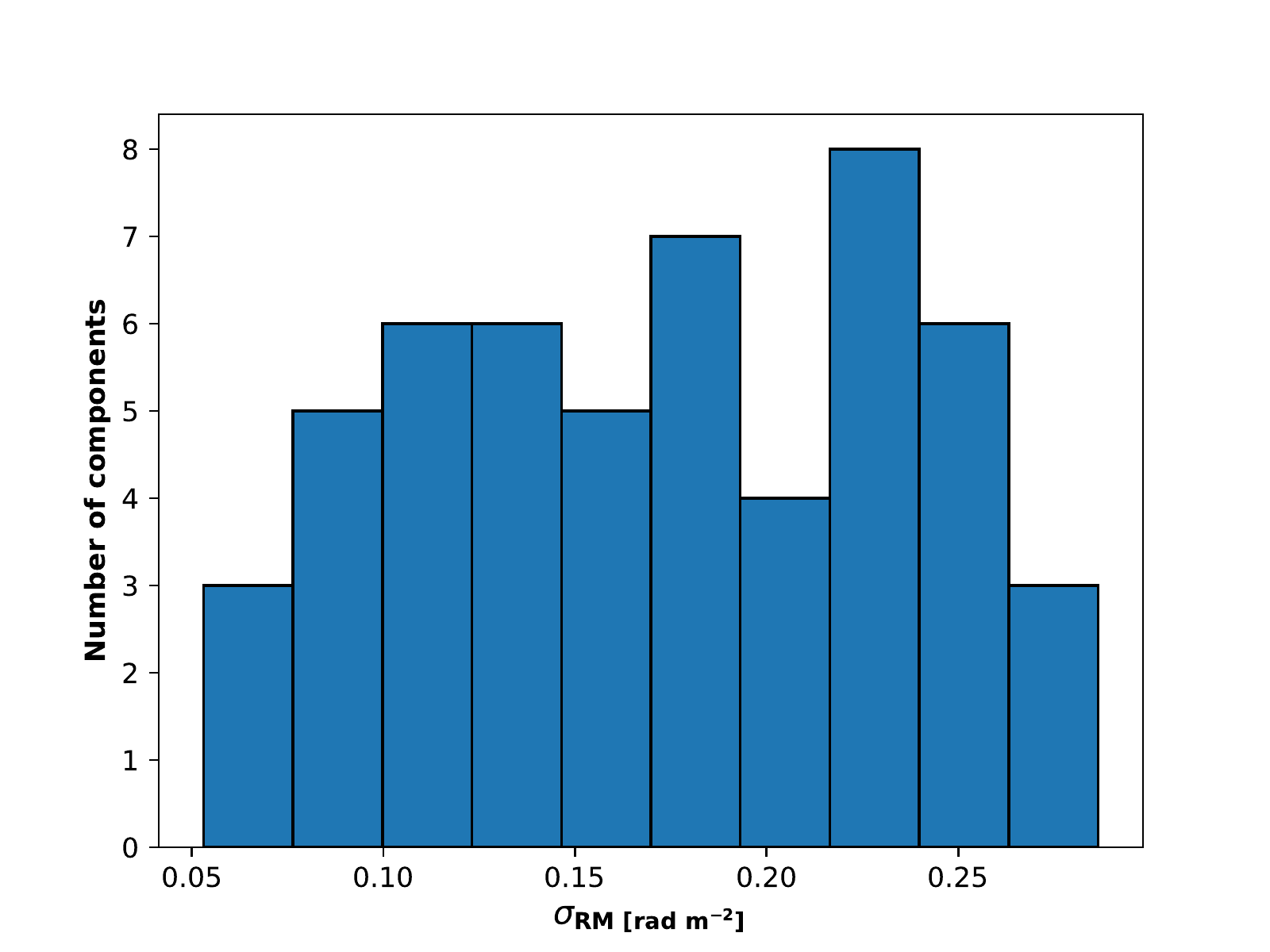}
 \caption{Distribution of RM dispersion values obtained using an external Faraday screen model.}
 \label{fig:sigma_RM}
\end{figure}

Using \df\ in Eq.~\ref{eq:depol} we can compute $\sigma_{\rm RM}$. The distribution of $\sigma_{\rm RM}$ is shown in Fig.~\ref{fig:sigma_RM}. The maximum value is 0.29 rad m$^{-2}$. Given the small amount of depolarization it is important to consider that the residual error in the ionospheric RM correction within the 8 hours of the observation could account for $\sim0.1-0.3$ rad m$^{-2}$ \citep{VanEck18a}. In principle, this could explain most or all the depolarization observed but we can test this because the residual ionospheric correction error is subtracted out in the difference in depolarization between the two hotspots of the same radio galaxy. |$\Delta$ \df| represents a lower limit to the depolarization that leads to $\sigma_{\rm RM}$ values between 0.05 and 0.25 rad m$^{-2}$. These estimates are further discussed in Sec.~\ref{sec:discussion}.

We have tested the possibility that the closest foreground cluster was the main origin of the measured depolarization by plotting |$\Delta$ \df| versus the distance from the cluster in the bottom panel of Fig.~\ref{fig:vs_cluster}. However, we do not find a correlation between these quantities. 

\df is also not correlated with the distance from the host galaxy, probably because all the sources are very extended and already well beyond the host galaxy's halo \citep{Strom88}. The Laing-Garrington effect (i.e., the differential Faraday depolarization that causes the counter-lobe to be more depolarized than the lobe closer to us \citealt{Laing88,Garrington88}) is indeed not expected to be a strong effect in this case. We note that none of the GRGs show a prominent jet in the total intensity images (see Fig.~\ref{fig:GRG1}), which is in line with the expectation that these sources are observed at large angles to the line of sight.

\section{Discussion}
\label{sec:discussion}

Since both RM and depolarization are integrated effects along the line of sight (Eq.~\ref{eq:RM} and \ref{eq:sigma_rm}), in order to disentangle the contribution of the different Faraday rotation and depolarization screens one should have detailed information on the environment surrounding each radio galaxy, the foreground, and the geometry and physical properties of the lobes. This requires a detailed study of each single source. We instead investigated several possible origins of the RM difference and Faraday depolarization considering the correlation of \drm\ and \df\ with different physical quantities.

Several statistical analyses on the RMs of extra-galactic sources have been performed. Structure function studies verified the dependence of \drm\ on the angular separation originated by the Galactic magnetic field \citep[e.g.,][]{Simonetti84,Sun04,Stil11}. The presence of a growing contribution to the RM with redshift was investigated by \citet{Pshirkov15}. The RM variance of background sources was modeled to separate an extra-galactic contribution of 6-7 rad m$^{-2}$ from the Galactic one \citep[e.g.,][]{Schnitzeler10,Oppermann15}. Bringing together these works, \citet{Vernstrom19} studied the average \drmq\ as function of angular separation, redshift, spectral index, and fractional polarization using two large samples of physical and non-physical pairs in order to isolate the extra-galactic contribution. A difference of $\sim10$ rad m$^{-2}$ in the average \drmq\ between the two samples was attributed to the IGM to derive an upper limit on the extra-galactic magnetic field of 40 nG. A contribution from the local magnetic field, producing a larger variance for non-physical pairs, cannot be excluded. All these studies were performed at 1.4 GHz, thanks to the presence of the RM catalog produced with the NVSS \citep{Condon98,Taylor09}. With the advent of LOFAR, these kind of studies are also possible at low frequencies. With respect to the NVSS, the LoTSS allows a better resolution, sensitivity, and precision in the RMs determination.

In this work, the RM difference between the lobes was found to be marginally correlated with the angular distances of the lobes (Fig.~\ref{fig:delta_RM2_vs_sep}). Although the correlation is not strong (with a Spearman correlation coefficient of 0.35), we found the relation between \drmq\ and $\delta\theta$ to be consistent with the Galactic structure function found by \citet{Vernstrom19} for physical pairs. This strongly suggests a Galactic origin of the \drm\ between the lobes. The accuracy in the determination of the amplitude parameter is 250 times higher than the one obtained using NVSS measurements. The same trend observed with the angular separation dominates also the correlation between \drm\ and physical distance. This suggests that the local gas densities and magnetic fields, which should have a stronger effect on the RM variation for normal size galaxies, are not dominant in this sample. This would also explain the fact that, although consistent within the errors, the amplitude of the power-law at 144 MHz is one order of magnitude lower than the one at 1.4 GHz (see Fig.~\ref{fig:delta_RM2_vs_sep}). While in \citet{Vernstrom19} the physical size of the sources is no taken into account, our GRG sample constitutes a population where the local contribution to \drm\ is negligible. A selection of a source population with low local RM variance is an important requirement for future RM grid experiments \citep{Rudnick19}.

Recently, \citet{O'Sullivan20} applied the same method of \citet{Vernstrom19} to the RMs derived at 144 MHz from the LoTSS. This study resulted in an extra-galactic contribution of 0.4$\pm$0.3 rad m$^{-2}$ which yielded to an upper limit on the co-moving magnetic field of 2.5 nG. Since the magnetic field in the IGM is not expected to vary with frequency, the discrepancy between the results obtained at 1.4 GHz and 144 MHz was attributed to the Faraday depolarization effect. Since a high local RM variance can depolarize sources below the detection level at low frequencies, observations at 144 MHz selects sources with low RM variance which unveils the effect of weaker magnetic fields and lower thermal gas densities.

To measure and investigate the origin of the depolarization is thus complementary to the aforementioned studies. In this context, the depolarization is caused by RM variance on scales of the synthesized beam which consequently affect the measurement of the RM variance on the scale of the angular separation between the sources (or the sources' lobes). The dependence of the RM variance and depolarization on the physical size of classical double radio sources was investigated by \citet{Strom88} and \citet{Johnson95} to study the local magnetic field. \citet{Machalski06} extended this work by comparing normal size and giant radio galaxies, finding that the depolarization factor strongly correlates with the size of the sources. Within the GRG sample collected by \citet{Machalski06} the median depolarization factor between 4.9 GHz and 1.4 GHz is  1.04$\pm$0.05, with the majority of sources showing undetectable levels of depolarization. The RMs, obtained with a fit between the two frequencies and thus subject to the $n\pi$ ambiguity, are also consistent with 0 within the large uncertainties. The wavelength at which substantial depolarization occurs increases with the size of the sources. The depolarization caused by a $\sigma_{\rm RM}\sim 0.3$ rad m$^{-2}$ would be undetected at GHz frequencies. Low-frequency observation are thus necessary to measure the small amount of depolarization experienced by the lobes of GRGs in order to constrain the magneto-ionic properties of their environment. 

While RM differences between the lobes probe magnetic field fluctuations on large scales (i.e., $\sim$1 Mpc) the depolarization is sensitive to angular scales below the 45$\arcsec$ resolution. This implies scales of 80-480 kpc in the redshift range of the sources. In the most common model of external Faraday dispersion, the depolarization roughly scales as 1/$\sqrt{N}$ where $N$ is the number of Faraday cells within the beam \citep{Sokoloff98}. A model of random magnetic field fluctuations in $N$=25 cells is able to explain the median \df=0.2 and it implies a magnetic field reversal scale of 3-25 kpc. 

The depolarization observed is thus most likely occurring in a very local environment. This is also supported by Fig.~\ref{fig:detrate_linsize} which shows an increasing detection rate with larger distances from the host galaxy and thus from the local enhancement of gas density. A simple model of constant thermal electron density of $\sim10^{-5}$ cm$^{-3}$ and magnetic field of $\sim0.1 \ \mu$G tangled on scales of 3-25 kpc could explain the values of $\sigma_{\rm RM}$ observed using Eq.~\ref{eq:sigma_rm} with an integration length < 100 kpc. Sub-$\mu$G magnetic fields and thermal electron densities of a few times $10^{-5}$ cm$^{-3}$ are consistent with the findings from detailed studies on single giant radio galaxies \citep[e.g.,][]{Willis78b,Laing06}. From the study of five well known GRGs, \citet{Mack98} also concluded that the density estimates in the environments of these sources are one order of magnitude lower than within clusters of galaxies. This is the typical environment that polarization observations with LOFAR allow us to study, since larger $\sigma_{\rm RM}$ would completely depolarize the emission. This automatically excludes all the source lying within dense cluster environment, as confirmed by the fact the all 21 GRGs known to reside in clusters are undetected in polarization. Sources residing in such under-dense environment are thus the dominant population of physical pairs also in the work by \citet{O'Sullivan20}.

We note that the $\sigma_{\rm RM}$ values shown in Fig.~\ref{fig:sigma_RM} were derived assuming external depolarization (Eq.~\ref{eq:depol}). With measurements at only 144 MHz and 1.4 GHz, we can not exclude other depolarization models \citep[e.g.,][]{Sokoloff98, Tribble91, O'Sullivan18b}. A detailed depolarization analysis with a larger wavelength-square coverage would be needed. For example, in the case in which the polarized emission at 144 MHz originates from an unresolved region within the 45$\arcsec$ beam across which the RM gradient is effectively zero and the rest of the polarized structure is completely depolarized by RM fluctuations, our $\sigma_{\rm RM}$ estimates are not applicable. This would imply that the true $\sigma_{\rm RM}$ of the local environment could be much higher but that our measurements at 144 MHz cannot detect this emission.

\subsection{The influence of foreground Galaxy Clusters}

Having investigated the Galactic and local Faraday effects on \drm\ and \df\ and their implication for present and future polarization studies with LOFAR, we shift our attention to the possible presence of Faraday screens in the foreground of our targets. Several statistical studies of the Faraday rotation of background sources have demonstrated the presence of magnetic field in clusters of galaxies \citep[e.g.,][]{Lawler82,Clarke01,Bohringer16}. The scatter in the RMs was found to be enhanced by the cluster magnetic field up to 800 kpc from the cluster center \citep{Johnston-Hollitt04}. The majority of the double detected sources in our study lies outside R$_{500}$ of foreground clusters (see Fig.~\ref{fig:vs_cluster}). Therefore, it is not surprising that the correlation between \drmq\ and the distance from the closest foreground cluster is rather weak (Spearman correlation coefficient of 0.11). In any case, because of LOFAR’s high sensitivity to small RMs, LOFAR allows us to explore regions far outside galaxy clusters which are traced by the lobes of GRGs.

We can use a $\beta$-model \citep{Cavaliere76} to describe the gas density profile in clusters: $n(r)=n_0(1+r^2/r_c^2)^{-3\beta/2}$, where we assume the central gas density, $n_0\sim10^{-3}$ cm$^{-3}$, the core radius, $r_c\sim200$ kpc, and $\beta$=0.7. We assume that the magnetic field strength scales with the gas density: $B(r)=B_0(n(r)/n_0)^{0.7}$ and that $B_0\sim3 \ \mu$G \citep{Dolag01,Bonafede10,Govoni17}. The choice of these parameters is somewhat arbitrary but they can reasonably describe galaxy cluster environments. Less massive clusters have a lower electron column density along the line of sight for a given radius scaled by $R_{500}$ and in our sample $R_{500}$ ranges between 0.56 and 1.01 Mpc. Considering a median $R_{500}\sim800$ kpc, outside the projected distance of 4 times $R_{500}$ the thermal electron density is < $3\times10^{-6}$ cm$^{-3}$ and the magnetic field strength < 0.05 $\mu$G. Assuming a large magnetic field fluctuation scale of 500 kpc, the mean RM from Eq.~\ref{eq:RM} is $<0.06$ rad m$^{-2}$ (where we used $B_{\parallel}=B/\sqrt{3}$). For GRGs with $\delta\theta^{\rm min}_{\rm cluster} > 4 \theta_{R_{500}}$ the foreground clusters cannot be the dominant origin of the RM difference since their signature would be too weak even for LOFAR RM accuracy. Therefore, the effect of foreground clusters and large-scale IGMF asymmetries to the RM difference is disfavored but it is still non-negligible for some of the GRGs in our sample.

Three double-detected GRGs lie within R$_{500}$ of the closest foreground cluster, namely GRG\,2, GRG\,91, and GRG\,120. We computed for each of them $\delta\theta^{\rm min}_{\rm cluster}$ and $\delta\theta^{\rm max}_{\rm cluster}$, i.e., the distances of the two lobes from the cluster. GRG\,91 is associated with a compact foreground cluster with $R_{500}$ of 570 kpc . While for GRG\,2 the two lobes are, respectively, at $\sim$2 and $\sim$0.95 R$_{500}$, the distance of both lobes of GRG\,120 from the foreground cluster is $\sim$0.96 R$_{500}$. Using the simplified galaxy cluster model previously assumed we would expect a \drmq\ of $\sim$20 rad m$^{-2}$ for GRG\,2 and $\sim$0.1 rad m$^{-2}$ for GRG\,120. Although this model overestimates the observed values, it is able to explain more the two order of magnitude difference between the two sources. This suggests that both the source distance and the difference of distances of the two lobes from foreground clusters can in principle play a role in determining \drm. For other sources, i.e., GRG\,0 and GRG\,87, which lies more than 4R$_{500}$ away from the closest foreground cluster, the enhanced RM difference could also be influenced by the presence of large-scale structure filaments, as proposed for GRG\,85 \citep{O'Sullivan19}. A detailed study of the local environment and of the foreground of the GRGs is required in this cases. Such a study may be addressed in future work. A complementary approach that was used by Mahatma et al. (2020, in prep.) is to invoke a universal pressure profile to predict the distributions of RM toward the population of radio galaxies with local and large-scale contributions

The fractional polarization, and thus depolarization factor, is also known to scale with the distance from the cluster center. \citet{Bonafede11} performed a study of the polarization fraction of sources in the background of galaxy clusters and found that the median fractional polarization at 1.4 GHz decreases toward the cluster center. The trend is observed up to $\sim5$ core radii (that, in the framework of the simple cluster model described above, corresponds to 1.25R$_{500}$) while far outside the median fractional polarization reaches a constant value of $\sim5 \ \%$. Fig.~\ref{fig:vs_cluster} (bottom panel) and Fig.~\ref{fig:cluster_hist} show that, while the depolarization does not correlate with the distance from foreground clusters, the presence of the latter disfavors the detection of the sources in polarization. This is consistent with the value of \df\ depending mostly on the magneto-ionic properties of the local environment of each GRG. Within R$_{500}$, the higher RM variance due to the turbulence in the foreground ICM influences the fractional polarization at GHz frequencies and depolarizes the radio emission at 144 MHz below the LoTSS detection limit. It is plausible that only under particular condition some background sources can be detected, for example, when the foreground cluster is poor and/or the polarized emission originates in a very compact region of the source. Thus, the detection rate at 144 MHz is strongly reduced up to 2-2.5 R$_{500}$. This highlights the presence of magnetic field at larger distances from galaxy clusters than was shown by previous studies at higher frequencies \citep{Clarke04}. This has also the important consequence that future RM grid studies using the LoTSS will mainly sample lines of sight in the extreme peripheries of galaxy clusters, through filaments and voids.

\section{Conclusions}
\label{sec:conclusions}

In this work we used data from the LOFAR Two-Metre Sky Survey to perform a polarization analysis of a sample of giant radio galaxies selected by \citet{Dabhade20}. Our aims were to {\it (i)} study the typical magnetic field in the environment of this class of sources which is unveiled by their polarization properties at low-frequencies {\it (ii)} understand how GRGs can be used in a RM grid to derive important information on foreground magnetic fields. We measured the linear polarization, Faraday rotation measure, and depolarization between 1.4 GHz and 144 MHz of the 37 sources detected in polarization. Compared to previous studies at GHz frequencies, this study allowed us to measure the small amount of Faraday rotation and depolarization experienced by these sources. The high precision in the RM determination ($\sim0.05$ rad m$^{-2}$) enables the detection of very small difference between the lobes of the GRGs (\drm) that we studied against the angular and physical separation and the distance from foreground galaxy clusters. Since the Faraday depolarization has a strong impact on the detection rate at 144 MHz, the latter was also used as a tool to investigate the presence of depolarizing screens. Our results are summarized as follows:

1. Among the 179 giant radio galaxies observed at $20\arcsec$ resolution with flux density above 50 mJy, the polarization detection rate is $20\ \%$ above an 8$\sigma_{QU}$ detection threshold. A comparison with the polarized point-source catalog by \citet{VanEck18a} indicates that sources with large angular size have a much greater chance of being detected. Our study suggests that this class of source preferentially resides in very rarefied environments experiencing low levels of  depolarization. GRG represent thus a good sample for targeted polarization studies of the magneto-ionized foreground medium.

2. The RM variation on scales below 40$\arcmin$ was investigated using the RM difference between the lobes of the same galaxy. Our study supports the idea that the main contribution to \drm\ on scales between 2$\arcmin$ and 40$\arcmin$ comes from the Milky Way foreground as obtained by \citet{Vernstrom19}. With respect to previous studies performed at GHz frequencies, our investigation provided two order of magnitude higher precision in the determination of \drm. A larger sample of sources would be needed to confirm this trend. Local and foreground galaxy cluster contributions to \drm\ are subdominant but non-negligible for some of the sources. 

3. Using NVSS archival data, we studied the depolarization between 1.4 GHz and 144 MHz. We detected Faraday depolarization caused by a Faraday dispersion of up to $\sim$0.3 rad m$^{-2}$. Such small amounts of depolarization cannot be detected at higher frequencies. It may occur in the local environment of the lobe/hotspot, due to small-scale (few tens kpc) magnetic field fluctuations. A factor of 10 better ionospheric RM correction would be needed to constrain the true astrophysical depolarization of each source. 

4. From our analysis, we observed that the environment of the detected giant radio galaxies is extremely rarefied, with thermal electron densities $< 10^{-5}$ cm$^{-3}$ and magnetic fields below $\sim0.1 \ \mu$G. This is likely the typical environment of the majority of sources that LOFAR can detect in polarization. Studies of the extra-galactic magnetic field performed with LoTSS \citep[e.g.,][]{O'Sullivan19} need to take into account a lower local contribution than studies performed at higher frequencies.

5. Furthermore, at LOFAR frequencies the chance of detecting a giant radio galaxy for background RM studies of galaxy clusters is 3 times larger outside 2$R_{500}$ than within it. This indicates that the magnetic field in the outskirts of galaxy clusters has an impact on the polarization of background sources at larger distances than previously observed \citep{Bonafede11}. 

This work shows the polarization and RM properties of the largest class of sources detected by LOFAR in polarization, and highlights the potential of their use to study the magneto-ionic properties of large-scale structures. A denser RM grid is needed to constrain the extra-galactic contribution to the RM variance. Future studies, on the basis of thousands of RMs with known redshifts detected by the LoTSS, will enable us to probe the weak signature of the intergalactic magnetic field both in the peripheries of, and far outside, galaxy cluster environments.


\begin{acknowledgements}

C.S. and A.B. acknowledge support from the ERC-StG DRANOEL, n. 714245. A.B. acknowledges support from the MIUR grant FARE SMS. The J\"ulich LOFAR Long Term Archive and the German LOFAR network are both coordinated and operated by the J\"ulich Supercomputing Centre (JSC), and computing resources on the supercomputer JUWELS at JSC were provided by the Gauss Centre for Supercomputing e.V. (grant CHTB00) through the John von Neumann Institute for Computing (NIC). This work made extensive use of the cosmological calculator of \citet{Wright06}, of the Python packages Astropy \citep{Astropy13}, Matplotlib \citep{Hunter07} and APLpy \citep{Robitaille12}, of TOPCAT \citep{Taylor05}, of the Aladin sky atlas \citep{Bonnarel00} and of SAOImageDS9 \citep{Joye03}. We thank the referee for the useful comments.

\end{acknowledgements}


\begin{appendix}

\section{Images}
\label{appendix}

\begin{figure*}
 \includegraphics[width=\textwidth]{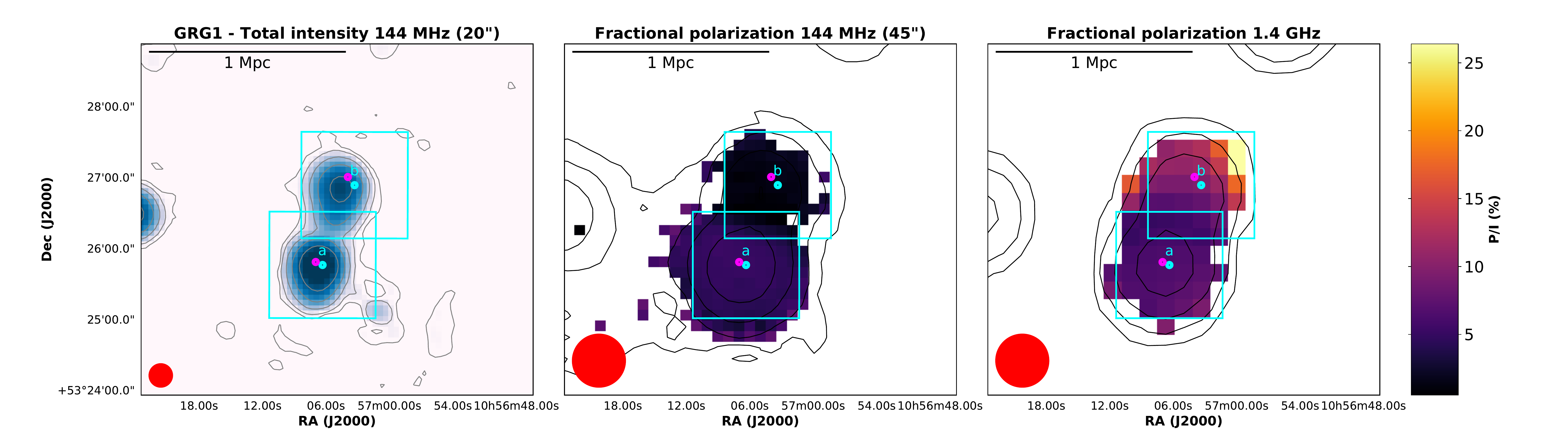}
 \includegraphics[width=\textwidth]{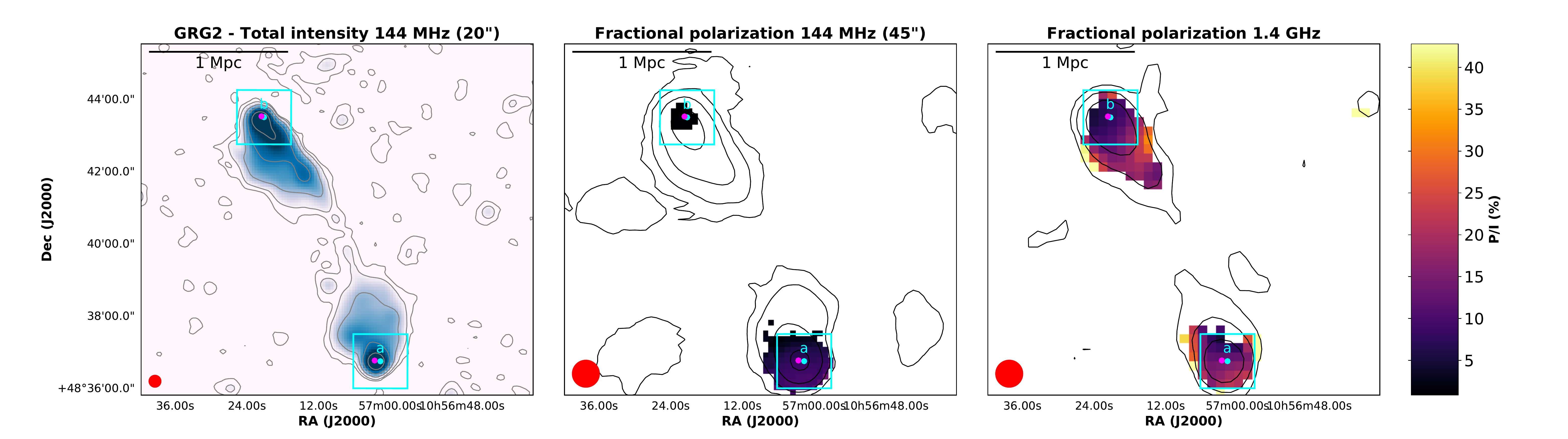}
 \includegraphics[width=\textwidth]{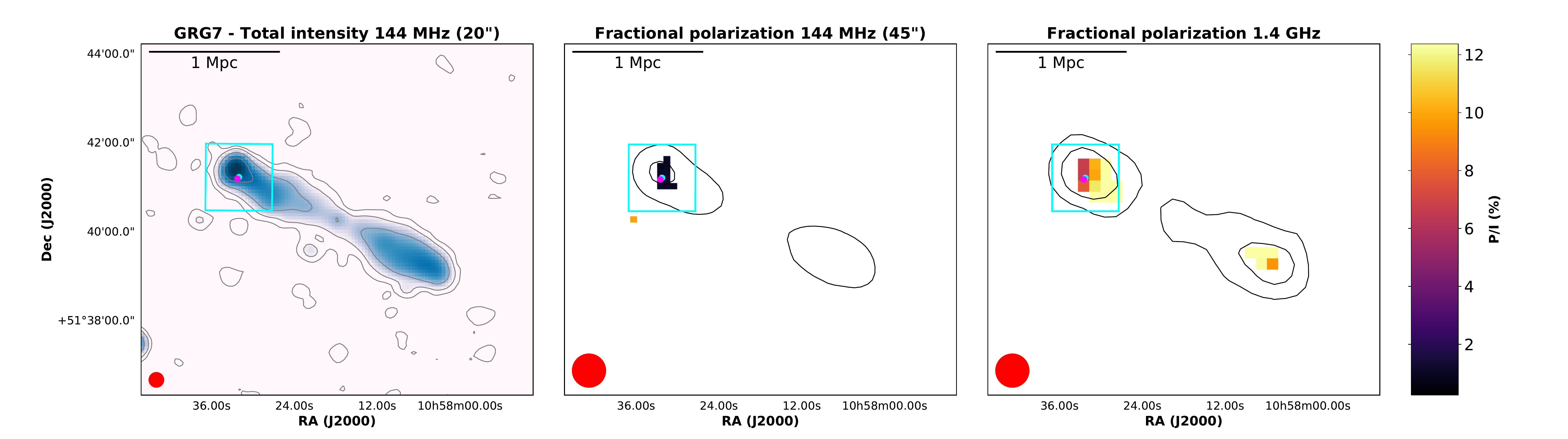}
 \includegraphics[width=\textwidth]{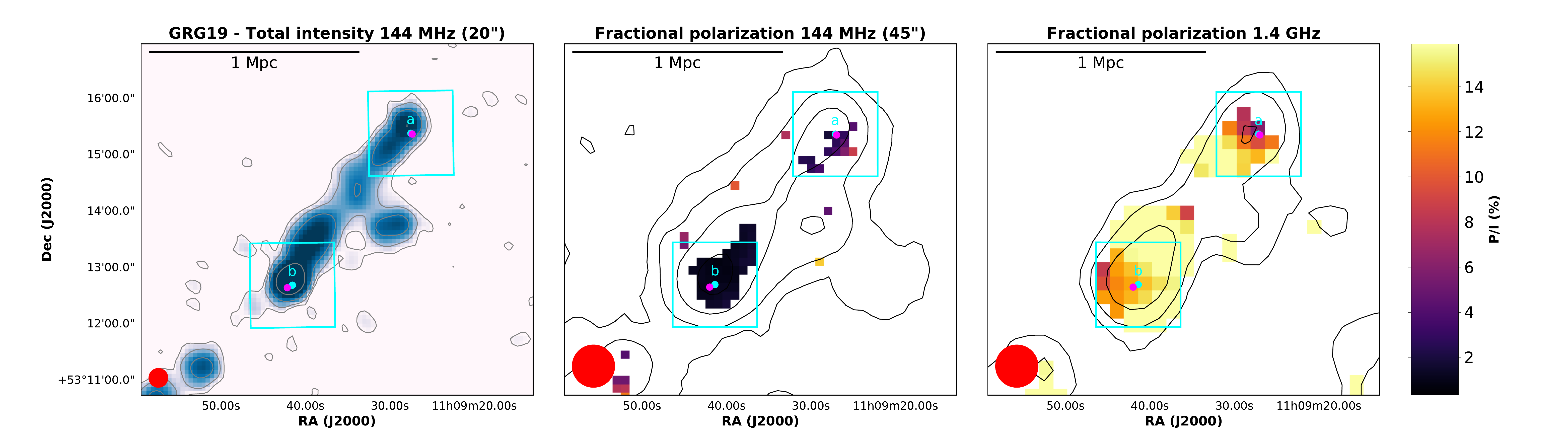}
 \caption{Images of the GRGs detected in polarization. Left: LoTSS total intensity image at 20$\arcsec$ resolution with contours overlaid. Contours start at 3$\sigma$ noise level and are spaced by a factor of four (with $\sigma$ ranging between 0.09 and 0.9 mJy/beam). Center: LoTSS fractional polarization at 45$\arcsec$ resolution with total intensity contours overlaid. Contours start at 3$\sigma$ noise level and are spaced by a factor of four (with $\sigma$ ranging between 0.1 and 8 mJy/beam). Only pixels above the $8\sigma_{QU}$ detection threshold in polarization are shown (except for GRG\,78, for which the threshold is at 7 times $\sigma_{QU}$, and GRG\,80 and GRG\,87, for which it is 6$\sigma_{QU}$). Right: NVSS fractional polarization with total intensity contours overlaid. Contours start at 3$\sigma$ noise level and are spaced by a factor of four (with $\sigma$ ranging between 0.2 and 0.7 mJy/beam). Only pixels with signal-to-noise higher than three in polarization are shown. The color scale and limits are the same in both $P$/$I$ images for each source. The cyan squares mark the component detected at 20$\arcsec$, cyan points mark the peak of polarized intensity at 20$\arcsec$ (RM and fractional polarization values at this position are listed in Tab.~\ref{tab:pol_double} and Tab.~\ref{tab:pol_single}) while magenta points mark the position where we computed the depolarization factors. Letters mark the two components listed in Tab.~\ref{tab:pol_double} for double-lobed detected sources.}
 \label{fig:GRG1}
\end{figure*}

The images of all the GRGs detected in polarization are shown in Fig.~\ref{fig:GRG1}. We show the total intensity images at 20$\arcsec$ resolution and the fractional polarization images at $45\arcsec$ compared with the NVSS fractional polarization images at 1.4 GHz. In some cases the detected regions appear as a few scattered pixels that are not beam-shaped. This is a consequence of having peak polarized intensities very close to the detection threshold cutoff.

\renewcommand{\thefigure}{A\arabic{figure} (continued)}
\addtocounter{figure}{-1}
\begin{figure*}
\includegraphics[width=\textwidth]{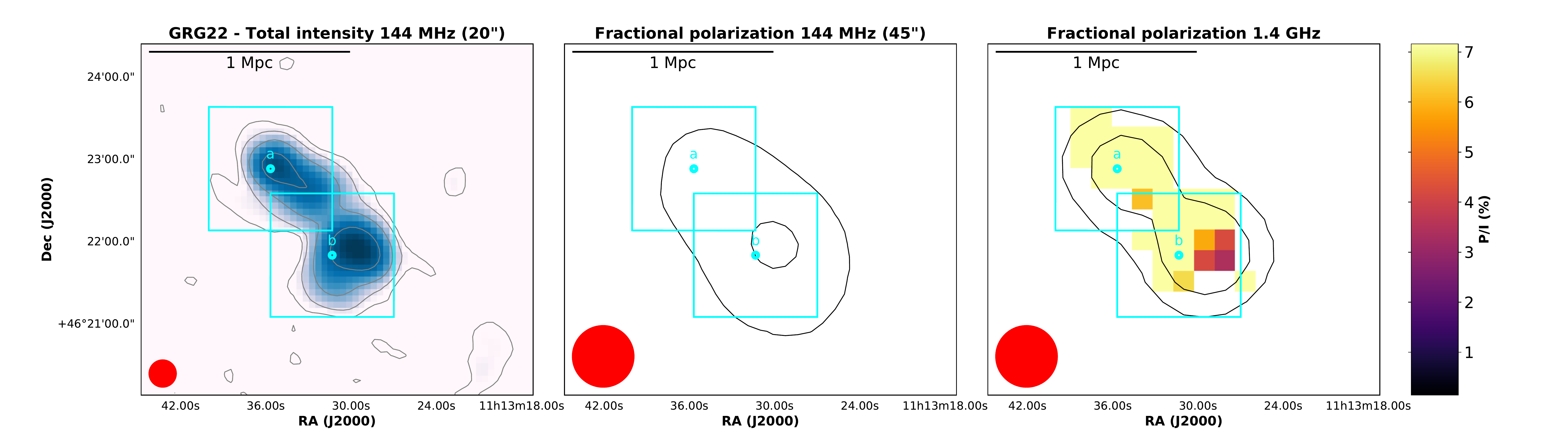}
\includegraphics[width=\textwidth]{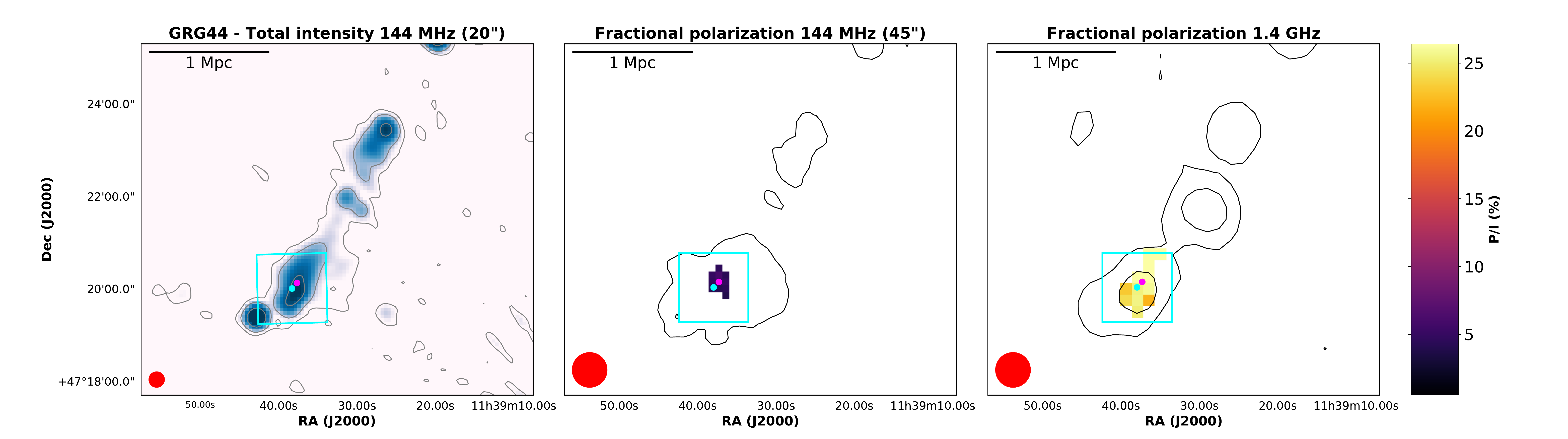}
\includegraphics[width=\textwidth]{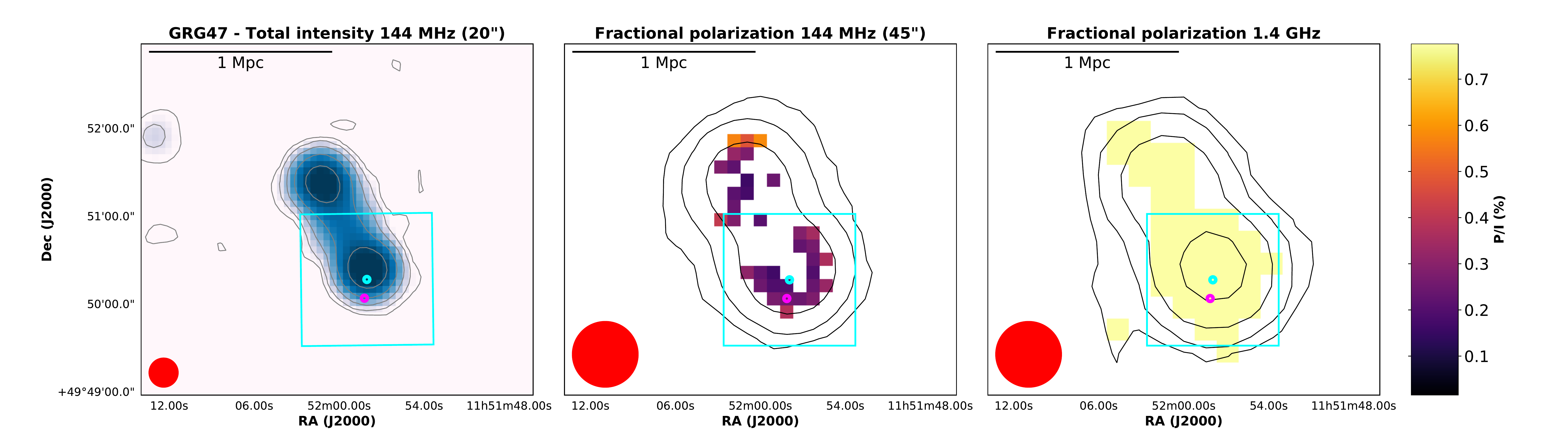}
\includegraphics[width=\textwidth]{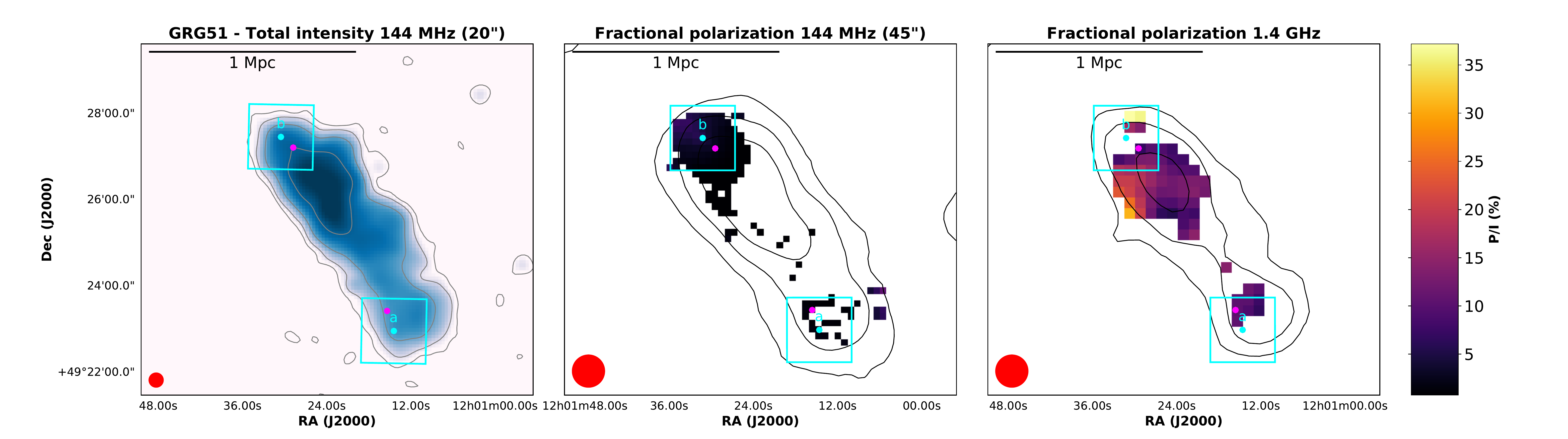}
 \caption{}
 \label{fig:GRG2}
\end{figure*}
\renewcommand{\thefigure}{A\arabic{figure} (continued)}
\addtocounter{figure}{-1}
\begin{figure*}
\includegraphics[width=\textwidth]{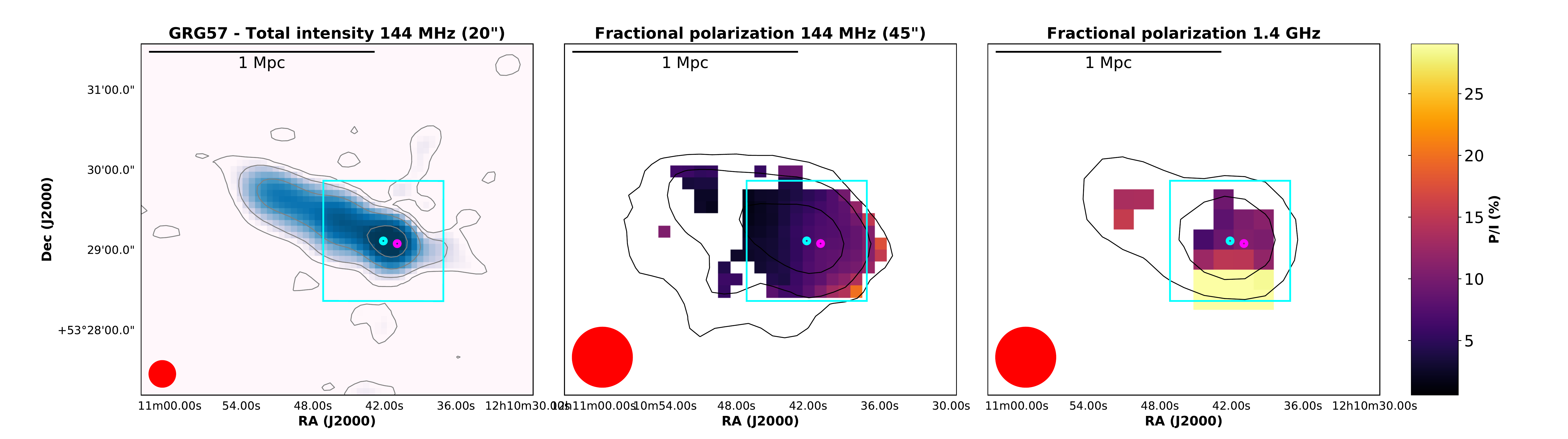}
\includegraphics[width=\textwidth]{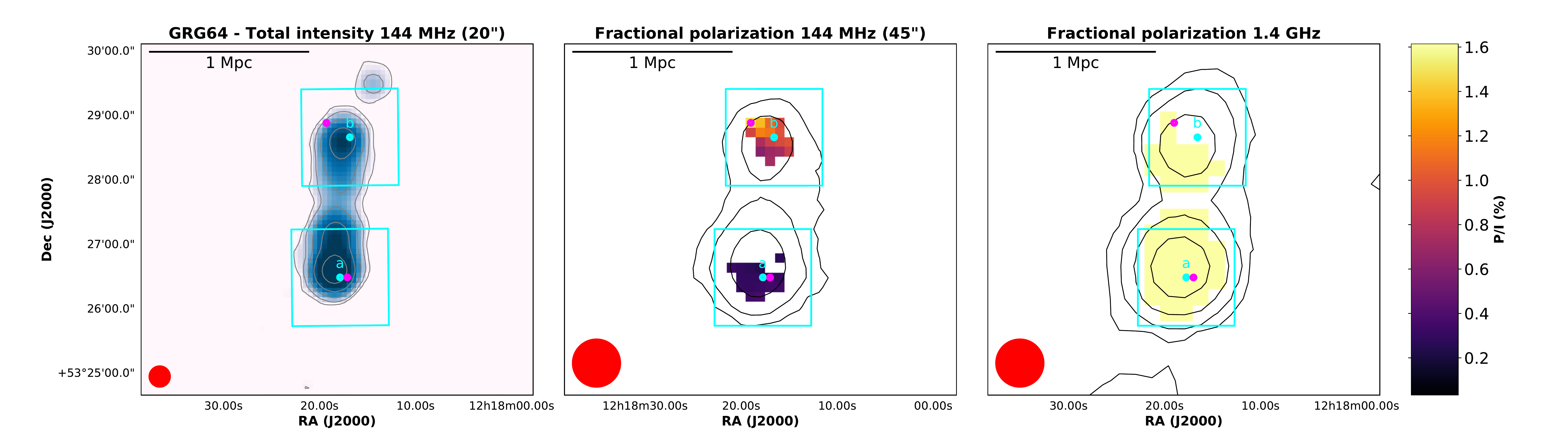}
\includegraphics[width=\textwidth]{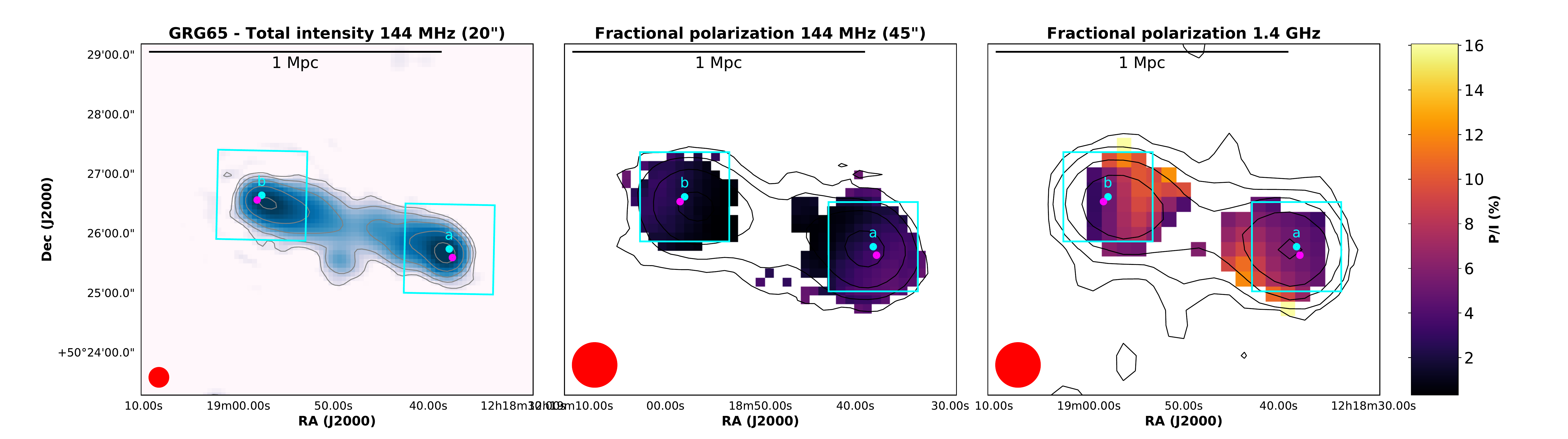}
\includegraphics[width=\textwidth]{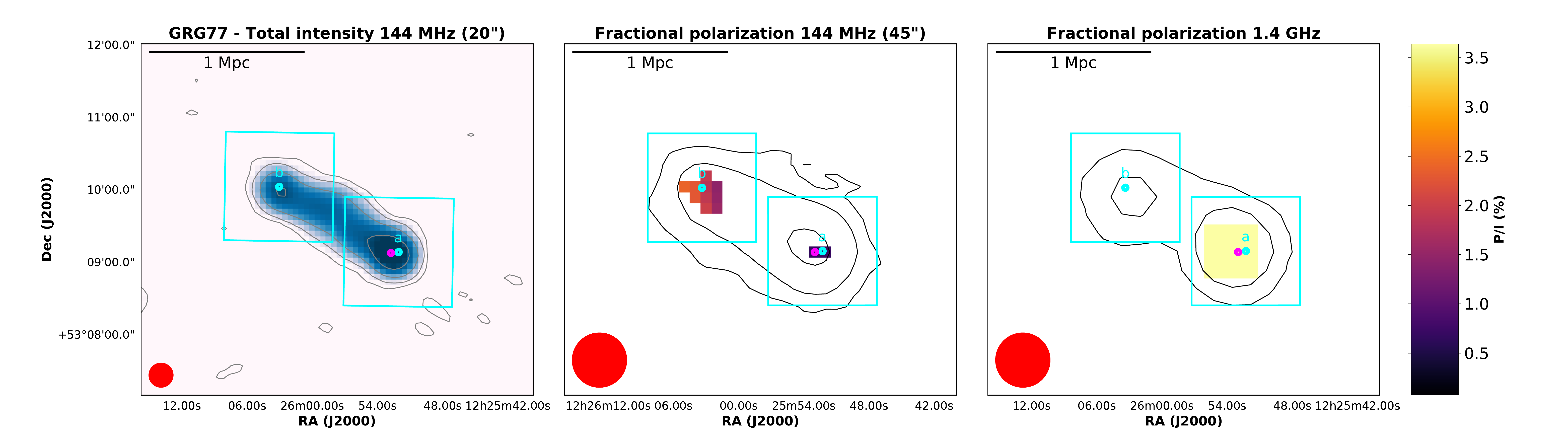}
 \caption{}
 \label{fig:GRG3}
\end{figure*}
\renewcommand{\thefigure}{A\arabic{figure} (continued)}
\addtocounter{figure}{-1}
\begin{figure*}
\includegraphics[width=\textwidth]{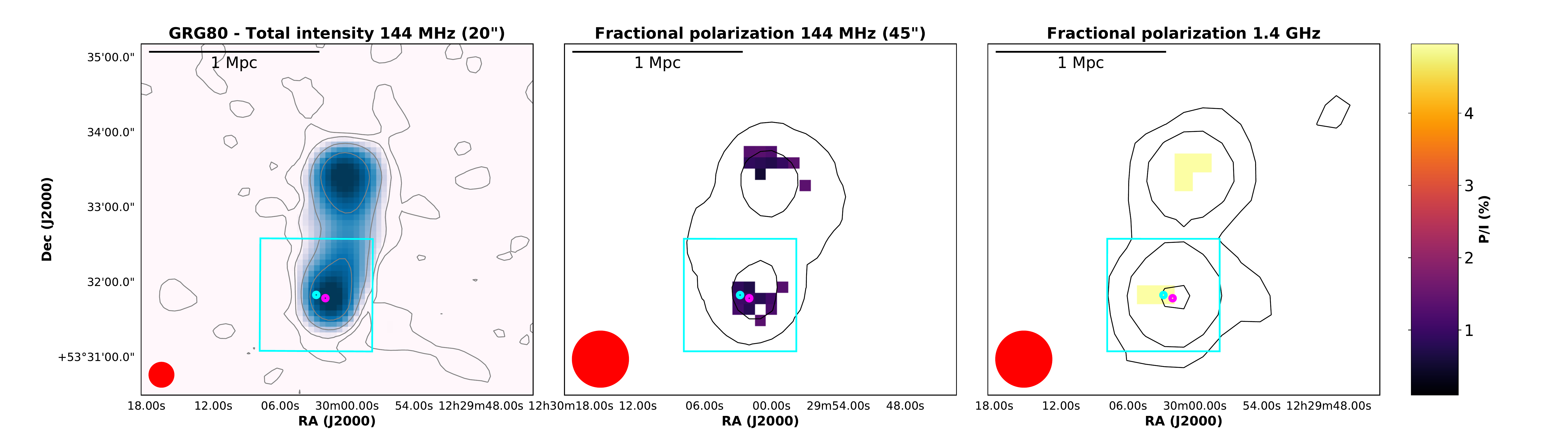}
\includegraphics[width=\textwidth]{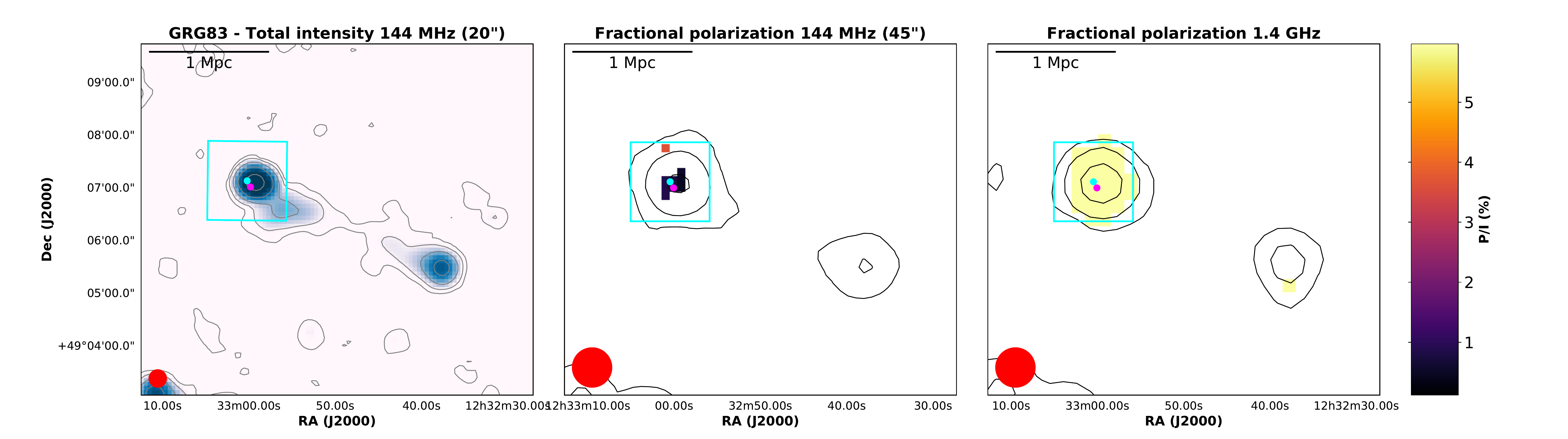}
\includegraphics[width=\textwidth]{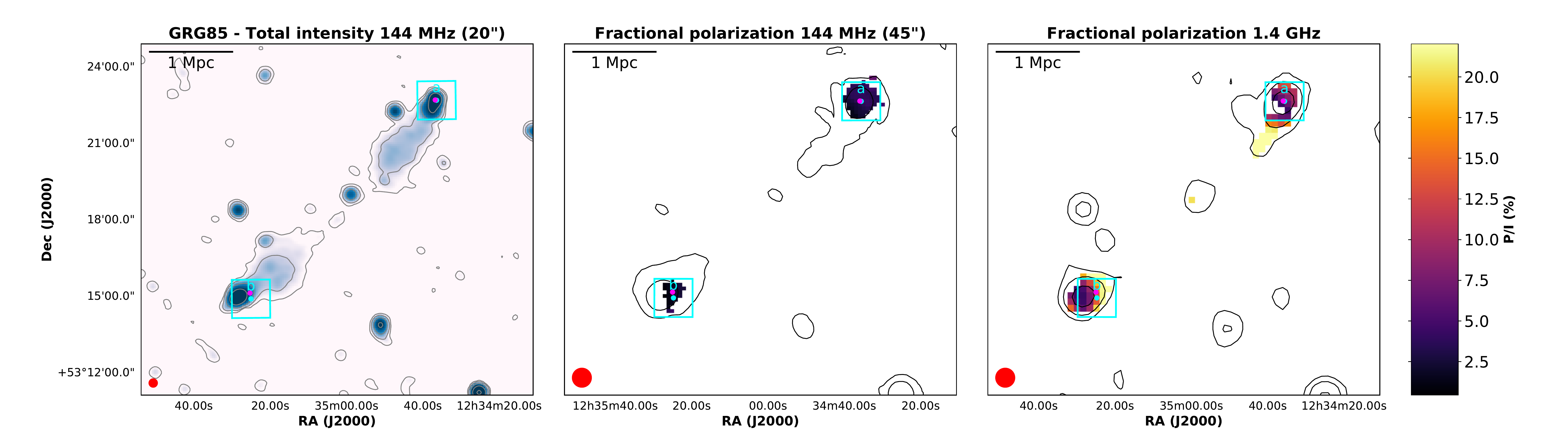}
 \includegraphics[width=\textwidth]{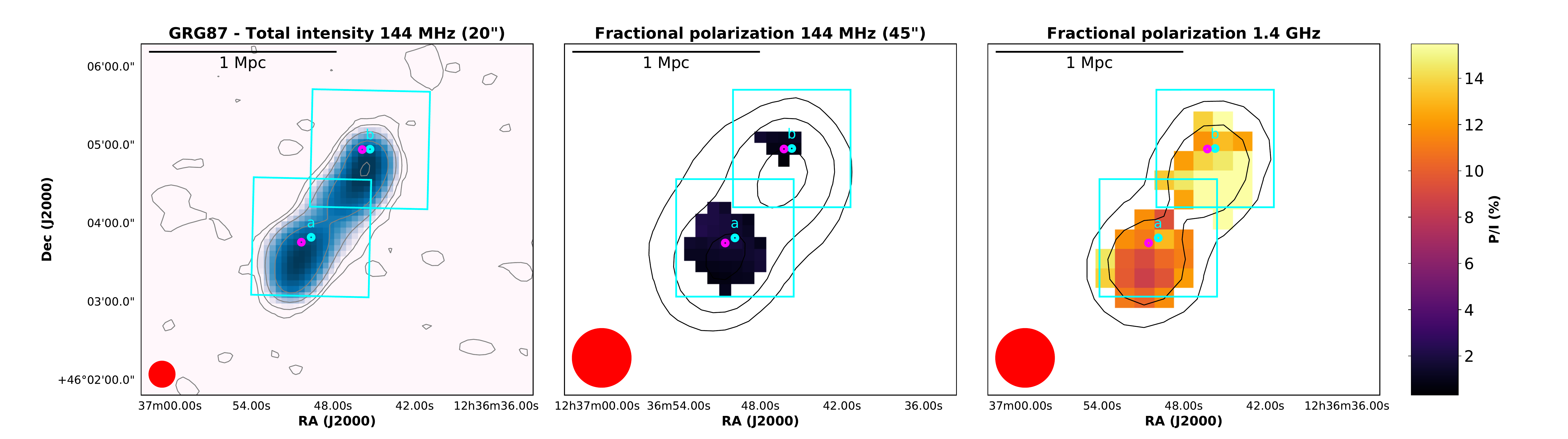}
 \caption{}
 \label{fig:GRG4}
\end{figure*}
\renewcommand{\thefigure}{A\arabic{figure} (continued)}
\addtocounter{figure}{-1}
\begin{figure*}
 \includegraphics[width=\textwidth]{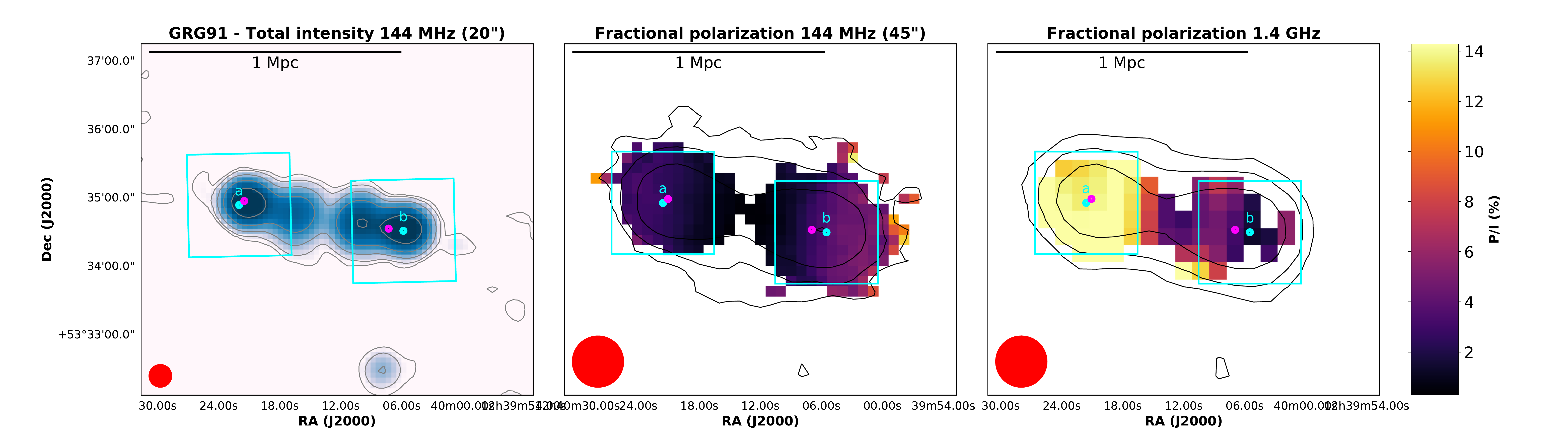}
 \includegraphics[width=\textwidth]{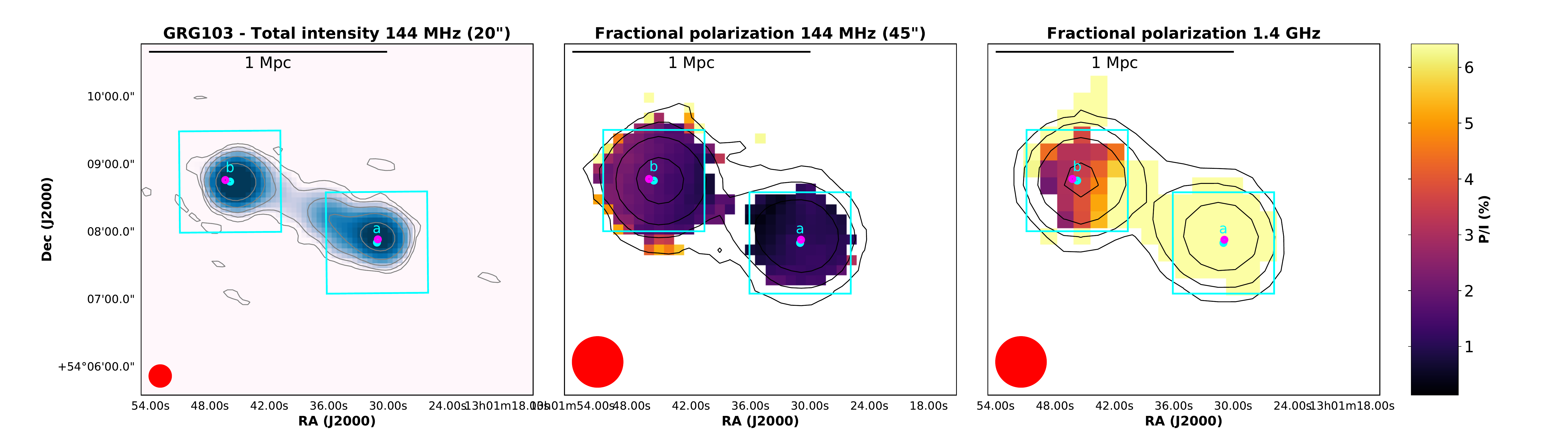}
 \includegraphics[width=\textwidth]{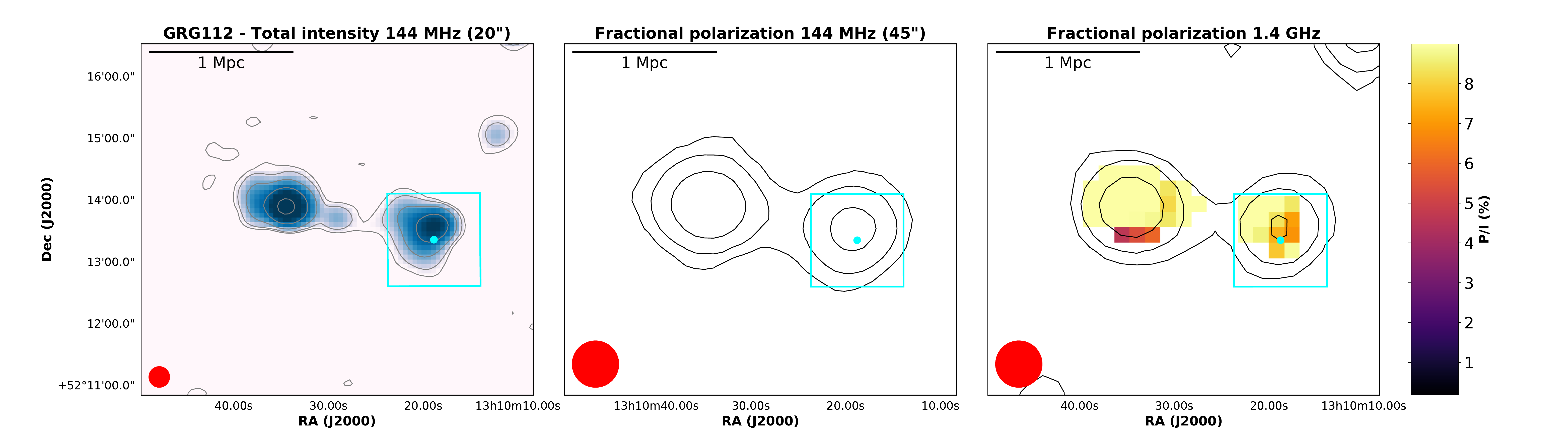}
 \includegraphics[width=\textwidth]{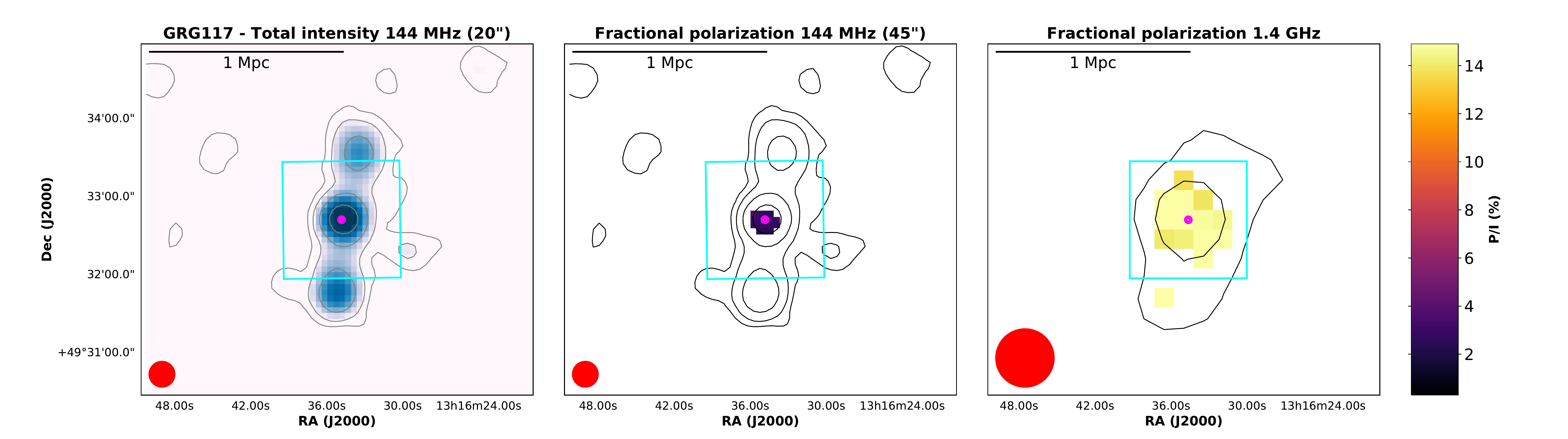}
 \caption{}
 \label{fig:GRG5}
\end{figure*}
\renewcommand{\thefigure}{A\arabic{figure} (continued)}
\addtocounter{figure}{-1}
\begin{figure*}
  \includegraphics[width=\textwidth]{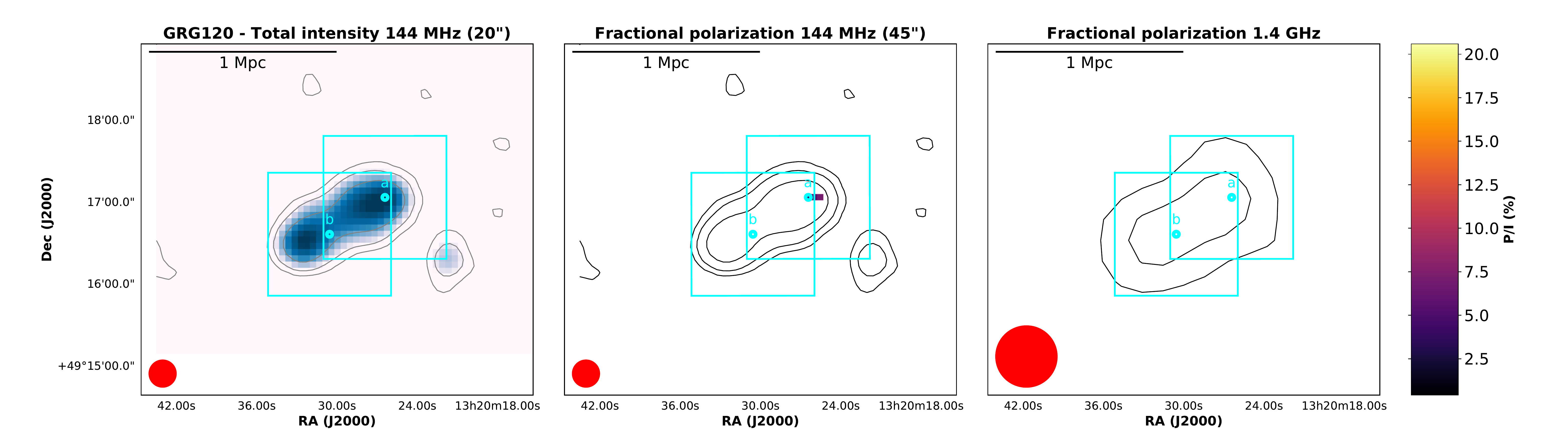}
  \includegraphics[width=\textwidth]{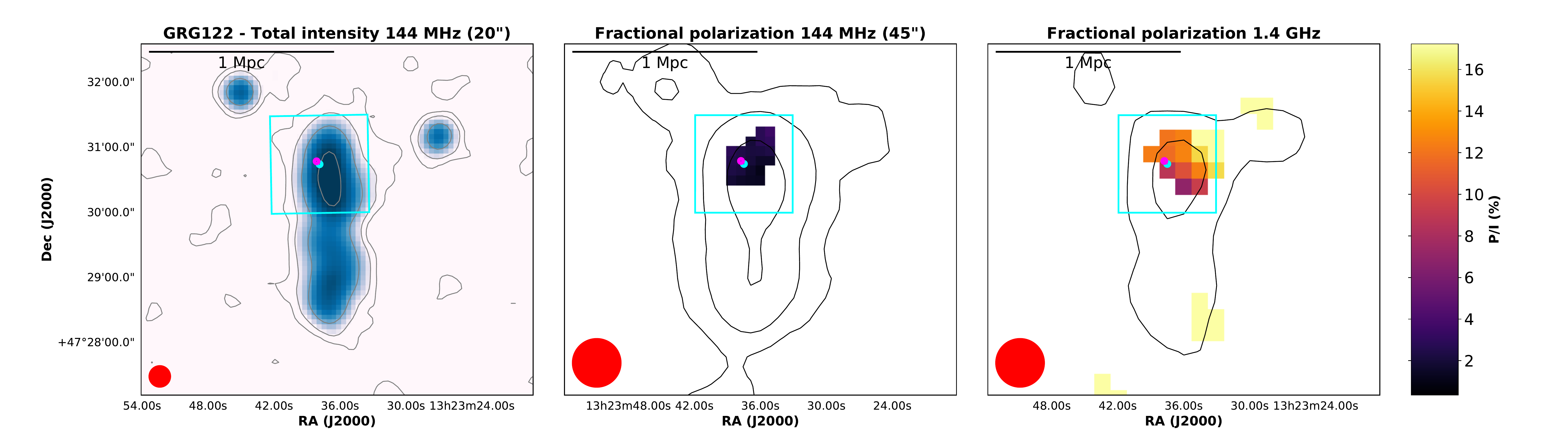}
  \includegraphics[width=\textwidth]{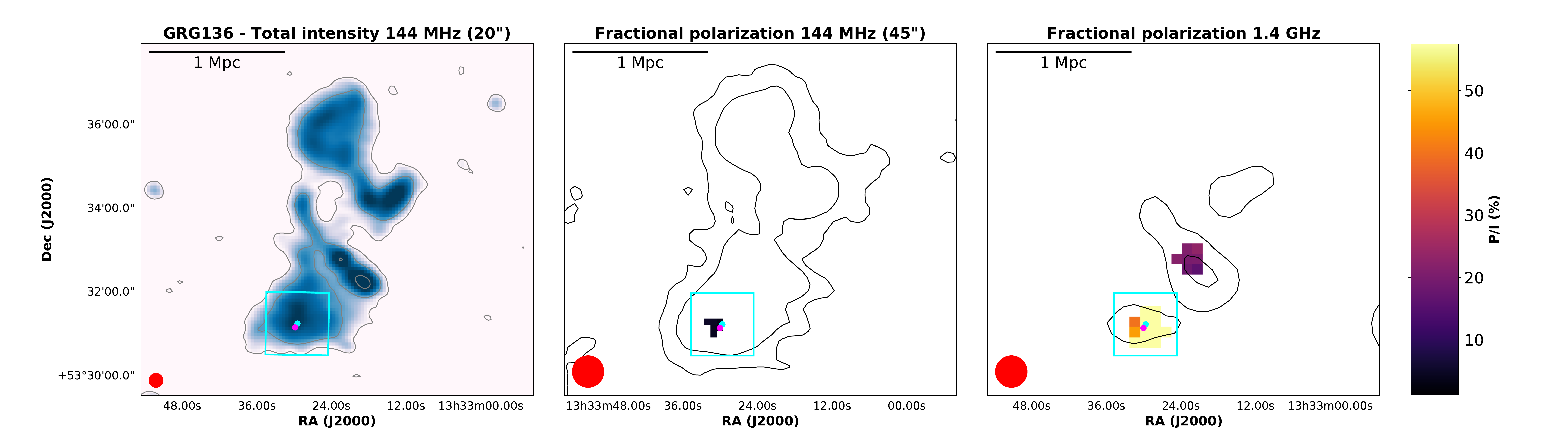}
  \includegraphics[width=\textwidth]{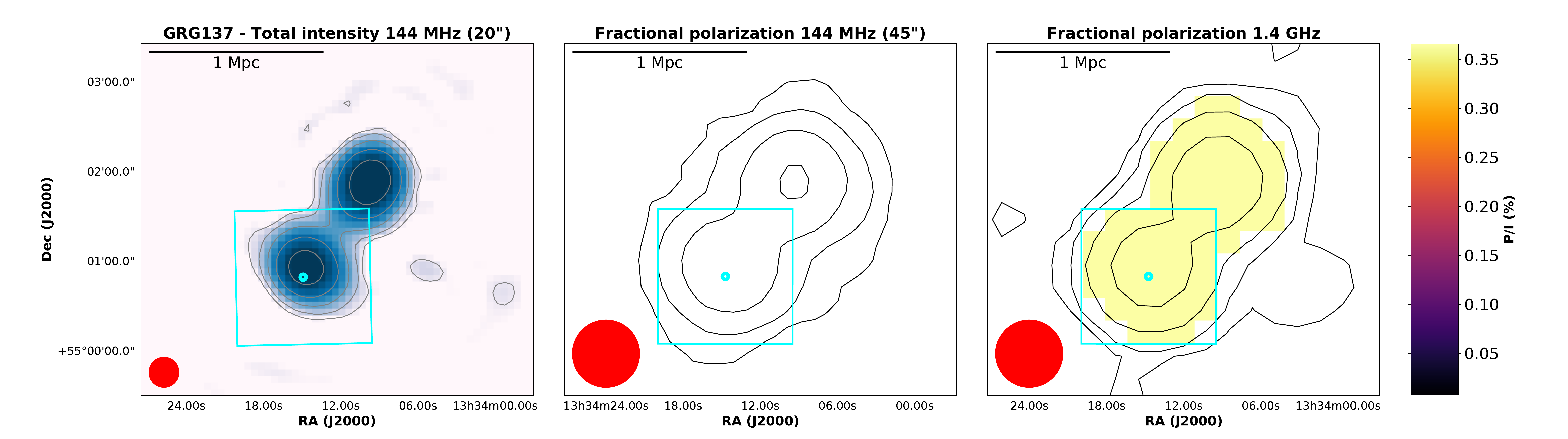}
  \caption{}
 \label{fig:GRG6}
\end{figure*}
\renewcommand{\thefigure}{A\arabic{figure} (continued)}
\addtocounter{figure}{-1}
\begin{figure*}
 \includegraphics[width=\textwidth]{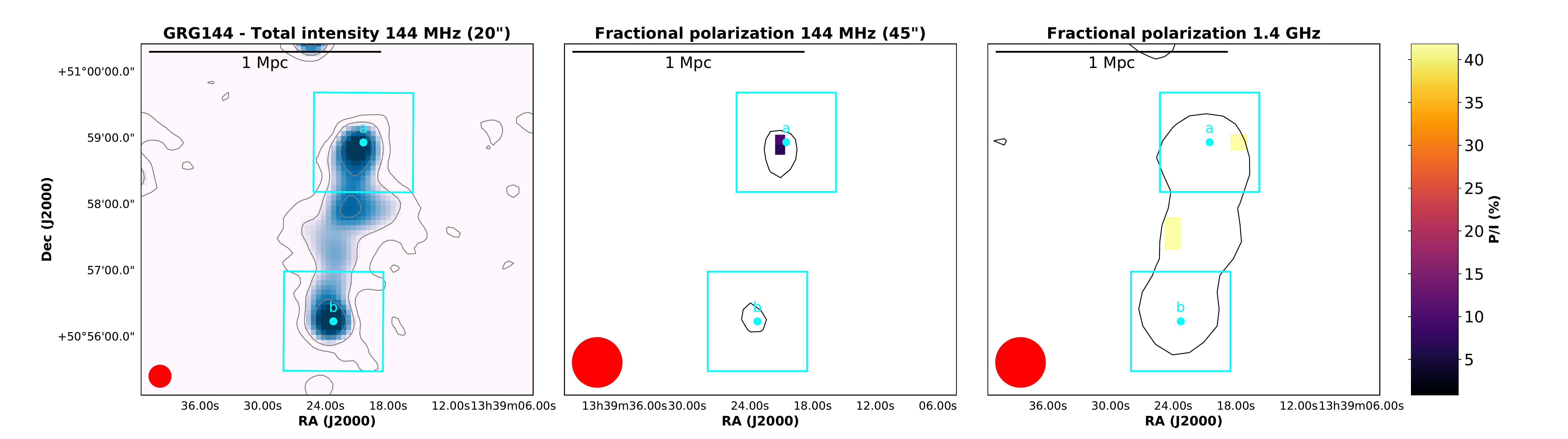}
   \includegraphics[width=\textwidth]{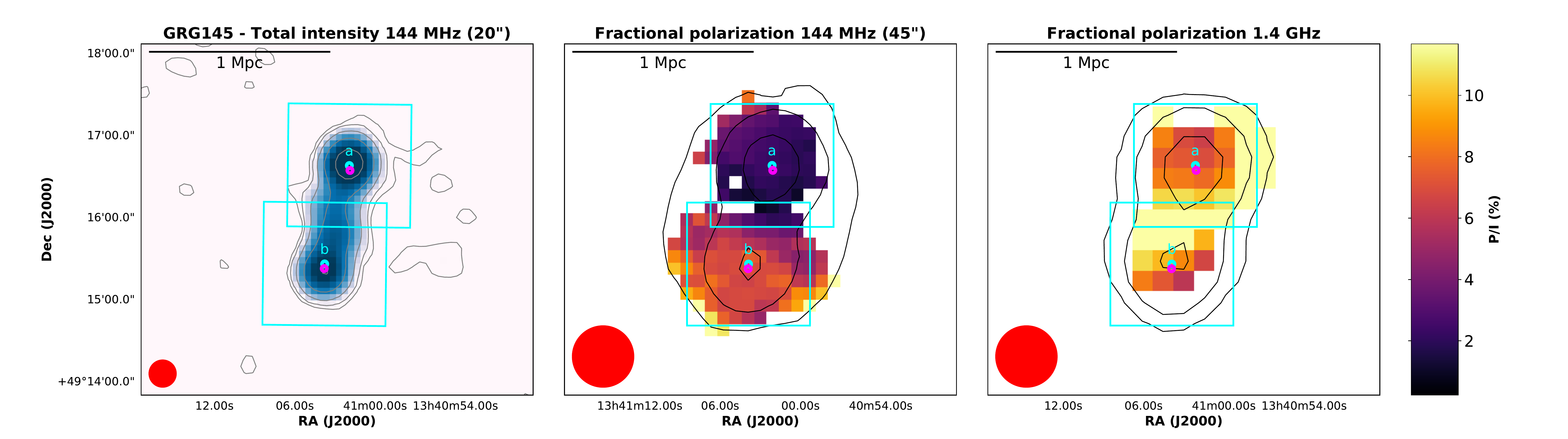}
   \includegraphics[width=\textwidth]{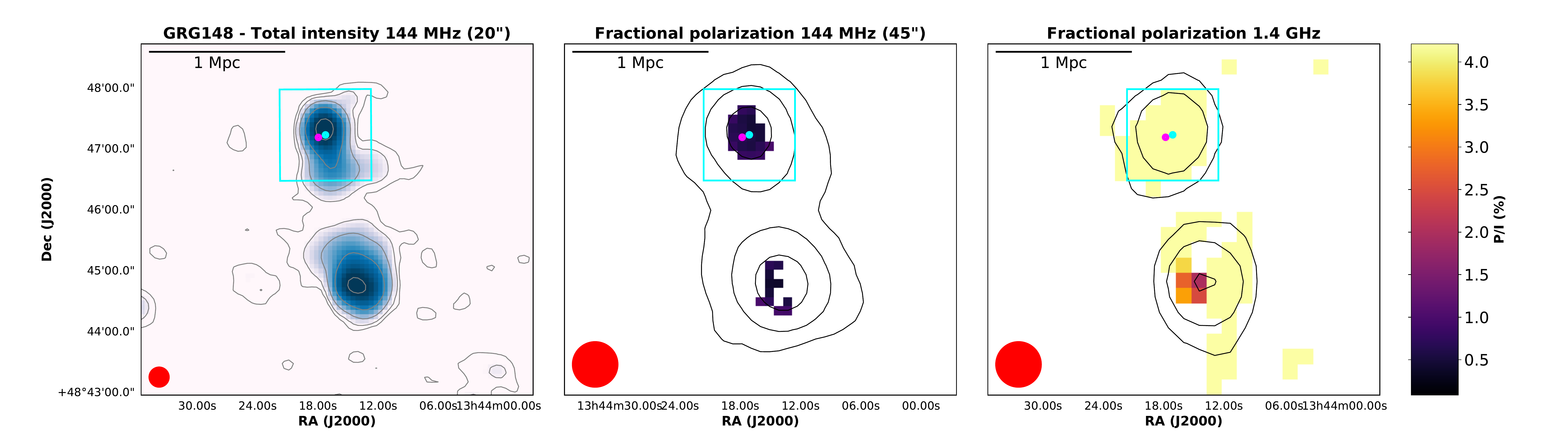}
   \includegraphics[width=\textwidth]{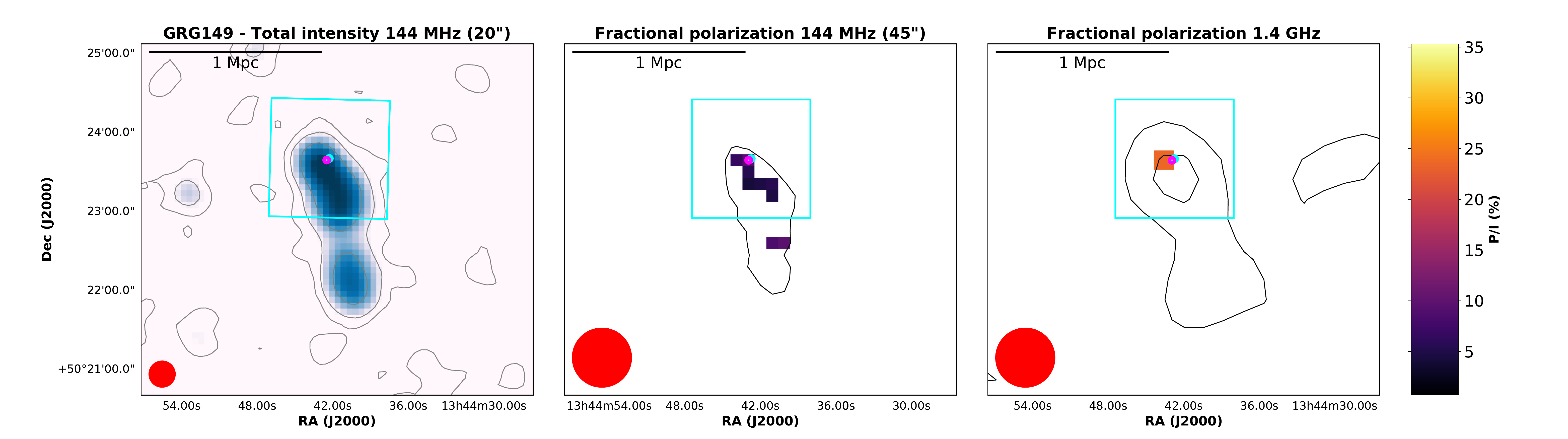}
    \caption{}
 \label{fig:GRG7}
\end{figure*}
\renewcommand{\thefigure}{A\arabic{figure} (continued)}
\addtocounter{figure}{-1}
\begin{figure*}
   \includegraphics[width=\textwidth]{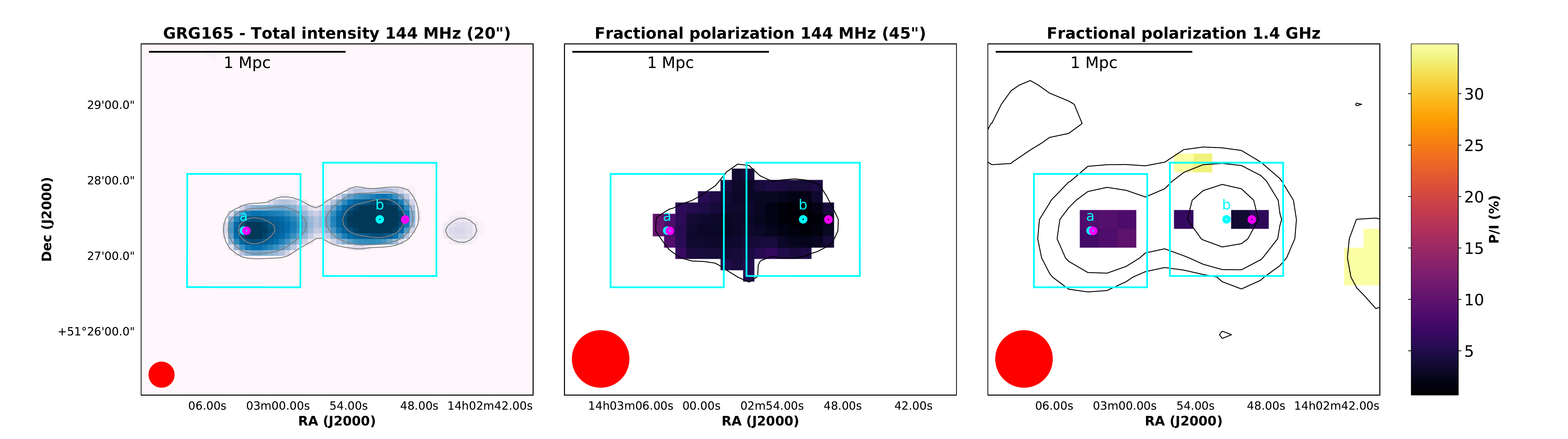}
    \includegraphics[width=\textwidth]{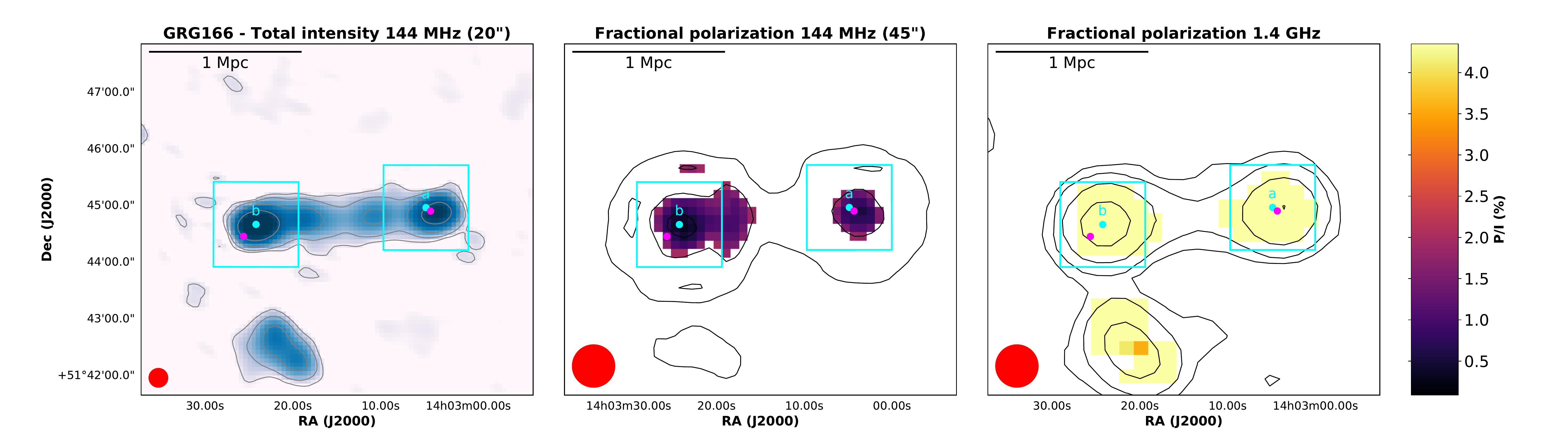}
    \includegraphics[width=\textwidth]{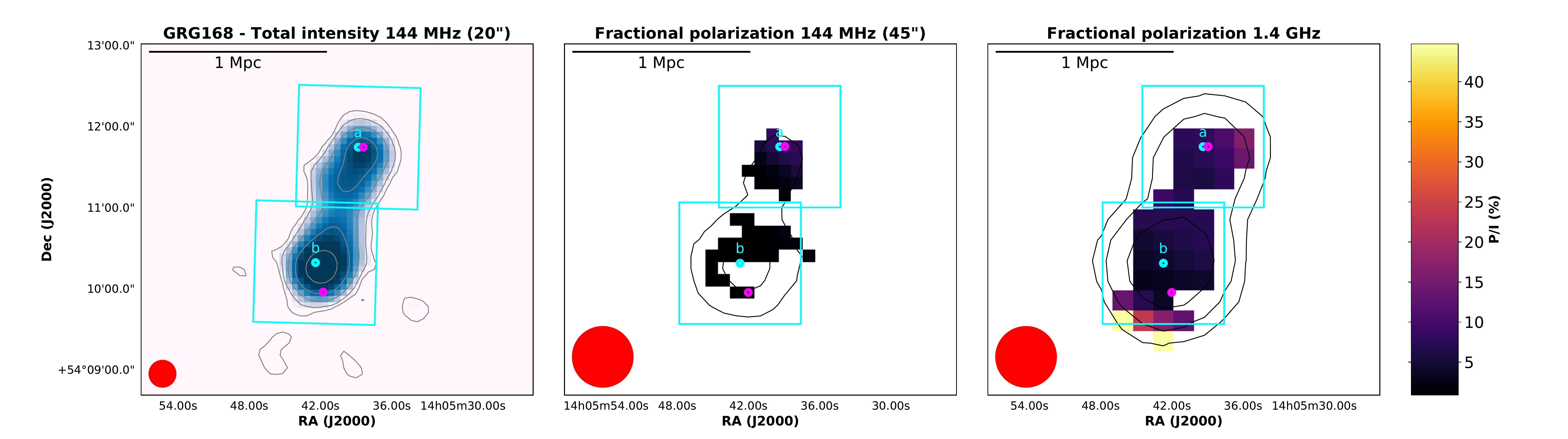}
    \includegraphics[width=\textwidth]{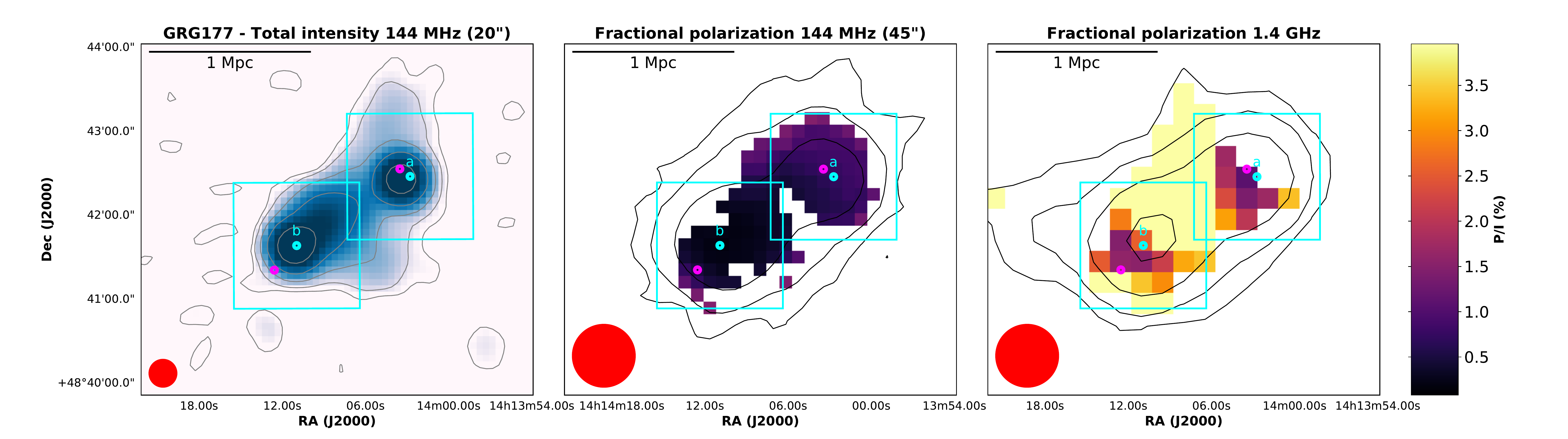}
     \caption{}
 \label{fig:GRG8}
\end{figure*}
\renewcommand{\thefigure}{A\arabic{figure} (continued)}
\addtocounter{figure}{-1}
\begin{figure*}
    \includegraphics[width=\textwidth]{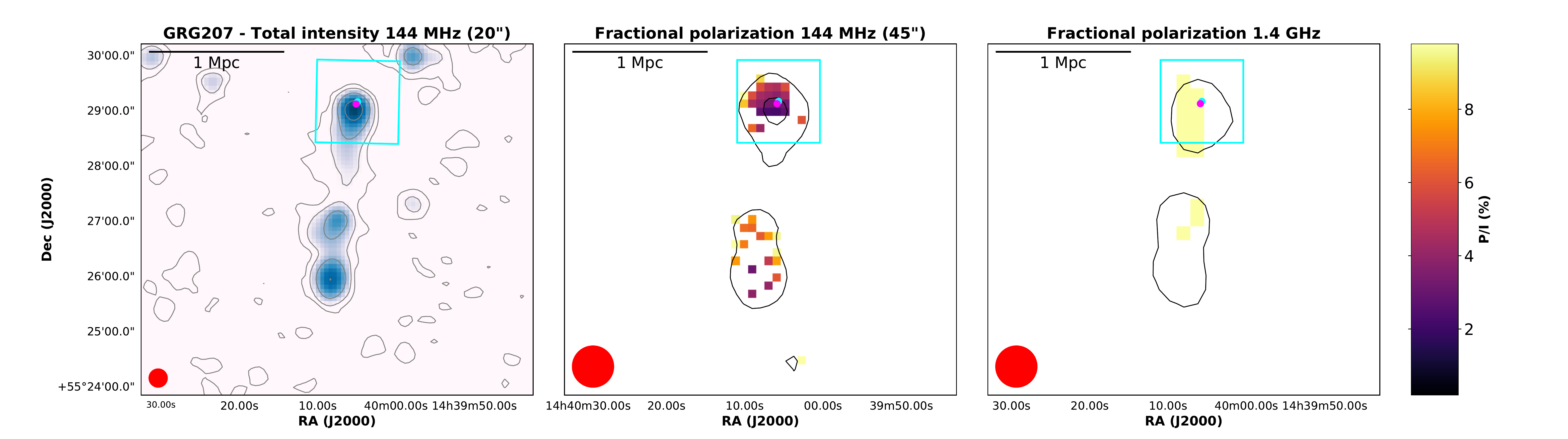}
     \includegraphics[width=\textwidth]{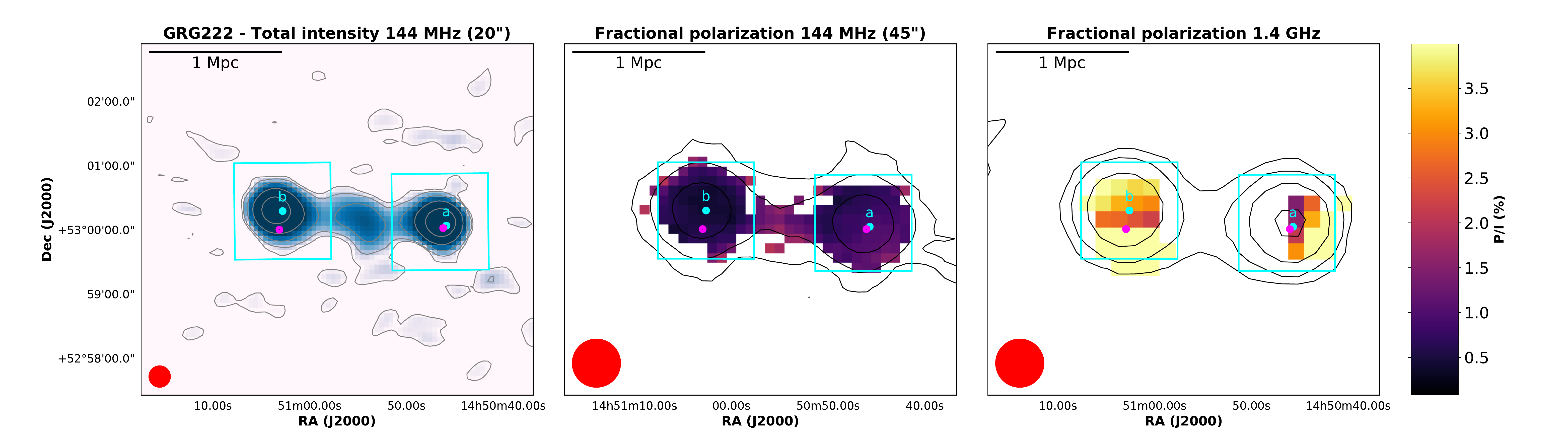}
     \includegraphics[width=\textwidth]{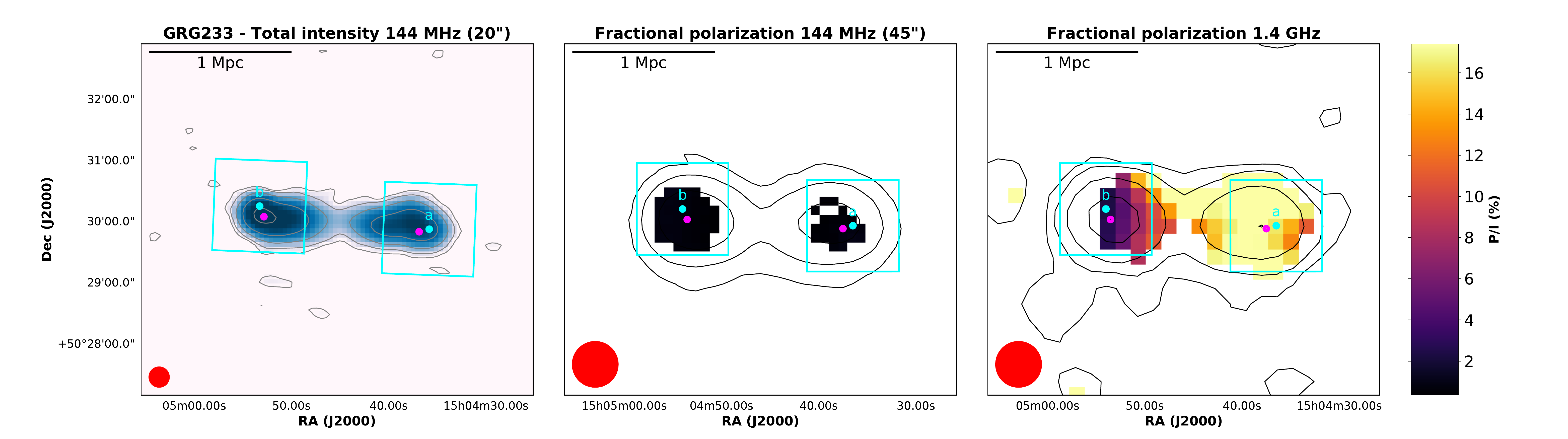}
     \includegraphics[width=\textwidth]{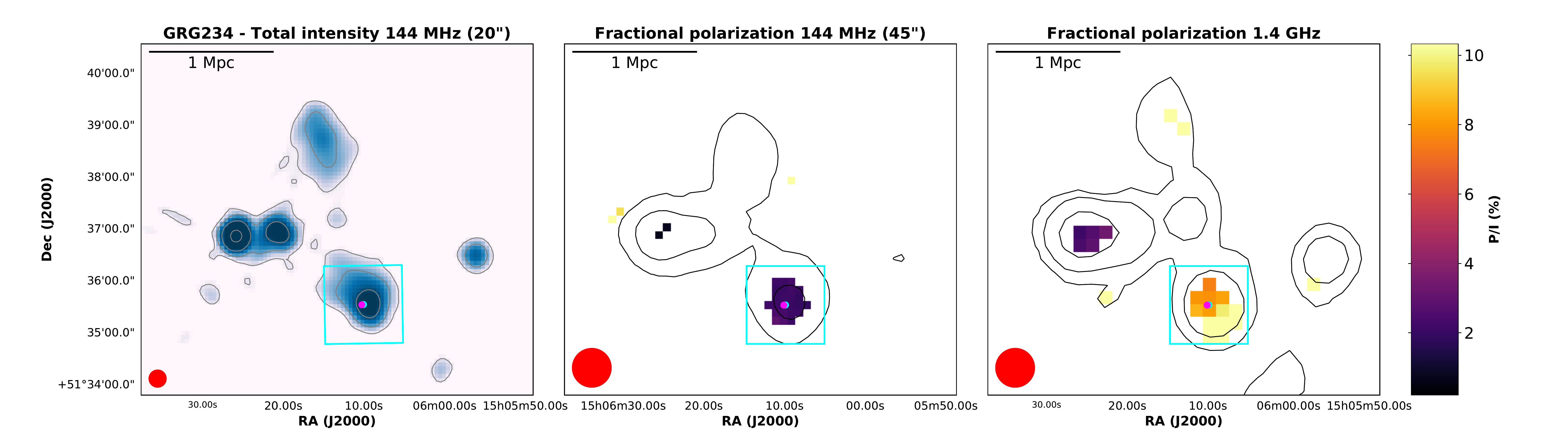}
\caption{}
 \label{fig:GRG9}
\end{figure*}
\renewcommand{\thefigure}{A\arabic{figure} (continued)}
\addtocounter{figure}{-1}
\begin{figure*}
     \includegraphics[width=\textwidth]{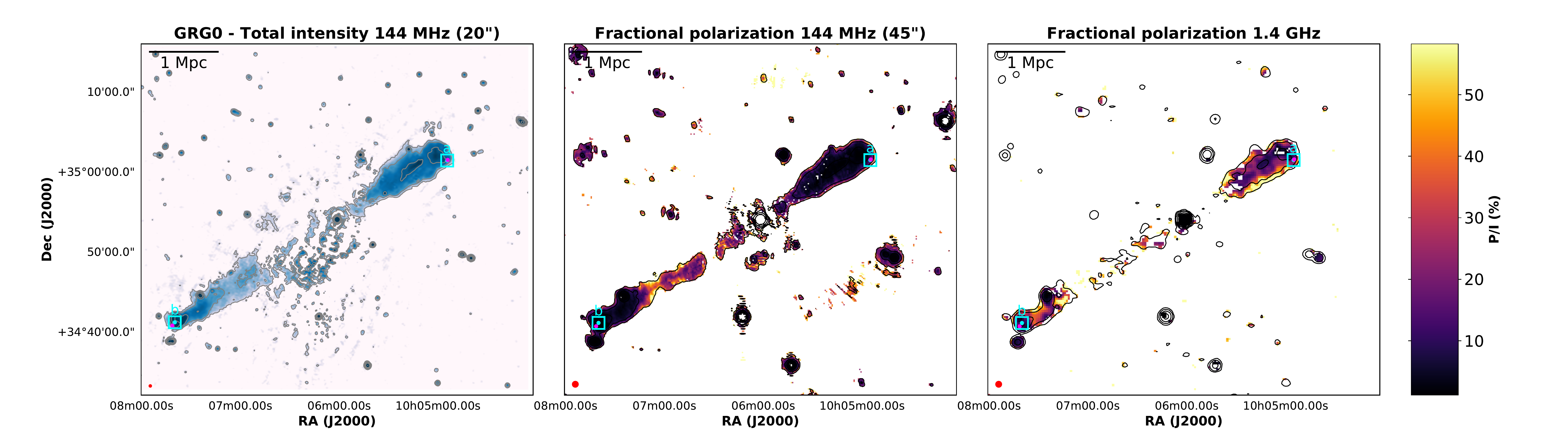}
          \caption{}
 \label{fig:GRG}
\end{figure*}
\renewcommand{\thefigure}{\arabic{figure}}
\end{appendix}



\bibliographystyle{aa}
\bibliography{GRG_pol} 


\end{document}